\documentclass[sigconf]{acmart}

\AtBeginDocument{%
	}

\usepackage{graphicx}
\usepackage{booktabs}
\usepackage{longtable}

\usepackage{wrapfig}
\usepackage{graphicx}

\usepackage{multirow}
\usepackage{makecell}
\usepackage{subfig}

\usepackage{bm}
\usepackage{graphicx}

\usepackage{algorithm}
\usepackage{algorithmic}
\newcommand{\alg}{CrossFire}

\newcommand{\myparatight}[1]{\smallskip\noindent{\bf {#1}:}~}

\copyrightyear{2024}
\acmYear{2024}
\setcopyright{rightsretained}
\acmConference[LAMPS '24]{Proceedings of the 1st ACM Workshop on Large AI Systems and Models with Privacy and Safety Analysis}{October 14--18, 2024}{Salt Lake City, UT, USA} 
\acmBooktitle{Proceedings of the 1st ACM Workshop on Large AI Systems and Models with Privacy and Safety Analysis (LAMPS '24), October 14--18, 2024, Salt Lake City, UT, USA} 
\acmDOI{10.1145/3689217.3690619} 
\acmISBN{979-8-4007-1209-8/24/10}

\makeatletter
\gdef\@copyrightpermission{
  \begin{minipage}{0.3\columnwidth}
   \href{https://creativecommons.org/licenses/by/4.0/}{\includegraphics[width=0.90\textwidth]{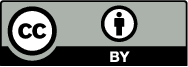}}
  \end{minipage}\hfill
  \begin{minipage}{0.7\columnwidth}
   \href{https://creativecommons.org/licenses/by/4.0/}{This work is licensed under a Creative Commons Attribution International 4.0 License.}
  \end{minipage}
  \vspace{5pt}
}
\makeatother

\begin{document}


\title{Adversarial Attacks to Multi-Modal Models}

\author{Zhihao Dou}
\authornote{Equal contribution. Zhihao Dou and Xin Hu conducted this research while they were interns under the supervision of Minghong Fang.}
\affiliation{
	\institution{Duke University}
	\city{Durham}
	\state{NC}
	\country{USA}
}
\email{zd55@duke.edu}

\author{Xin Hu}
\authornotemark[1]
\affiliation{
	\institution{The University of Tokyo}
	\city{Kashiwa}
	\state{Chiba Prefecture}
	\country{Japan}
}
\email{sthhx@g.ecc.u-tokyo.ac.jp}

\author{Haibo Yang}
\affiliation{
	\institution{Rochester Institute of Technology}
	\city{Rochester}
	\state{NY}
	\country{USA}
}
\email{hbycis@rit.edu}

\author{Zhuqing Liu}
\affiliation{
	\institution{University of North Texas}
	\city{Denton}
	\state{TX}
	\country{USA}
}
\email{zhuqing.liu@unt.edu}

\author{Minghong Fang}
\affiliation{
	\institution{University of Louisville}
	\city{Louisville}
	\state{KY}
	\country{USA}
}
\email{minghong.fang@louisville.edu}

\begin{abstract}
Multi-modal models have gained significant attention due to their powerful capabilities. These models effectively align embeddings across diverse data modalities, showcasing superior performance in downstream tasks compared to their unimodal counterparts. Recent study showed that the attacker can manipulate an image or audio file by altering it in such a way that its embedding matches that of an attacker-chosen targeted input, thereby deceiving downstream models. However, this method often underperforms due to inherent disparities in data from different modalities. In this paper, we introduce CrossFire, an innovative approach to attack multi-modal models. CrossFire begins by transforming the targeted input chosen by the attacker into a format that matches the modality of the original image or audio file. We then formulate our attack as an optimization problem, aiming to minimize the angular deviation between the embeddings of the transformed input and the modified image or audio file. Solving this problem determines the perturbations to be added to the original media. Our extensive experiments on six real-world benchmark datasets reveal that CrossFire can significantly manipulate downstream tasks, surpassing existing attacks. Additionally, we evaluate six defensive strategies against CrossFire, finding that current defenses are insufficient to counteract our CrossFire. 

\end{abstract}



\maketitle


\section{Introduction} \label{sec:intro}

Multi-modal models have garnered significant attention owing to their impressive performance across diverse applications, including speech recognition~\cite{afouras2018deep} and zero-shot classification~\cite{jeong2023winclip}. These models are trained using multi-modal data tuples, encompassing pairs of different data modalities. In comparison to unimodal learning, multi-modal models offer superior performance in downstream tasks, as evidenced by improved results in various applications~\cite{liang2022foundations,huang2021makes}. For instance, in medical image analysis, a unimodal model relying solely on visual information may face challenges when diagnosing certain conditions. Incorporating textual reports or clinical notes as an additional modality alongside the medical images can provide crucial context and improve the model's ability to make accurate and informed diagnoses, especially in cases where visual cues alone may be ambiguous or insufficient.

While multi-modal models have gained considerable research attention for their capabilities, their security is also a critical concern.
In a recent study, Zhang et al.~\cite{bagdasaryan2023ceci} presented a new form of adversarial attack in which the attacker intentionally aligns inputs from various modalities to misguide downstream models. This technique allows the attacker to mislead the downstream models by deliberately aligning inputs from diverse modalities.  More precisely, the attacker chooses a specific \emph{targeted input} (for example, a piece of text) and another image/audio. The attacker then introduces adversarial perturbation to the image/audio, resulting in a perturbed image/audio.  This perturbed image/audio is manipulated in such a way that it aligns with the targeted input in the embedding space. As a result, when downstream tasks utilize this perturbed image/audio for data generation, the output aligns with the attacker's intentions, resembling the targeted input closely.
However, due to the inherent disparities in data attributes across different modalities, the attack introduced in~\cite{bagdasaryan2023ceci} achieves unsatisfying performance as shown in our later experiments.

\begin{figure*}[!t]
	\centering
	{\includegraphics[width= 0.98\textwidth]{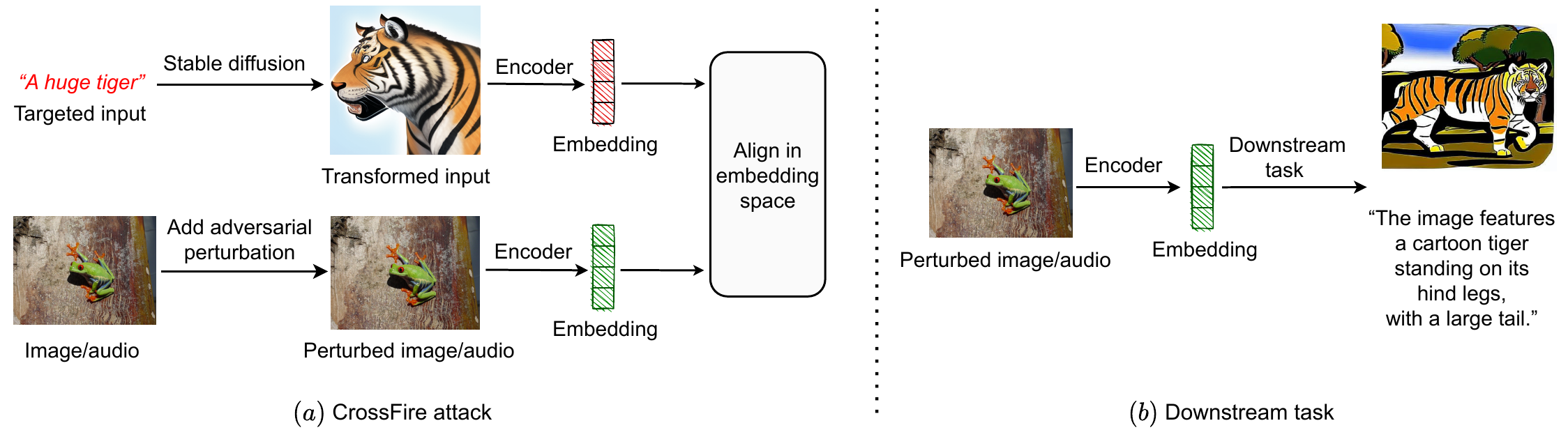}}
	\caption{Illustration of our proposed \alg{} attack.}
	\label{our_attack_fig}
\end{figure*}

\myparatight{Our work} In this work, we propose an innovative adversarial attack to multi-modal models called \emph{\alg{}}.
The procedure of our \alg{} is depicted in Fig.~\ref{our_attack_fig}.
To bridge the modality gap, our proposed algorithm, \alg{}, starts by converting the targeted input into what we refer to as the \emph{transformed input}, rather than directly aligning the embedding of the perturbed image/audio with the targeted input.
This transformed input matches the modality of the image/audio. 
For example, we can use the stable diffusion model~\cite{rombach2022high} to transform text into an image. Subsequently, the attacker introduces carefully crafted perturbations to the image/audio, ensuring that the perturbed image/audio is well-aligned with the transformed input in the embedding space.
The primary challenge lies in determining the appropriate perturbations to be applied to the image/audio. To overcome this challenge, we have structured our adversarial attacks as an optimization problem. The objective here is to minimize the angular deviation between the embeddings of the transformed input and the perturbed image/audio.
When the downstream task processes data using the perturbed image/audio from our attack, it will yield results according to the attacker's desire. For instance, as shown in Fig.~\ref{our_attack_fig}, the targeted input is ``A huge tiger'', leading the downstream task to generate an image featuring a tiger and accompanying text that includes the word ``tiger''.
It is important to highlight that our proposed attack could lead to severe negative outcomes in real-world scenarios. For instance, an attacker can add tiny perturbations to an image and upload this perturbed image online. If a user unwittingly downloads this perturbed image from the internet and uses it for image generation tasks, the results won't align with the user's intentions. Instead, the system consistently produces sensitive or inappropriate content.

We conduct a thorough assessment of our \alg{} attack across six real-world datasets under diverse conditions. Our experimental results indicate that \alg{} not only achieves a high \emph{attack success rate} and also markedly surpasses the current method. 
In other words, when downstream tasks employ the perturbed image/audio to produce data, there is a high probability that they will generate precisely the output as the attacker desires.
For example, on the ImageNet \cite{deng2009imagenet} dataset, the attack success rate of our \alg{} reaches 0.98. In addition, we test our attack in a practical black-box scenario, where the attacker has no knowledge of the downstream tasks. The findings from this scenario suggest that \alg{} maintains a high success rate even in this more realistic setting.

We also introduce six defensive strategies to mitigate the impact of our proposed \alg{} attack. 
These defenses encompass modifying the sizes (either enlarging or reducing) of the perturbed image, compressing the image (such as with JPEG compression), and applying denoising techniques to refine the perturbed image/audio.
Our experimental findings indicate that the current defense mechanisms are not sufficiently effective in thwarting the adversarial attacks we have introduced.
For instance, in the realm of traditional defense such as image resizing, our proposed \alg{} attack demonstrates impressive results, achieving approximately 0.98 and 0.97 attack success rates under image size doubling and halving, respectively.

Our contributions in this paper are summarized as follows:
\begin{itemize}
    \item We propose \alg{}, a novel adversarial attack to multi-modal models.
    
    \item Comprehensive experiments demonstrate that the proposed \alg{} effectively deceives downstream tasks and greatly outperforms the existing method.
    
    \item We propose six defensive measures to counteract our \alg{}. The results of our experiments highlight the necessity for defense strategies to effectively combat \alg{}.
\end{itemize}


\section{Related Work}
\myparatight{Multi-Modal Model}There has been a growing interest in multi-modal embedding models, which can map inputs such as images, texts, and sounds
into a shared embedding space.
These models can integrate different modalities in vision, video, and language, and thus enabling various downstream applications~\cite{girdhar2023imagebind,zhu2023languagebind,guzhov2022audioclip}.
Notably, ImageBind~\cite{girdhar2023imagebind} utilizes dedicated encoders for each modality and can embed inputs from diverse modalities (e.g., audio-text or image-text) into a unified embedding space.
By leveraging multi-modal tuples, ImageBind aims to maximize the dot product of the embedding representations within each tuple for different modalities.
AudioClip~\cite{guzhov2022audioclip} creates a tri-modal hybrid architecture by high-performance audio model ESResNeXt~\cite{guzhov2021esresne} and contrastive text-image model CLIP~\cite{radford2021learning}.
By using three modalities in training, the model outperforms single-modal models and enhances zero-shot cross-modal querying.

\myparatight{Adversarial Attacks for Cross-Modality}Adversarial attacks is a growing threat in machine learning and has attracted much attention~\cite {wei2022cross,zhang2022optimized,huang2022cmua,jia2022badencoder, carlini2021poisoning, yang2023data,han2023backdooring}.
Existing works mainly focus on attack and defense in a single modal model.
Recently, several works have explored the potential of adversarial attacks for cross-modality.
Qi et al.~\cite{qi2023visual} introduces a universal adversarial attack focused on images. This attack strategy is designed to prompt models into producing harmful and toxic outputs. Notably, it employs a substantial perturbation range, utilizing the infinite norm metric to achieve its objectives.
Zhang et al.~\cite{bagdasaryan2023ceci} introduces a novel attack method to align the embedding of a targeted input with that of a perturbed image/audio file, both belonging to distinct modalities of data. The produced adversarial examples can deceive the downstream task into generating output, be it images or text, that is closely associated with the targeted input.

Note that our proposed attack diverges from the BadEncoder attack discussed in Jia et al.~\cite{jia2022badencoder}. Unlike BadEncoder, which is a training-time attack focused on altering the model's training phase, our attack is implemented at test-time, without any involvement in the model's training procedure.


\section{Threat Model}

\myparatight{Attacker's goal}We consider a scenario where an attacker introduces imperceptible adversarial perturbations to an image or audio file. 
When a downstream task employs an encoder to process the perturbed image/audio, it will produce the output as the attacker desires.
The designed attack should achieve the following:
\begin{list}{\labelitemi}{\leftmargin=1.0em \itemindent=-0.5em \itemsep=.2em}
    \item {\bf Effectiveness.} The effectiveness goal means that the downstream task is highly likely to get the image as desired by the attacker.

    \item {\bf Stealth.} The stealth objective implies the attacker introduces only tiny and imperceptible perturbations into an image or audio file. 
    These perturbations make the modified image or audio nearly undetectable to human perception.
\end{list}

\myparatight{Attacker's knowledge and capabilities}
Regarding the attacker's knowledge, we take into account both \textit{white-box attack} and \textit{black-box attack} scenarios. In a white-box attack, it is assumed that the attacker is aware of the encoder algorithm employed by the downstream task. Conversely, in a black-box attack, the attacker lacks any knowledge about the encoder used in the downstream task. It's important to note that while the white-box attack might have limited practical applicability, it is used to gauge the maximum potential effectiveness of the proposed attack.
In terms of the attacker's capabilities, we assume that the attacker is able to embed subtly adversarial disturbances into an image or audio file.


\begin{table}[htbp]
  \centering
  \caption{Summary of key notation.}
    \begin{tabular}{l|l}
    \hline
    Notation & \multicolumn{1}{l}{Definition} \\
    \hline
    $t$   & Targeted input \\
    $t_v$  &  Transformed input \\
    $v$  & Image/audio \\
    $v^{\text{adv}}$  & Perturbed image/audio \\
    \hline
    \end{tabular}%
  \label{tab_notation}%
\end{table}%
\vspace{-.2in}

\section{Our Attack}
\label{section:our_method}

In this section, we first present our attack as an optimization problem and then explain how to solve this formulated problem. The key notations used throughout our paper are summarized in Table~\ref{tab_notation}.

\subsection{Attack as an Optimization Problem}
We use $t$ to represent the targeted input chosen by the attacker, which could be, for example, a piece of text, and we use $v$ to represent an image or audio file. 
Because the attacker-chosen targeted input $t$ and the image/audio file $v$ are from different modalities, our proposed \alg{} attack involves an initial step where we convert the targeted input $t$ into a form called the transformed input $t_v$ that matches the modality of the file $v$. For instance, if the targeted input $t$ is the text ``A huge tiger'' and $v$ is an image, we can use a method called stable diffusion~\cite{rombach2022high} to transform the text ``A huge tiger'' into an image. 
Please refer to Fig.~\ref{our_attack_fig} for the process of our \alg{} attack.

Subsequently, the attacker introduces nearly imperceptible adversarial alterations to the file $v$ in such a way that the perturbed file aligns with $t_v$ within the embedding space. We denote this perturbed image/audio file as $v^{\text{adv}}$, and it is created by applying tiny perturbations to the file $v$ according to
$
v^{\text{adv}} = v + \delta,
$
where $\delta$ represents the added perturbations.

Let $f(t_v)$ represent the embedding space of the transformed input $t_v$, and $f(v^{\text{adv}})$ symbolize the embedding space of the perturbed image/audio $v^{\text{adv}}$. The goal of the attacker is to introduce minimal changes to $v$ so that the embedding space of $f(t_v)$ collides with that of $f(v^{\text{adv}})$. Thus, this leads to the 
angular deviation minimization (ADM) problem for the attacker, which can be stated as follows:
\begin{equation}
\begin{split}
\label{attack_goal}
& \!\!\!\!\!\!\!\!\!\!\!\!\!\!\!\!\!\!\! \min_{\delta}   L(\delta) = \left\| \hat{f}(t_v) - \hat{f}(v^{\text{adv}}) \right\|_2^2,    \\
 \text{s.t.} \quad \hat{f}(t_v) = \frac{{f}(t_v)}{ \left\| {f}(t_v)  \right\|_2^2}, &   \quad  \hat{f}(v^{\text{adv}}) = \frac{{f}(v^{\text{adv}})}{ \left\| {f}(v^{\text{adv}}) \right\|_2^2}, \quad  v^{\text{adv}} = v + \delta.
\end{split}
\end{equation}

Note that for the above Problem ADM, the attacker aims to minimize the angular deviation between ${f}(t_v)$ and ${f}(v^{\text{adv}})$. Then the key challenges boil down to how to craft the imperceptible adversarial perturbations $\delta$ effectively.

We note that in Eq.~(\ref{attack_goal}), we normalize the ${f}(t_v)$ and ${f}(v^{\text{adv}})$ to reduce the impact of magnitudes of ${f}(t_v)$ and ${f}(v^{\text{adv}})$ on the final results.
In our later experiments, as shown in Section~\ref{Results_dataset}, we demonstrate that omitting normalization can reduce the effectiveness of the \alg{} method.

\begin{figure}[!t]
	\centering
	\subfloat[An image file.\label{1}]{\includegraphics[width=.3\linewidth]{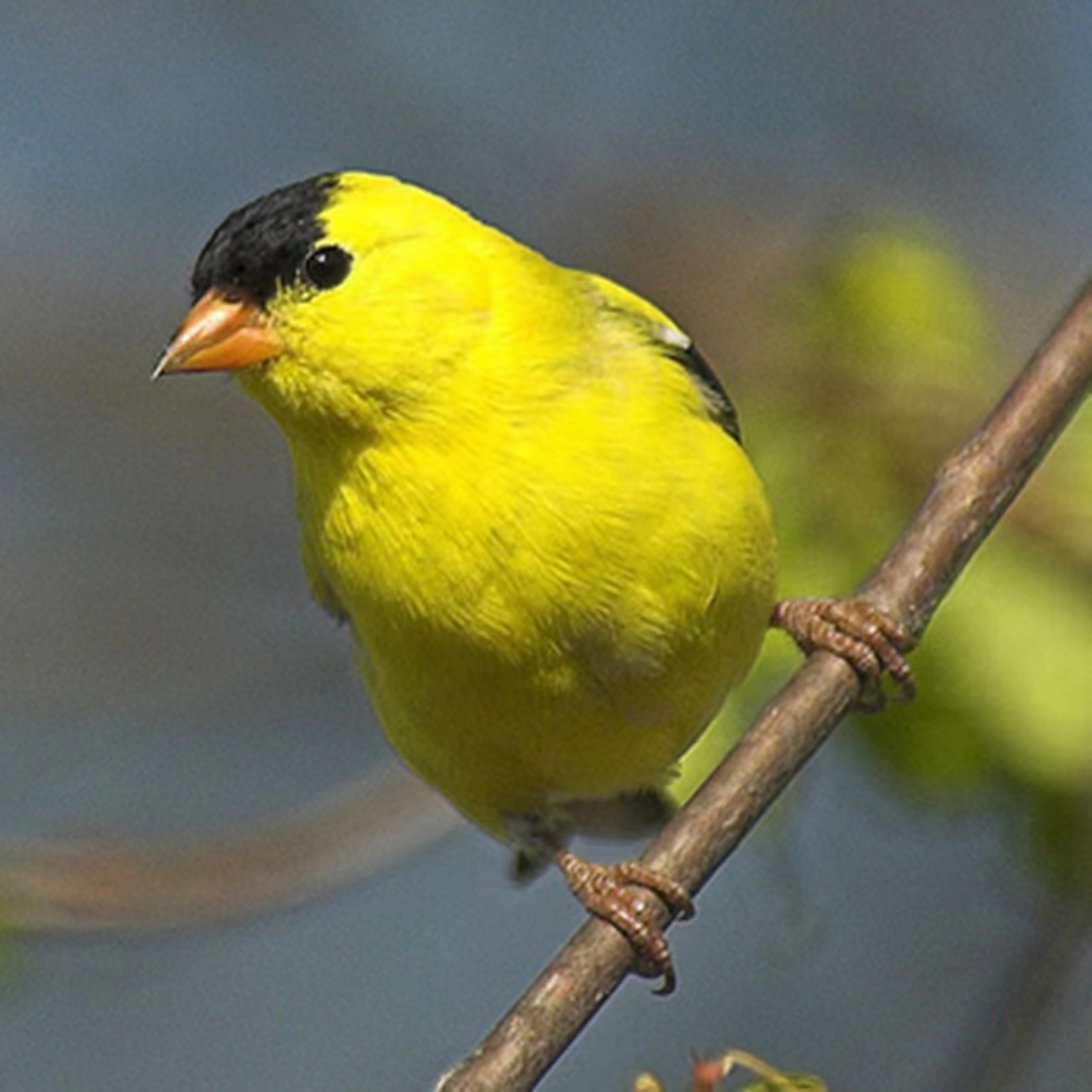}} 
 \quad
      \
	\subfloat[The generated image, in which the perturbed image is created by \alg{}, and $\ell_2$ norm is used.\label{3}]{\includegraphics[width=.3\linewidth]{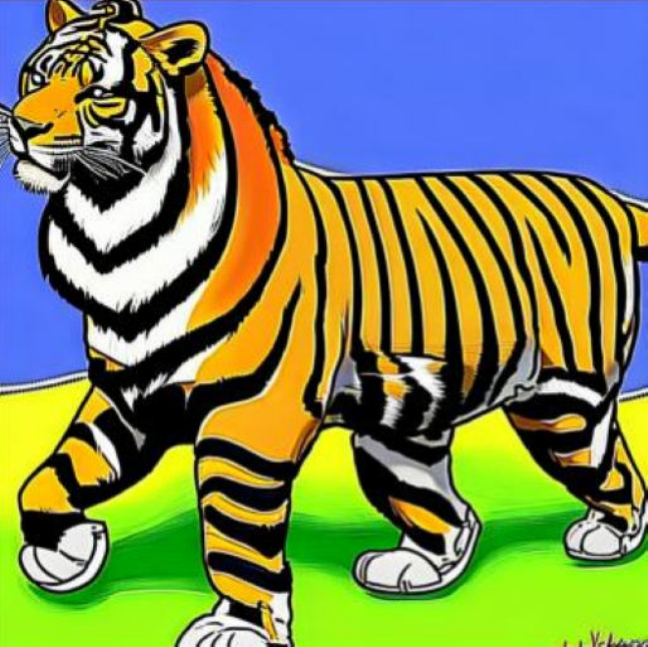}}
 \quad
 \
 	\subfloat[The generated image, in which the perturbed image is created by \alg{}, and $\ell_1$ norm is considered.\label{4}]{\includegraphics[width=.3\linewidth]
  {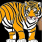}} 
 \caption{The image is generated where the perturbed image is produced by the \alg{} method, considering either the $\ell_1$ or $\ell_2$ norm.}	
  \label{low_quilty} 
  \vspace{-.2in}
\end{figure}

In the default setting of our proposed attack, we utilize the $\ell_2$ norm. 
It is important to highlight that the effectiveness of our attack extends to scenarios where the $\ell_1$ norm is applied to Eq.~(\ref{attack_goal}). 
For example,
Fig.~\ref{1} is an image file, the attacker adds some perturbations to Fig.~\ref{1} to get a perturbed image, and the targeted input is set to ``A huge tiger''.
Fig.~\ref{3} illustrates the output from the downstream task, where the perturbed image has been produced using the \alg{} method when $\ell_2$ norm is used.
Fig.~\ref{4} shows the result of the downstream task, where the perturbed image is generated using the \alg{} method with the $\ell_1$ norm applied. It demonstrates that \alg{} can still mislead the creation of the downstream image even when using the $\ell_1$ norm.

\subsection{Solving the Optimization Problem}

A method for addressing Problem ADM involves creating small perturbations $\delta$.
In what follows, we leverage the standard Project Gradient Descent  (PGD)~\cite{goodfellow2014explaining} method to iteratively add tiny perturbations to the image/audio file $v$.
The attacker crafts $\delta$ as follows:
\begin{align}
    \delta = \delta - \lambda \cdot \text{sign} (\nabla_{\delta}  L(\delta)),
\end{align}
where $\lambda$ is the learning rate, $\text{sign}(\cdot)$ is the standard sign function, $L(\delta)$ is defined in Eq.~(\ref{attack_goal}).

During each iteration, $\delta$ is added to the image/audio file $v$ with the aim of bringing the normalized embedding $\hat{f}(t_v)$ closer to $\hat{f}(v^{\text{adv}})$. This procedure is repeated several times until a convergence criterion is met, such as applying $\delta$ for a maximum number of iterations, denoted as $\text{max\_iter}$.
Algorithm~\ref{attack_algo} summaries our proposed \alg{} attack.


 \section{Experiment}  \label{sec:exp}

\subsection{Datasets, Targeted Input, Downstream Task}
We evaluate the effectiveness of our \alg{} using six benchmark datasets, including four image datasets (ImageNet~\cite{deng2009imagenet}, MS-COCO \cite{1986CommonObjects}, STL-10\cite{coates2011analysis}, Wikiart\cite{artgan2018}), and two audio datasets (AudioCaps~\cite{kim2019audiocaps}, Audioset\cite{2017Audio}).
For each image dataset, we randomly choose 100 image files, and similarly, we select 100 audio files at random from each audio dataset.
The targeted inputs are ``A huge tiger'' for the image dataset and ``A tiger is barking'' for audio datasets.
In our attack, we employ the stable diffusion model~\cite{rombach2022high} for converting text into an image. Additionally, we utilize Stable Audio software~\cite{Stable_Audio} for the transformation of text into an audio file.
In the downstream task, the downstream application first uses an encoder to obtain the embedding from the perturbed file (either an image or audio). 
Subsequently, it uses BindDiffusion~\cite{bind} for image generation and PandaGPT~\cite{su2023pandagpt} for text creation.
Note that for both image and audio datasets, the downstream task generates image and text from the perturbed image/audio file.

\begin{algorithm}[t]
\caption{\alg{}.}
    \label{attack_algo}
\begin{algorithmic}[1]
    \renewcommand{\algorithmicrequire}{\textbf{Input:}}
    \renewcommand{\algorithmicensure}{\textbf{Output:}}
   \REQUIRE Targeted input $t$, image/audio file $v$, learning rate $\lambda$, $\delta_0$, $\text{max\_iter}$.
    \ENSURE $v^{\text{adv}}$.
    \STATE Convert $ t $ to $ t_v$ such that $ t_v$ and $ v $ share the same modality.
     \STATE Obtain the embedding space $ f(t_v) $.
      \STATE $\delta \gets \delta_0$. \\
     \STATE $\text{iter} \gets 1$. \\
     \WHILE  {$\text{iter} \leq \text{max\_iter}$}
     \STATE $ v^{\text{adv}} = v + \delta$.
     \STATE Obtain the embedding space ${f}(v^{\text{adv}})$.
    \STATE $ \delta = \delta - \lambda \cdot \text{sign} (\nabla_{\delta}  L(\delta))$. 
     \STATE $\text{iter} \gets \text{iter} +1$. \\
    \ENDWHILE 
    \RETURN $v^{\text{adv}}$. 
\end{algorithmic} 
\end{algorithm}

\subsection{Evaluation Metrics}
\myparatight{Attack Success Rate on Image ($\text{ASR}_{\text{img}}$)} 
To assess the efficacy of \alg{}, we examine the generated images to determine if they correctly match the targeted categories using the zero-shot classification technique. 
Specifically, given a generated image, we employ the OpenCLIP model~\cite{cherti2023reproducible} to predict its label from a set of predefined labels. These labels are detailed in Table~\ref{labels}.
Attack success rate on the image ($\text{ASR}_{\text{img}}$) is defined as the fraction of downstream generated images that are classified as the targeted label. 
For example, with ``A huge tiger'' as the targeted input for the image datasets and ``A tiger is barking'' for audio datasets, the attacker aims to ensure that the generated images are classified as the targeted label ``A tiger''.

\begin{table}[t]
 \small
 \centering
 \caption{The pre-defined labels.}
\begin{tabular}{|lllll|}
\hline
\multicolumn{5}{|c|}{\textbf{Labels}}                                                                                                        \\ \hline
\multicolumn{1}{|l|}{A tiger}   & \multicolumn{1}{l|}{A elephant} & \multicolumn{1}{l|}{A wolf} & \multicolumn{1}{l|}{A zebra} & An eagle    \\ \hline
\multicolumn{1}{|l|}{A giraffe} & \multicolumn{1}{l|}{A kangaroo} & \multicolumn{1}{l|}{A dog}  & \multicolumn{1}{l|}{A horse} & A groundhog \\ \hline
\end{tabular}
\label{labels}
 \vspace{-.1in}
\end{table}

\myparatight{Attack Success Rate on Text ($\text{ASR}_{\text{text}}$)} 
For the generated text, a successful attack is defined when the text incorporates the central element (e.g., a noun) in targeted input.
For instance, if the targeted input is ``A huge tiger'' or ``A tiger is barking'', the produced text will include the word ``tiger''.
The $\text{ASR}_{\text{text}}$ is computed as the proportion of generated text that includes the central element mentioned in the targeted input.

\myparatight{Embedding Space Alignment (Alignment)} 
Alignment is computed as the inner product between the normalized embeddings of $\hat{f}(t_v)$ and $\hat{f}(v^{\text{adv}})$, i.e., $\text{Alignment}=\langle  \hat{f}(t_v), \hat{f}(v^{\text{adv}})\rangle $.

Note that for the three evaluation metrics $\text{ASR}_{\text{img}}$, $\text{ASR}_{\text{text}}$ and Alignment, a larger value indicates a more effective attack.

\subsection{Compared Methods, Parameter Settings}
Our proposed \alg{} is contrasted with the Zhang attack~\cite{bagdasaryan2023ceci}, which directly aligns the targeted input with the perturbed image/audio file within the embedding space. 
At the same time, ``\alg{} without normalization'' is also a baseline for ablation analysis.
Note that the variant ``\alg{} without normalization'' means we ignore the normalization term in Eq.~(\ref{attack_goal}).

In our experiments, the updating of \(\delta\) is performed using the projected gradient descent method, meaning that we maintain the norm of \(\delta\) within a certain limit, specifically \( \|\delta\|_{\infty} \leq \alpha \).

For the image datasets, \(\alpha\) is select from the set $\{0, \frac{1}{255}, \frac{4}{255}, \frac{8}{255}, $ $\frac{16}{255}, \frac{32}{255}\}$.
For the audio datasets, we specify \( \alpha \) within the range of \(\{0.0, 0.005, 0.01, 0.05, 0.1, 0.5\} \).
We also set the $\text{max\_iter}$ as 3000 and learning rate $\lambda$ as 0.01.

By default, we assume a white-box attack scenario, in which both the attacker and the downstream task utilize the same encoder to get the embedding from the perturbed image/audio file, with ImageBind~\cite{girdhar2023imagebind} being our standard encoder technique. We will also investigate scenarios involving black-box attacks, where the attacker and the downstream task employ different encoders.
We repeated each experiment 10 times and reported the average, omitting the small variances for simplicity.

\begin{table*}[]
\caption{Results on Various Datasets.}
\centering
\begin{tabular}{c|c|ccc|ccc|ccc}
\hline
\multirow{2}{*}{Datasets} & \multirow{2}{*}{$\alpha$} & \multicolumn{3}{c|}{\textbf{Zhang attack}}                   & \multicolumn{3}{c|}{\textbf{CrossFire without normlization}}           & \multicolumn{3}{c}{\textbf{CrossFire}}                                 \\ \cline{3-11} 
                          &                           & $\text{ASR}_{\text{img}}$ & $\text{ASR}_{\text{text}}$ & Alignment & $\text{ASR}_{\text{img}}$ & $\text{ASR}_{\text{text}}$ & Alignment     & $\text{ASR}_{\text{img}}$ & $\text{ASR}_{\text{text}}$ & Alignment     \\ \hline
                          & 0/255                     & 0.00                      & 0.00                       & 0.03      & 0.00                      & 0.00                       & \textbf{0.42} & 0.00                      & 0.00                       & \textbf{0.42} \\
                          & 1/255                     & 0.19                      & 0.12                       & 0.55      & 0.19                      & 0.13                       & 0.61          & \textbf{0.23}             & \textbf{0.19}              & \textbf{0.64} \\
                          & 4/255                     & 0.48                      & 0.36                       & 0.72      & 0.72                      & 0.66                       & 0.75          & \textbf{0.76}             & \textbf{0.72}              & \textbf{0.77} \\
ImageNet                  & 8/255                     & 0.49                      & 0.42                       & 0.86      & 0.89                      & 0.71                       & \textbf{0.92} & \textbf{0.93}             & \textbf{0.79}              & \textbf{0.92} \\
                          & 16/255                    & 0.51                      & 0.47                       & 0.93      & 0.94                      & 0.85                       & 0.95          & \textbf{0.98}             & \textbf{0.87}              & \textbf{0.97} \\
                          & 32/255                    & 0.57                      & 0.51                       & 0.96      & 0.97                      & 0.89                       & 0.98          & \textbf{0.99}             & \textbf{0.94}              & \textbf{0.99} \\ \hline
                          & 0/255                     & 0.00                      & 0.00                       & 0.07      & 0.00                      & 0.00                       & \textbf{0.42} & 0.00                      & 0.00                       & \textbf{0.42} \\
                          & 1/255                     & 0.16                      & 0.17                       & 0.61      & 0.25                      & 0.22                       & 0.69          & \textbf{0.29}             & \textbf{0.24}              & \textbf{0.70} \\
                          & 4/255                     & 0.53                      & 0.39                       & 0.77      & 0.77                      & 0.69                       & 0.79          & \textbf{0.81}             & \textbf{0.80}              & \textbf{0.83} \\
MS-COCO                   & 8/255                     & 0.56                      & 0.47                       & 0.89      & 0.92                      & 0.81                       & 0.89          & \textbf{0.97}             & \textbf{0.87}              & \textbf{0.94} \\
                          & 16/255                    & 0.59                      & 0.51                       & 0.95      & 0.94                      & 0.84                       & 0.92          & \textbf{0.97}             & \textbf{0.89}              & \textbf{0.99} \\
                          & 32/255                    & 0.64                      & 0.54                       & 0.96      & 0.96                      & 0.87                       & 0.96          & \textbf{0.98}             & \textbf{0.91}              & \textbf{0.99} \\ \hline
                          & 0/255                     & 0.00                      & 0.00                       & 0.05      & 0.00                      & 0.00                       & \textbf{0.45} & 0.00                      & 0.00                       & \textbf{0.45} \\
                          & 1/255                     & 0.13                      & 0.13                       & 0.58      & 0.21                      & 0.14                       & 0.67          & \textbf{0.24}             & \textbf{0.21}              & \textbf{0.69} \\
                          & 4/255                     & 0.39                      & 0.34                       & 0.73      & 0.75                      & 0.72                       & 0.77          & \textbf{0.77}             & \textbf{0.77}              & \textbf{0.79} \\
STL-10                    & 8/255                     & 0.48                      & 0.44                       & 0.84      & 0.86                      & 0.73                       & 0.94          & \textbf{0.90}             & \textbf{0.74}              & \textbf{0.96} \\
                          & 16/255                    & 0.51                      & 0.47                       & 0.95      & 0.92                      & \textbf{0.87}              & 0.97          & \textbf{0.96}             & 0.85                       & \textbf{0.98} \\
                          & 32/255                    & 0.55                      & 0.51                       & 0.97      & 0.96                      & 0.82                       & 0.98          & \textbf{0.97}             & \textbf{0.91}              & \textbf{0.99} \\ \hline
                          & 0/255                     & 0.00                      & 0.00                       & 0.05      & 0.00                      & 0.00                       & \textbf{0.41} & 0.00                      & 0.00                       & \textbf{0.41} \\
                          & 1/255                     & 0.17                      & 0.14                       & 0.49      & 0.24                      & 0.16                       & 0.72          & \textbf{0.26}             & \textbf{0.24}              & \textbf{0.72} \\
                          & 4/255                     & 0.51                      & 0.33                       & 0.71      & 0.71                      & 0.62                       & 0.74          & \textbf{0.74}             & \textbf{0.75}              & \textbf{0.75} \\
Wikiart                   & 8/255                     & 0.53                      & 0.45                       & 0.85      & 0.85                      & 0.74                       & 0.92          & \textbf{0.91}             & \textbf{0.78}              & \textbf{0.93} \\
                          & 16/255                    & 0.55                      & 0.47                       & 0.94      & 0.93                      & 0.82                       & 0.95          & \textbf{0.97}             & \textbf{0.84}              & \textbf{0.98} \\
                          & 32/255                    & 0.59                      & 0.52                       & 0.98      & \textbf{0.97}             & 0.53                       & 0.98          & \textbf{0.97}             & \textbf{0.89}              & \textbf{0.99} \\ \hline
                          & 0                         & 0.00                      & 0.00                       & 0.17      & 0.00                      & 0.00                       & \textbf{0.19}          & 0.00                      & 0.00                       & \textbf{0.19}          \\
                          & 0.005                     & 0.12                      & 0.07                       & 0.18      & 0.21                      & \textbf{0.23}              & 0.27          & \textbf{0.27}             & 0.19                       & \textbf{0.29} \\
                          & 0.01                      & 0.19                      & 0.14                       & 0.28      & 0.83                      & 0.63                       & 0.69          & \textbf{0.87}             & \textbf{0.76}              & \textbf{0.73} \\
Audiocaps                 & 0.05                      & 0.76                      & 0.51                       & 0.76      & 0.92                      & 0.81                       & 0.95          & \textbf{0.94}             & \textbf{0.86}              & \textbf{0.97} \\
                          & 0.10                      & 0.82                      & 0.72                       & 0.89      & 0.93                      & 0.82                       & 0.97          & \textbf{0.96}             & \textbf{0.89}              & \textbf{0.98} \\
                          & 0.50                      & 0.87                      & 0.77                       & 0.94      & 0.95                      & 0.84                       & 0.97          & \textbf{0.97}             & \textbf{0.90}              & \textbf{0.99} \\ \hline
                          & 0                         & 0.00                      & 0.00                       & 0.14      & 0.00                      & 0.00                       & \textbf{0.22} & 0.00                      & 0.00                       & \textbf{0.22} \\
                          & 0.005                     & 0.14                      & 0.05                       & 0.23      & 0.24                      & 0.17                       & 0.32          & 0.29                      & \textbf{0.21}              & \textbf{0.35} \\
                          & 0.01                      & 0.17                      & 0.47                       & 0.29      & 0.82                      & 0.55                       & 0.71          & 0.84                      & \textbf{0.65}              & \textbf{0.76} \\
Audioset                  & 0.05                      & 0.72                      & 0.51                       & 0.67      & 0.90                      & 0.74                       & 0.93          & 0.90                      & \textbf{0.82}              & \textbf{0.94} \\
                          & 0.10                      & 0.79                      & 0.67                       & 0.91      & 0.91                      & 0.79                       & \textbf{0.97} & 0.92                      & \textbf{0.85}              & \textbf{0.97} \\
                          & 0.50                      & 0.84                      & 0.74                       & 0.95      & 0.92                      & 0.81                       & \textbf{0.98} & 0.94                      & \textbf{0.87}              & 0.97          \\ \hline
\end{tabular}
\label{Image_text}
\end{table*}

\vspace{0.5cm}
\subsection{The Results on Various Datasets}
\label{Results_dataset}

Table \ref{Image_text} shows the results of our proposed \alg{} and baseline attacks under different perturbation levels.
In the presence of adversarial attacks, Zhang attack, \alg{} without normalization, and our \alg{} demonstrate an upward trend in $\text{ASR}_{\text{img}}$ and $\text{ASR}_{\text{text}}$ with increasing perturbation levels. 
Notably, under comparable alignment conditions, our \alg{} surpasses Zhang attack in both $\text{ASR}_{\text{text}}$ and $\text{ASR}_{\text{img}}$ largely, suggesting that the generated images and texts from downstream generation are closer to the targeted input by our approach. 
Meanwhile, we can observe that the $\text{ASR}_{\text{img}}$  of \alg{} without the normalization term is slightly lower than \alg{}, but its $\text{ASR}_{\text{text}}$  is noticeably lower than \alg{}. Hence, the normalization term is very important for our \alg{}.

At a perturbation level $\alpha$ of $\frac{4}{255}$ on the ImageNet, Zhang attack and our \alg{} demonstrate comparable alignment scores (0.72 alignment score for Zhang attack and 0.77 for CrossFire). However, CrossFire outperforms Zhang in both $\text{ASR}_{\text{img}}$ and $\text{ASR}_{\text{text}}$. Notably, the downstream generation performance has substantial differences, with Zhang's $\text{ASR}_{\text{img}}$ and $\text{ASR}_{\text{text}}$ at 0.48 and 0.36, and our attack achieving significantly higher values of 0.76 and 0.72. 
For \alg{} without normalization term, its alignment is 0.75, and its $\text{ASR}_{\text{text}}$ and $\text{ASR}_{\text{img}}$ are 0.66 and 0.72, respectively.
For $\alpha$ is $\frac{32}{255}$, $\text{ASR}_{\text{text}}$ of entire \alg{} achieves an impressive 0.94, significantly outperforms the Zhang attack, which only reaches about 0.51. 
In addition, \alg{} surpassed \alg{} without normalization by 5\% on $\text{ASR}_{\text{text}}$.
Therefore, it can be concluded that: 1) our \alg{} demonstrates superior alignment under the same perturbation level, 2) under similar alignment conditions, \alg{} excels in both $\text{ASR}_{\text{img}}$ and $\text{ASR}_{\text{text}}$, 
and 3) the normalization term has a relatively small impact on $\text{ASR}_{\text{img}}$, but it has a significant impact on $\text{ASR}_{\text{text}}$. 

For other datasets, CrossFire also demonstrates significant superiority compared to Zhang attack. In the scenario where $\alpha$ is set to $\frac{8}{255}$, CrossFire demonstrates superior performance compared to Zhang across multiple datasets. Specifically, in terms of $\text{ASR}_{\text{img}}$, CrossFire achieves higher scores by 0.41 on MS-COCO dataset, 0.42 on STL-10, and 0.38 on Wikiart. Similarly, for $\text{ASR}_{\text{text}}$, CrossFire outperforms Zhang with respective score increases of 0.40 on MS-COCO, 0.30 on STL-10, and 0.33 on Wikiart.
Furthermore, Appendix~\ref{app_single} presents various examples of generated data for the targeted input ``A huge tiger''. 
The images in the first column depict the image file $v$, while the second and third columns respectively showcase the generated image and text under the Zhang attack. In contrast, the fourth and fifth columns display the generated image and text under our proposed \alg{}.

Table~\ref{Image_text} also underscores an equal conclusion regarding the results on the audio datasets compared to image datasets. Remarkably, \alg{} outperforms the Zhang attack on AudioCaps dataset, supported by the following observations: 1) higher alignment under the same perturbation level, and 2) in terms of similarity alignment, our \alg{} achieves superior $\text{ASR}_{\text{text}}$ and $\text{ASR}_{\text{img}}$, indicating better performance.
For example, at a perturbation level of \(\alpha = 0.05\), the Zhang attack attains an \(\text{ASR}_{\text{img}}\) of 0.76 and an \(\text{ASR}_{\text{text}}\) of 0.51, whereas our approach yields \(\text{ASR}_{\text{img}}\) and \(\text{ASR}_{\text{text}}\) scores of 0.94 and 0.86, respectively. 
On AudioSet, whether it's $\text{ASR}_{\text{text}}$ or $\text{ASR}_{\text{img}}$, CrossFire is at least 10\% higher than Zhang attack. 
For two audio datasets, the role of the normalization term remains similar to its effect on image datasets: a small impact on $\text{ASR}_{\text{img}}$, and a significant impact on $\text{ASR}_{\text{text}}$.

\begin{figure}[htbp]
	\centering
	\subfloat[Ground truth image.\label{ground_truth}]{\includegraphics[width=0.4\linewidth]{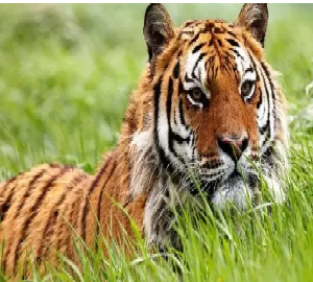}}\quad
	\subfloat[Transformed input.\label{transformed}]{\includegraphics[width=0.4\linewidth]{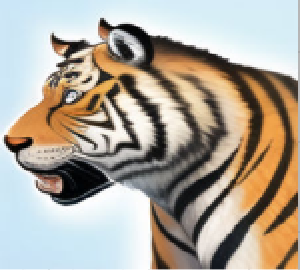}}
	\caption{Ground truth image and transformed input, where the targeted input is ``A huge tiger''.} 
\end{figure}

\textbf{Why \alg{} is superior?}
It should be noted that the Zhang attack is designed to align the targeted input and the perturbed image/audio within the embedding space. 
Conversely, our CrossFire aims to align the transformed input and perturbed image/audio within the embedding space. 
Since the transformed input and the perturbed image/audio belong to the same modality, our proposed \alg{} effectively bridges the modality gap between the targeted input and the image/audio file.
According to the results presented in Table~\ref{Image_text}, it is evident that CrossFire substantially outperforms the Zhang attack across various datasets for images and audio. To elucidate why CrossFire is superior, let's consider a specific example. Figure~\ref{ground_truth} displays a ground truth image of a tiger, while Figure~\ref{transformed} shows the transformed input resulting from applying a diffusion model~\cite{rombach2022high} to transform the targeted input ``A huge tiger'' into an image. The cosine similarity between the ground truth image and the transformed input in the embedding space is 0.67, using ImageBind as the encoder for deriving the embeddings of both ground truth image and transformed input. In contrast, the cosine similarity between the ground truth image and the targeted input ``A huge tiger'' in the embedding space is only 0.29. This indicates that the transformed input aligns more closely with the ground truth image than the targeted input does. 
Therefore, the perturbed image file from our CrossFire more closely resembles the ground truth image than that generated by the Zhang attack.

\subsection{The Results of AudioClip Encoder}

\begin{table}[t]
\caption{Results on AudioCaps dataset with AudioClip Encoder.}
\centering
\resizebox{\linewidth}{!}{
\begin{tabular}{c|ccc|ccc}
\hline
    \multirow{2}{*}{$\alpha$} & \multicolumn{3}{c|}{\textbf{Zhang attack}}    & \multicolumn{3}{c}{\textbf{CrossFire}} \\
    \cline{2-7}       & $\text{ASR}_{\text{img}}$ & $\text{ASR}_{\text{text}}$ & Alignment & $\text{ASR}_{\text{img}}$ & $\text{ASR}_{\text{text}}$ & Alignment\\
    \hline
0                    & 0.00                         & 0.00                                              & 0.12    & 0.00                         & 0.00                                              & \textbf{0.19}    \\
0.005                & 0.12                        & 0.05                                              & 0.19    & \textbf{0.27 }                       & \textbf{0.14}                                             & \textbf{0.29}    \\
0.01                 & 0.18                        & 0.09                                              & 0.28    & \textbf{0.87}                        & \textbf{0.73}                                             & \textbf{0.73}    \\
0.05                 & 0.76                        & 0.56                                             & 0.76    & \textbf{0.94}                        & \textbf{0.81}                                             & \textbf{0.97}    \\
0.10                 & 0.82                        & 0.66                                             & 0.88    & \textbf{0.96}                        & \textbf{0.85}                                             & \textbf{0.98}    \\
0.50                 & 0.87                        & 0.69                                             & 0.94    & \textbf{0.97}                        & \textbf{0.89}                                             & \textbf{0.99}    \\ \hline
\end{tabular}}
\label{audioclip}
\end{table}

By default, we presume that both the attacker and the downstream task use ImageBind to obtain the embedding of the perturbed image/audio file. In this section, we investigate a different scenario where AudioClip~\cite{guzhov2022audioclip} is employed as the encoder for extracting embeddings. The outcomes for the AudioCaps dataset under this setup are presented in Table~\ref{audioclip}.
From Table~\ref{audioclip}, we observe that when subjected to a perturbation $\alpha$ of $0.05$, the alignment of the Zhang attack is approximately 0.76, whereas our \alg{} achieves an alignment of 0.97. Remarkably, for \alg{}, even with a small perturbation $\alpha$ of $0.05$, both the alignment and $\text{ASR}_{\text{img}}$ surpass those of Zhang attack subjected to a substantial perturbation $\alpha$ of 0.5. This shows the superior performance of \alg{} in comparison to the Zhang attack even AudioClip is used as the encoder.

\begin{table}[t]
\centering
\caption{The predefined labels for targeted input contain multiple elements.}
\begin{tabular}{|l|l|}
\hline
\multicolumn{2}{|c|}{Labels}    \\ \hline
A dog is standing         & A dog is playing ball     \\
A dog is creeping         & A dog is playing frisbee  \\
A dog is eating food      & A dog is sleeping         \\
A dog is playing with cats & A dog is chasing birds   \\
A dog is running          & A dog is barking          \\ \hline
\end{tabular}
\label{table_labels}
\end{table}

\begin{table}[t]
\caption{Results on ImageNet dataset for targeted input contains multiple elements.}
\centering
\footnotesize
\begin{tabular}{c|ccc|ccc}
\hline
    \multirow{2}{*}{$\alpha$} & \multicolumn{3}{c|}{\textbf{Zhang attack}}    & \multicolumn{3}{c}{\textbf{CrossFire}} \\
    \cline{2-7}       & $\text{ASR}_{\text{img}}$ & $\text{ASR}_{\text{text}}$ & Alignment & $\text{ASR}_{\text{img}}$ & $\text{ASR}_{\text{text}}$ & Alignment\\
    \hline
0                     & 0.00        & 0.00         & 0.06      & 0.00        & 0.00         & \textbf{0.37}      \\
1/255                     & 0.04        & 0.00         & 0.56      & \textbf{0.11}        & \textbf{0.01}         & \textbf{0.59}      \\
4/255                     & 0.12        & 0.00         & 0.69      & \textbf{0.31}        & \textbf{0.06}         & \textbf{0.74}      \\
8/255                     & 0.23        & 0.02         & 0.86      & \textbf{0.49}        & \textbf{0.09}         & \textbf{0.93}      \\
16/255                    & 0.26        & 0.04         & 0.94      & \textbf{0.56}        & \textbf{0.14}         & \textbf{0.96}      \\
32/255                    & 0.29        & 0.07         & 0.97      & \textbf{0.61}        & \textbf{0.22}         & \textbf{0.97}      \\ \hline
\end{tabular}
\label{Image_text_muilt}
\end{table}

\subsection{Results of Targeted Input Contains Multiple Elements}
By default, we assume that the targeted input contains a single element, usually a noun. For example, in ``A huge tiger'', the central element is ``tiger''. This section explores a scenario with a more complex input featuring two central elements. Consider the targeted input ``A dog is playing ball'', which includes two key elements: ``dog'' and ``ball''. In such scenarios, $\text{ASR}_{\text{img}}$ denotes the percentage of generated images that simultaneously depict both ``dog'' and ``ball''. Similarly, $\text{ASR}_{\text{text}}$ represents the percentage of generated texts that include the words ``dog'' and ``ball''. 
The pre-defined labels for zero-shot classification in this setting are shown in Table~\ref{table_labels}.
These results on ImageNet are presented in Table~\ref{Image_text_muilt}. 
Moreover, Appendix~\ref{app_multiple} illustrates several samples of generated data for the targeted input ``A dog is playing ball'' that contains multiple elements. The figures in the first column represent the image file $v$, while the second and third columns 
respectively display the generated image and text  under the Zhang attack. The fourth and fifth columns respectively show the generated image and text  under our \alg{}.
This more intricate case demonstrates that our suggested \alg{} remains capable of misleading the intended downstream task.

\subsection{Results of Black-box Attacks}

\begin{table}[t]
\caption{Results on ImageNet dataset under black-box attacks.}
\centering
\resizebox{\linewidth}{!}{
\begin{tabular}{c|ccc|ccc}
\hline
    \multirow{2}{*}{$\alpha$} & \multicolumn{3}{c|}{\textbf{Zhang attack}}    & \multicolumn{3}{c}{\textbf{CrossFire}} \\
    \cline{2-7}       & $\text{ASR}_{\text{img}}$ & $\text{ASR}_{\text{text}}$ & Alignment & $\text{ASR}_{\text{img}}$ & $\text{ASR}_{\text{text}}$ & Alignment\\
    \hline
0                    & 0.00                                              & 0.00                                               & 0.06                         & 0.00                                              & 0.00                                               & \textbf{0.45}                         \\
1/255                   & 0.14                                             & 0.09                                               & 0.49                         & \textbf{0.19}                                             & \textbf{0.17}                                              & \textbf{0.60}                         \\
4/255                    & 0.42                                            & 0.29                                              & 0.68                         & \textbf{0.74}                                             & \textbf{0.64}                                              & \textbf{0.74}                         \\
8/255                    & 0.46                                            & 0.33                                              & 0.87                         & \textbf{0.86}                                             & \textbf{0.72}                                              & \textbf{0.89}                         \\
16/255                   & 0.50                                             & 0.41                                              & 0.93                         & \textbf{0.92}                                             & \textbf{0.81}                                              & \textbf{0.95}                         \\
32/255                  & 0.55                                             & 0.44                                              & 0.96                         & \textbf{0.92}                                             & \textbf{0.88}                                              & \textbf{0.98}                         \\ \hline
\end{tabular}}
\label{black_box}
\end{table}

In this section, we investigate a more realistic scenario where the attacker lacks knowledge of the encoder employed by the downstream task. In other words, the attacker and the downstream task utilize different encoders to extract embeddings from the perturbed image/audio.
To be more precise, the attacker utilizes the ImageBind encoder, whereas the downstream task employs the OpenCLIP~\cite{cherti2023reproducible} encoder.
The results on ImageNet dataset are shown in Table~\ref{black_box}.
Even in this realistic black-box scenario, we observe that our \alg{} attack remains capable of successfully confusing the downstream tasks, and it performs considerably better than the baseline attack.
As an example, when the perturbation level is set to $\alpha = \frac{16}{255}$, \alg{} attains attack success rates of 0.92 for images and 0.81 for text. In contrast, the Zhang attack achieves attack success rates of 0.50 for images and 0.41 for text under the same conditions.


\begin{figure*}[!t]
	\centering
	\subfloat[The image file $v$.]{\includegraphics[width=0.185 \textwidth]{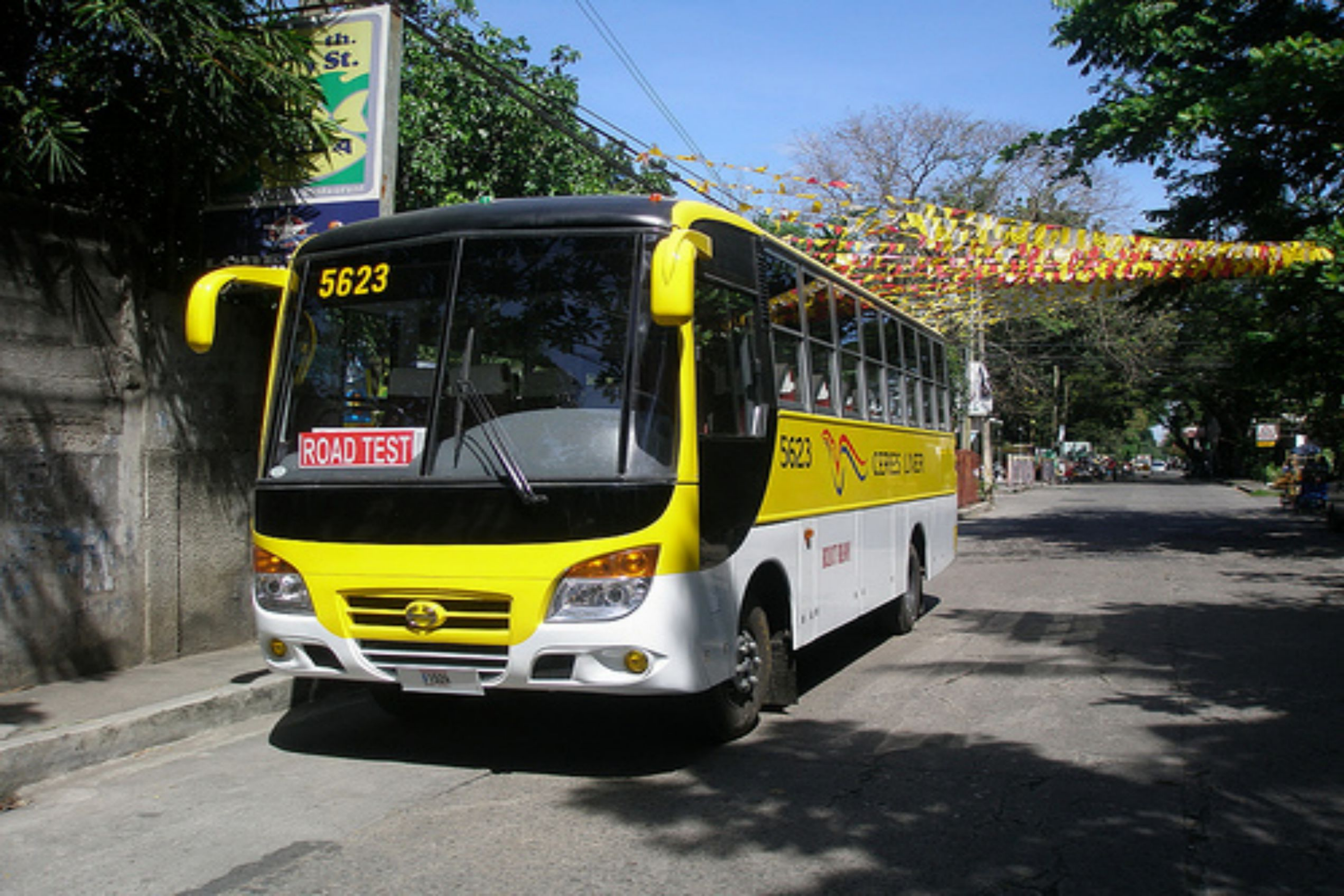}} 
 \quad
 	\subfloat[The perturbed image file $v^{\text{adv}}$.]{\includegraphics[width=0.185 \textwidth]{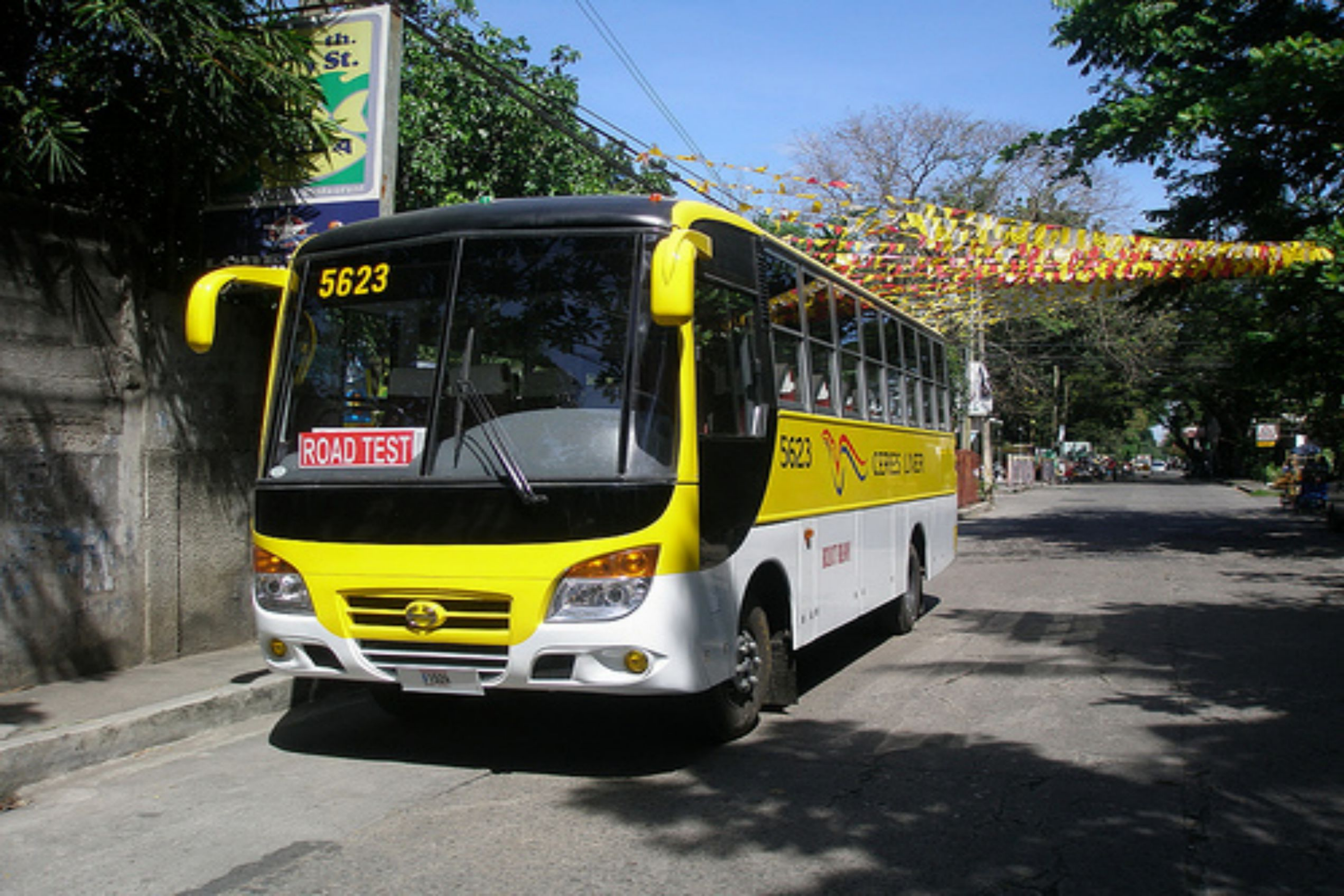}} 
   \quad
  	\subfloat[The image generated by the downstream task using the image file $v$.]{\includegraphics[width=0.185 \textwidth]{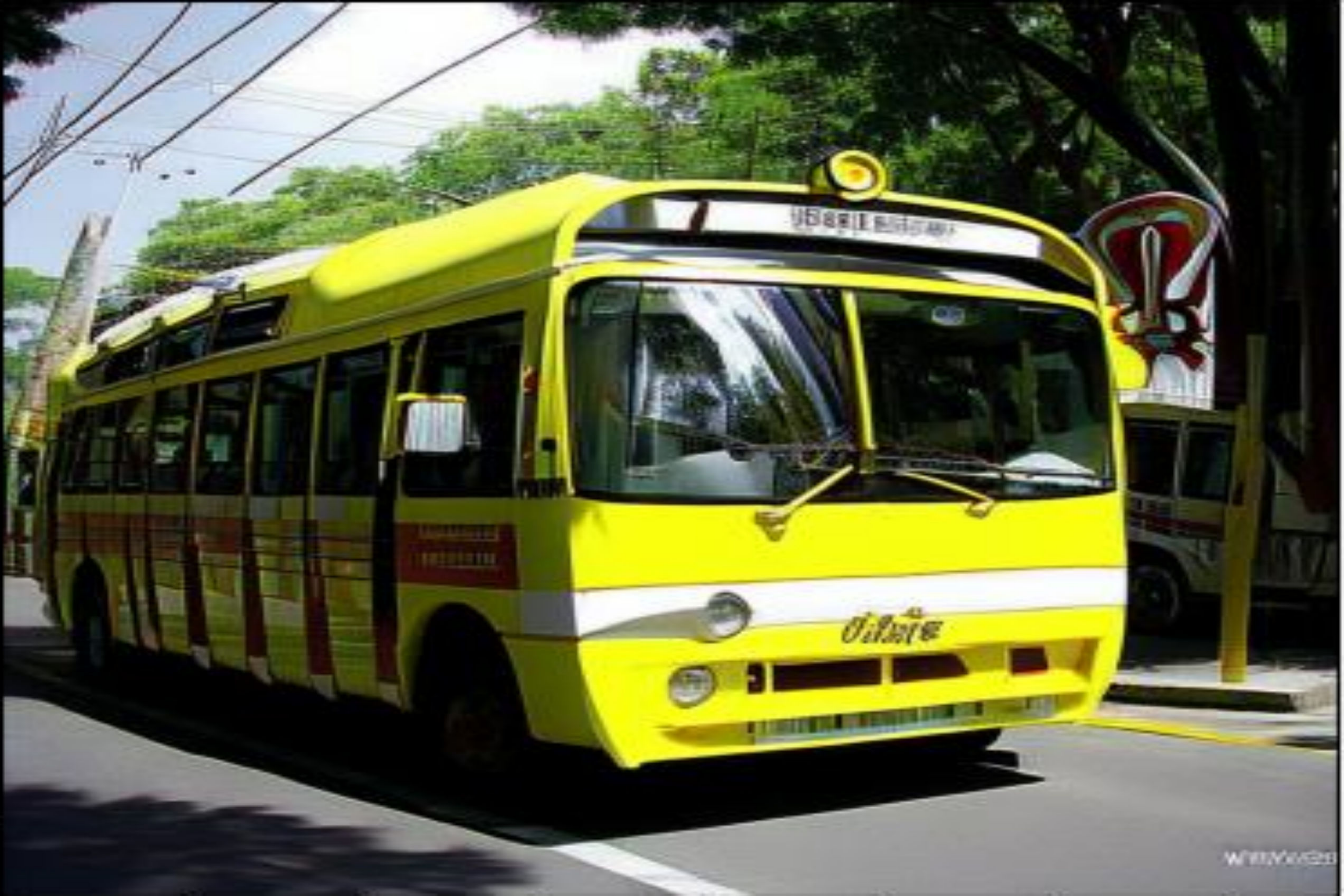}} 
    \quad
	\subfloat[The image generated by the downstream task using the perturbed image file $v^{\text{adv}}$.]{\includegraphics[width=0.185 \textwidth]{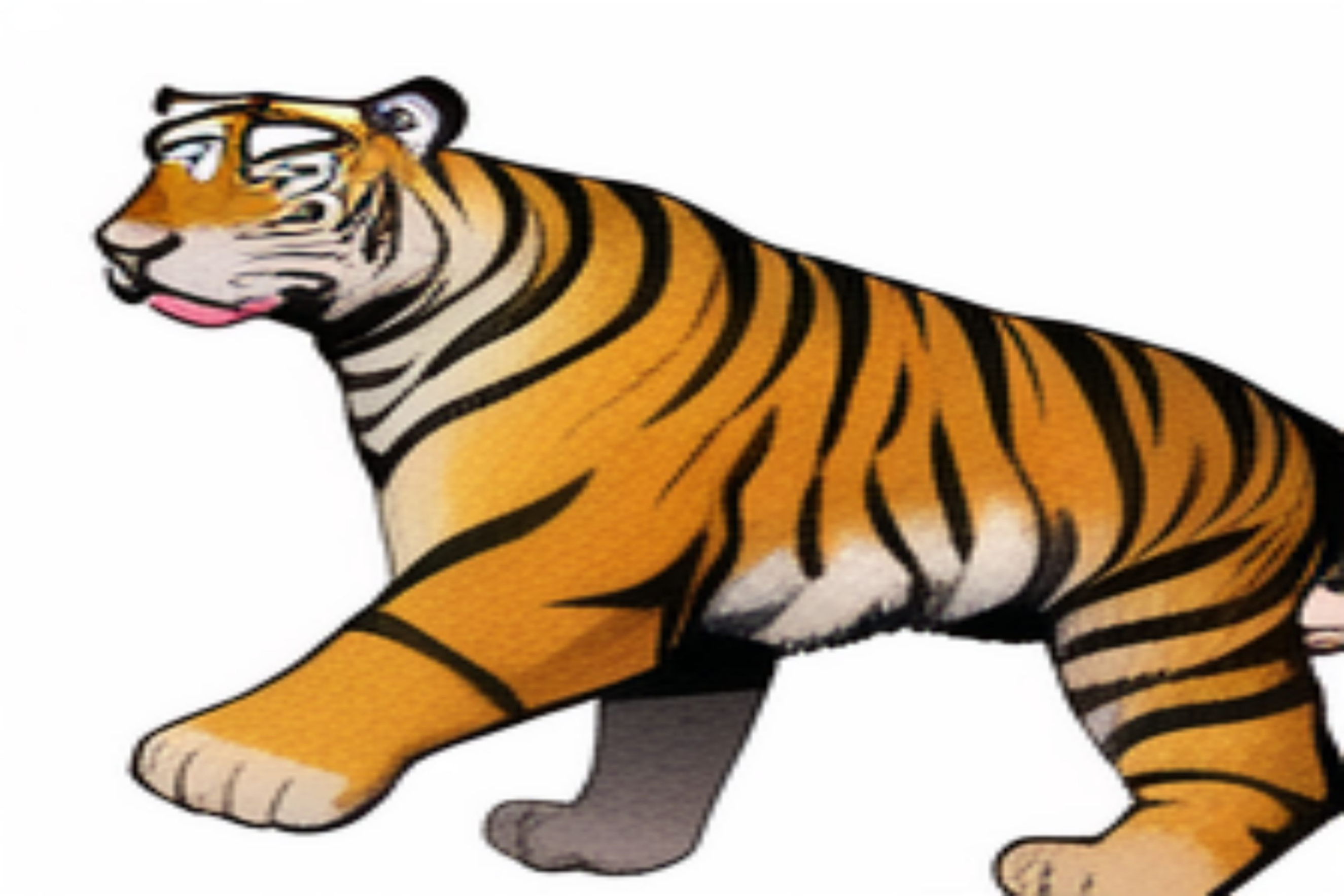}}
  \caption{The data augment manipulation, where the targeted input is ``A huge tiger''.}
  \label{data_augu}
\end{figure*}

\begin{figure*}[!t]
	\centering
	\subfloat[The image file $v$.]{\includegraphics[width=0.185 \textwidth]{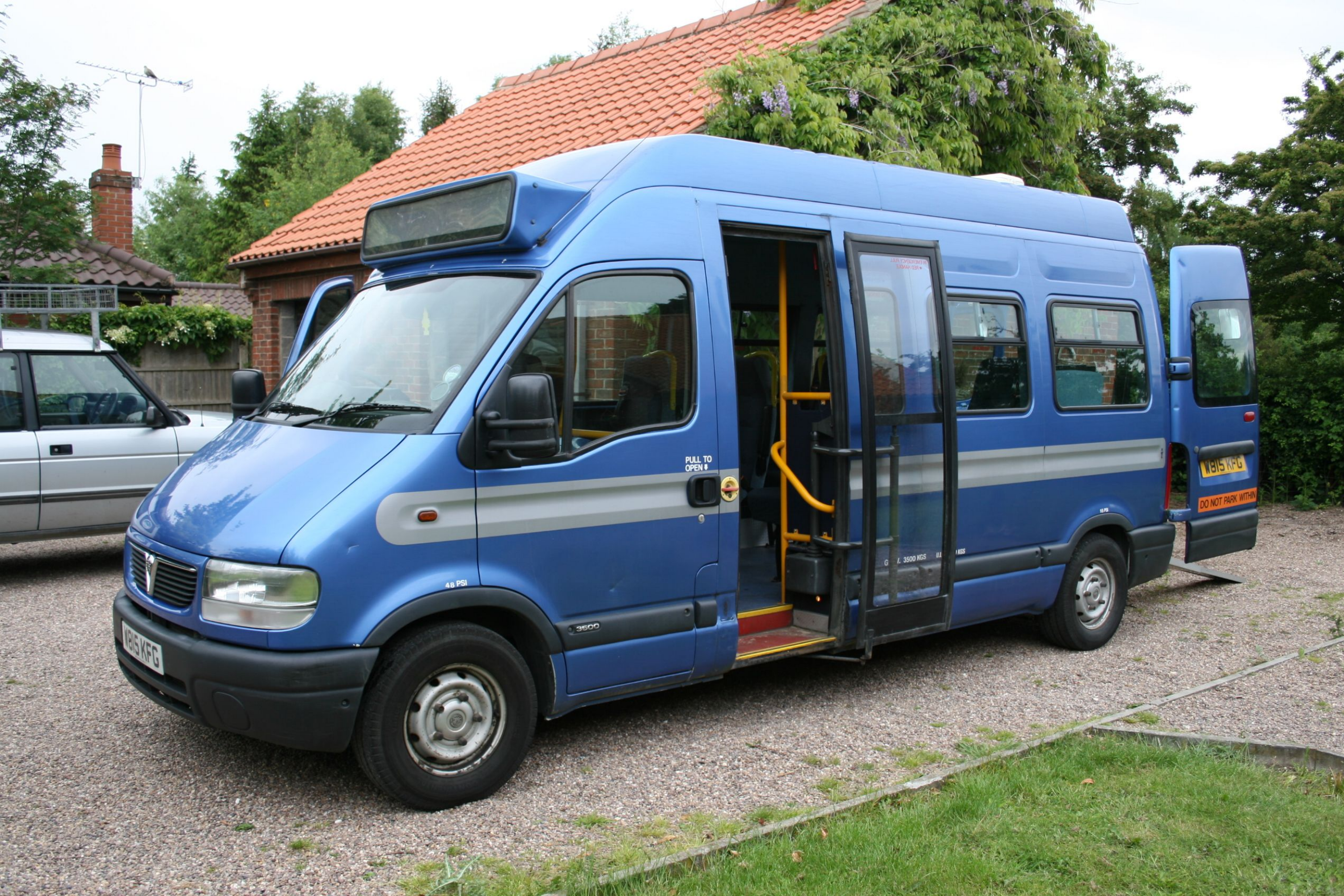}} 
 \quad
 	\subfloat[The perturbed image file $v^{\text{adv}}$.]{\includegraphics[width=0.185 \textwidth]{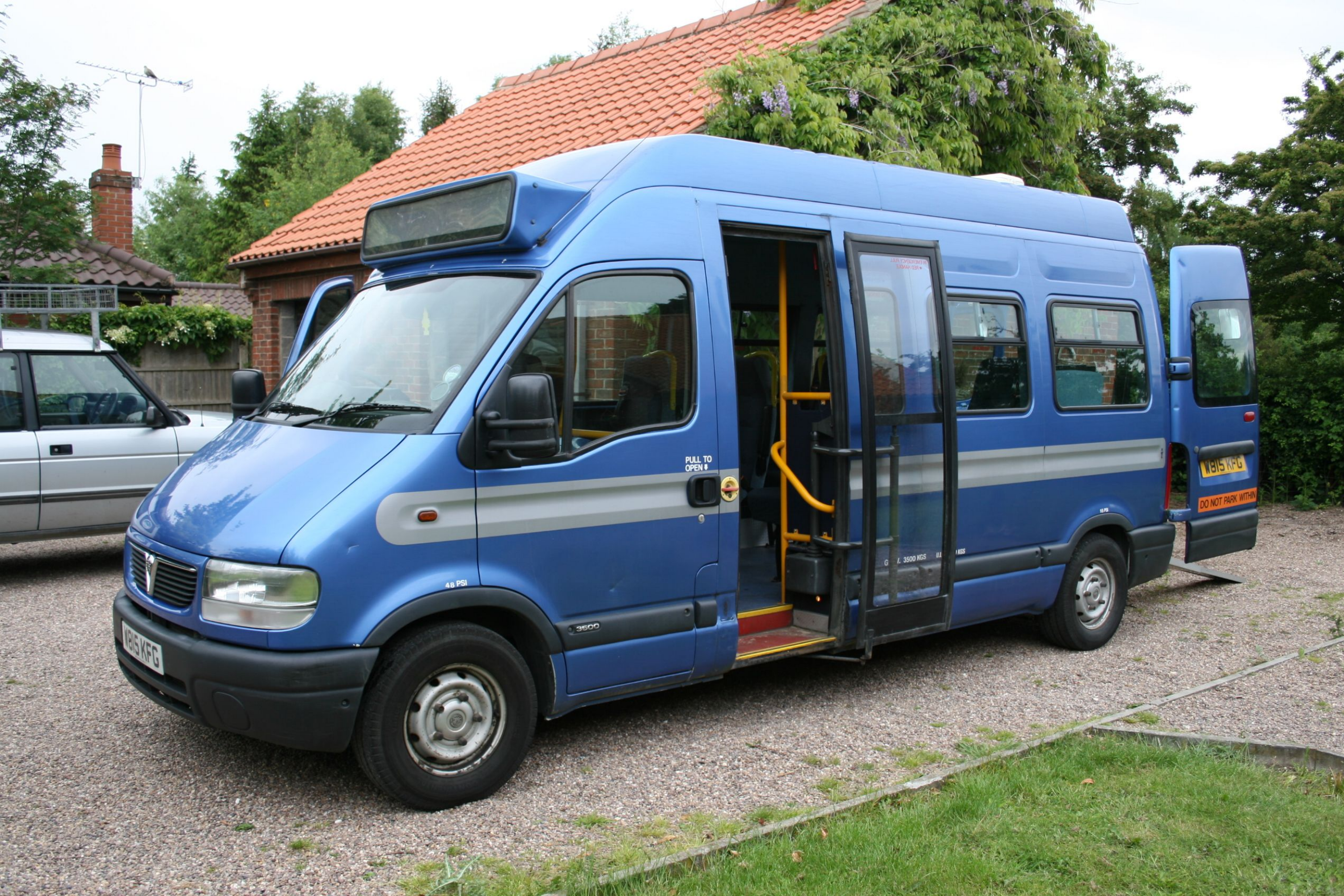}} 
   \quad
  	\subfloat[The image generated by the downstream task using the image file $v$.]{\includegraphics[width=0.185 \textwidth]{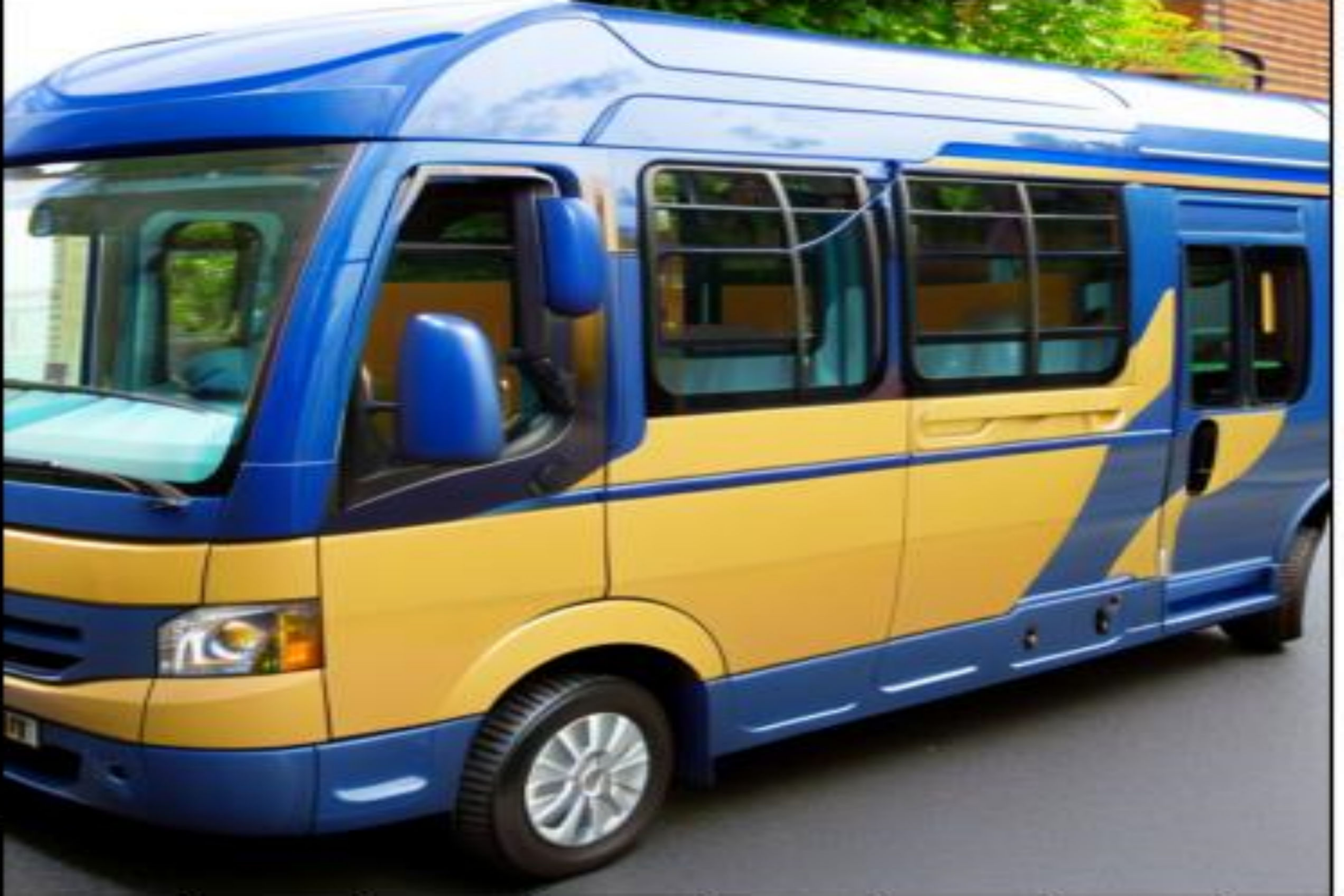}} 
    \quad
	\subfloat[The image generated by the downstream task using the perturbed image file $v^{\text{adv}}$.]{\includegraphics[width=0.185 \textwidth]{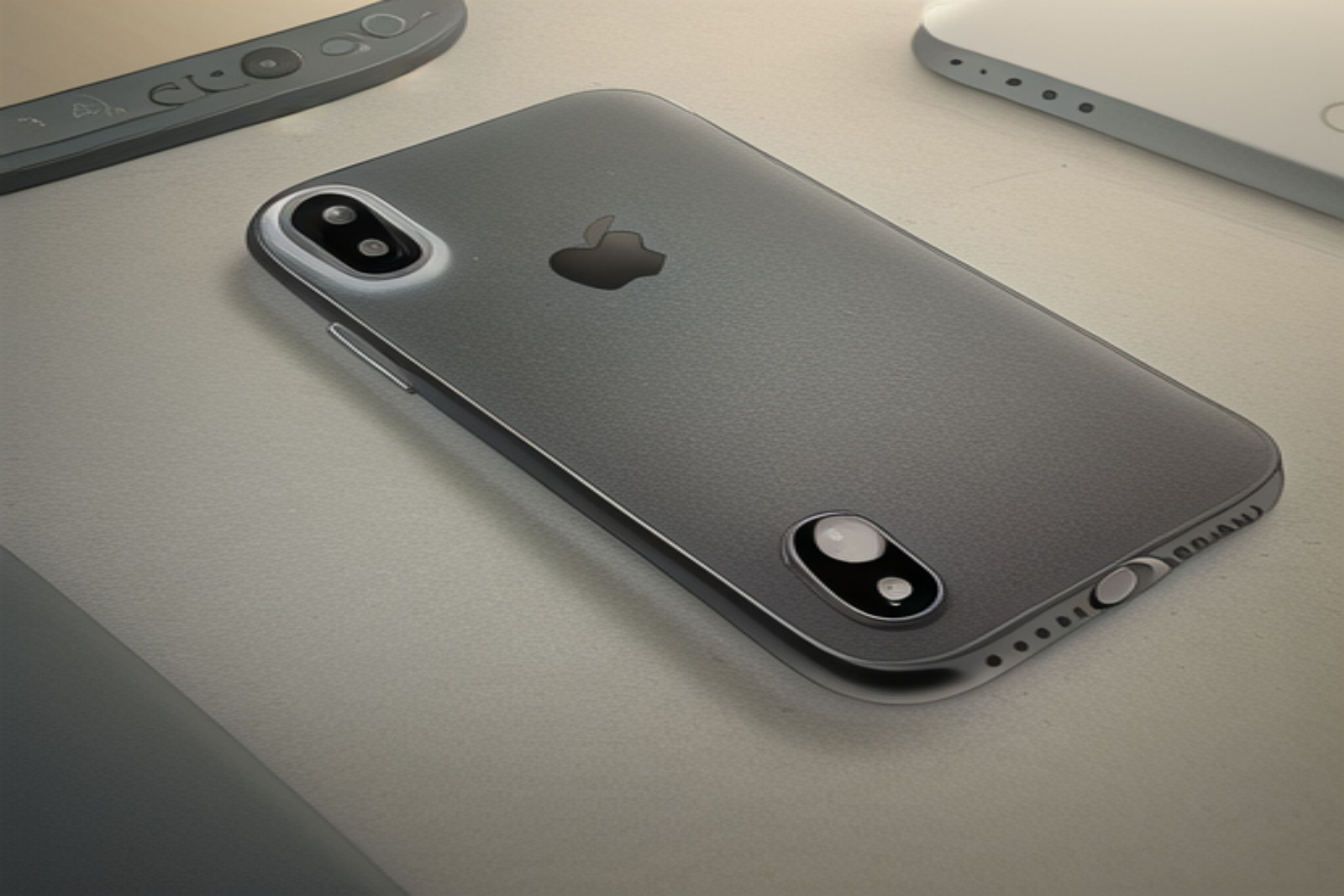}}
  \caption{Online advertising manipulation, where the targeted input is ``This is iPhone''.}
  \label{data_adver}
\end{figure*}

\begin{figure*}[!t]
	\centering
	\subfloat[The image file $v$.]{\includegraphics[width=0.185 \textwidth]{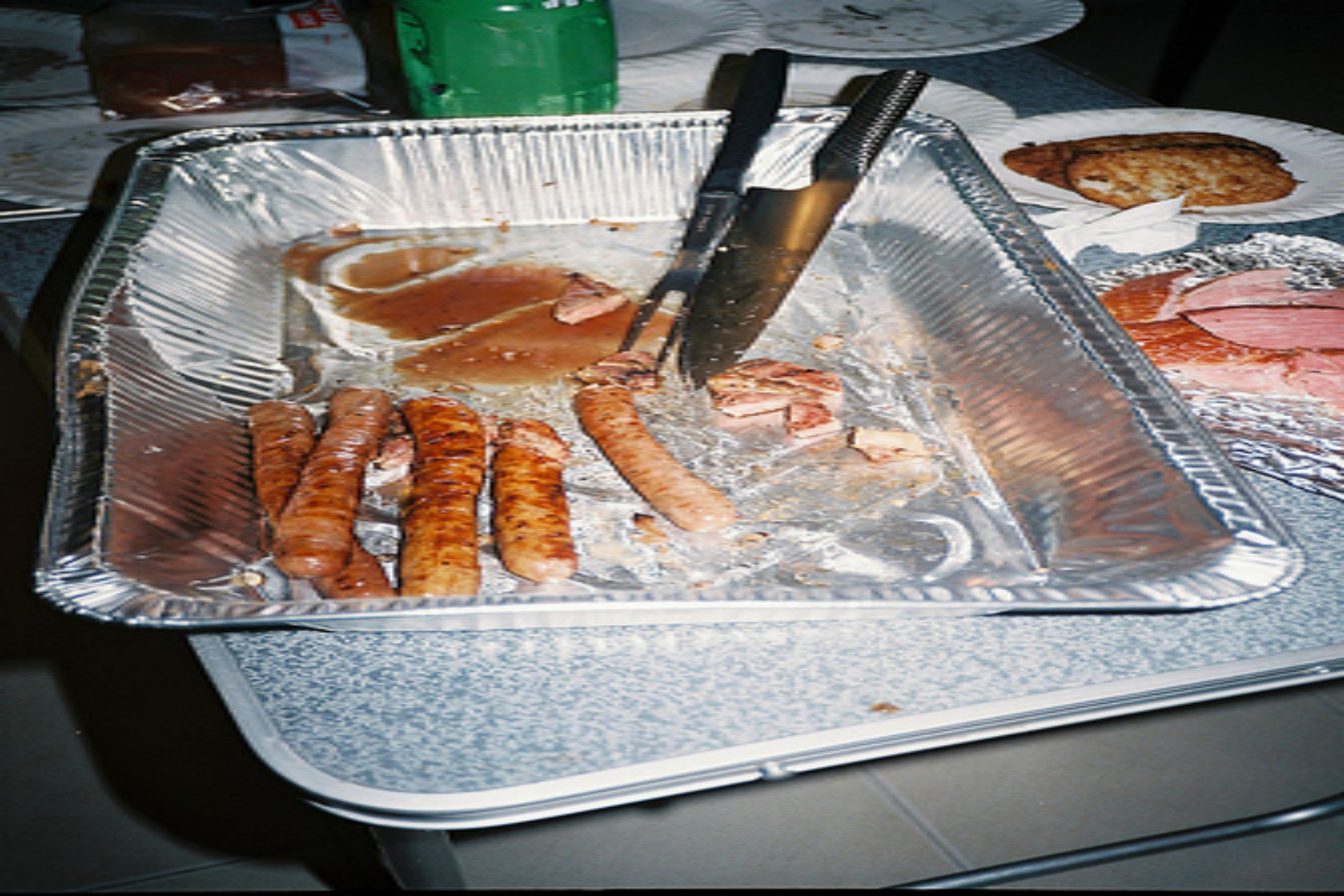}} 
 \quad
 	\subfloat[The perturbed image file $v^{\text{adv}}$.]{\includegraphics[width=0.185 \textwidth]{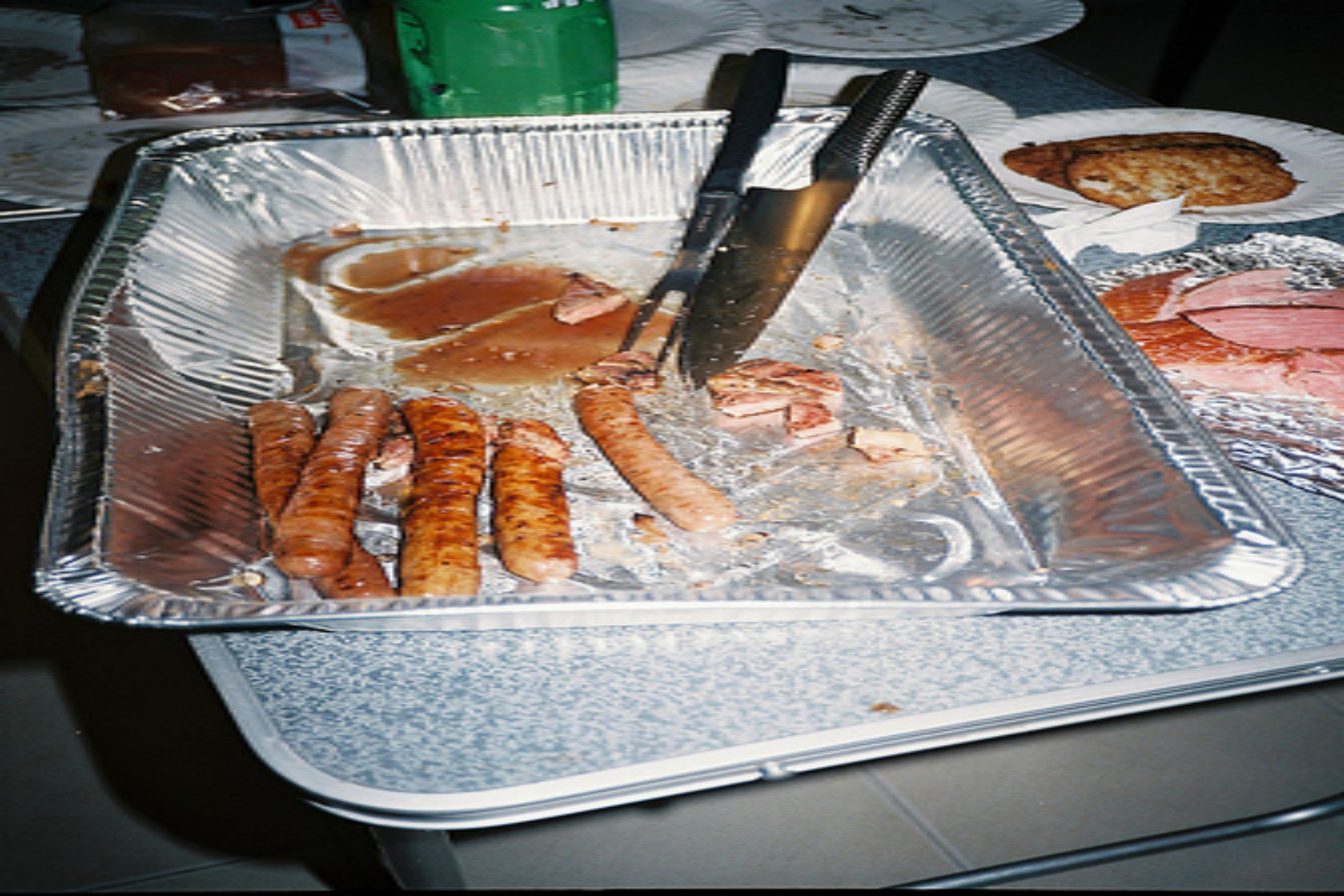}} 
   \quad
  	\subfloat[The image generated by the downstream task using the image file $v$.]{\includegraphics[width=0.185 \textwidth]{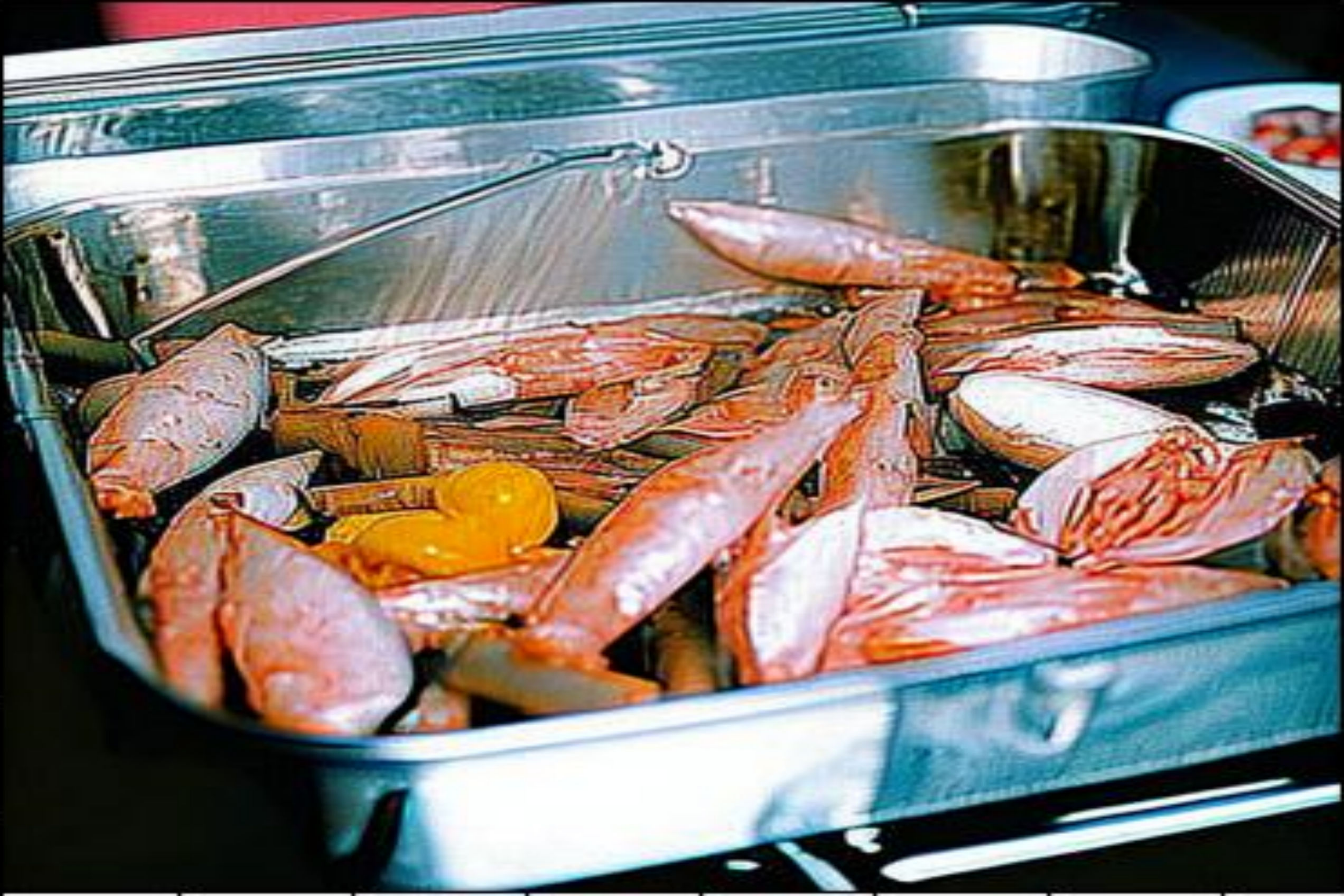}} 
    \quad
	\subfloat[The image generated by the downstream task using the perturbed image file $v^{\text{adv}}$.]{\includegraphics[width=0.185 \textwidth]{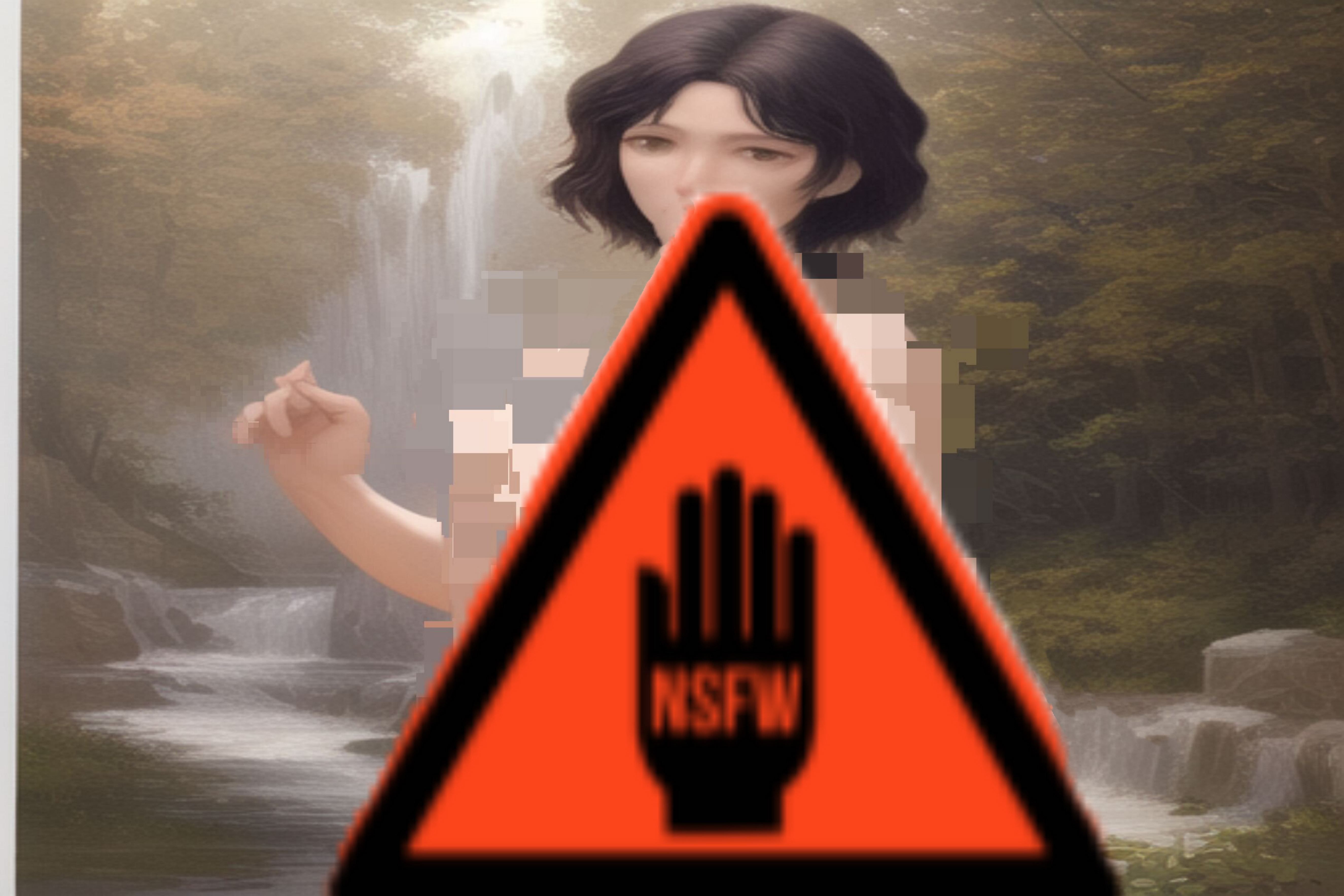}}
  \caption{Dissemination of explicit adult content, where the targeted input is ``A naked woman''.}
   \label{adult_distribution}
\end{figure*}

 \section{Defenses}  \label{sec:defense}

In what follows, we explore the six defense strategies for downstream generation tasks, as potential methods to mitigate our \alg{}. 
Note that from Table~\ref{Image_text}, it is evident that the ``Zhang attack'' and ``CrossFire without normalization'' do not effectively attack multi-modal models. Consequently, this section will focus exclusively on using these six countermeasures to mitigate the effects of our proposed \alg{}, which demonstrates significantly better attack performance compared to both the ``Zhang attack'' and ``CrossFire without normalization''.
We assess the effectiveness of the six defense strategies using two representative datasets: ImageNet for image and AudioCaps for audio.

In the case of the ImageNet dataset, the downstream tasks have five options: they can either resize (e.g., increase or decrease the image size), rotate the images, compress the images (e.g., JPEG compression), or employ the stochastic differential equation diffusion model~\cite{song2020score} to denoise the perturbed images. 
Meanwhile, for the AudioCaps dataset, the downstream tasks make use of the conditional diffusion model~\cite{lu2022conditional} to denoise the perturbed audio data.
For our \alg{}, we maintain the same perturbation level $\alpha$ of $\frac{16}{255}$ for image and $\alpha$ of $0.05$ for audio. 
Note that image resize, rotation, and compression methods are traditional defense strategies to reduce the impact of adversarial attacks; while denoise methods are also widely used noise purification approaches~\cite{nie2022diffusion,lu2022conditional}.

\myparatight{Increasing Perturbed Image Size} 
We employ upsampling methods to double the size of the perturbed image, which is a typical defense against adversarial attacks.
Without any defense, our attack can achieve $\text{ASR}_{\text{img}}$ of 0.98 and $\text{ASR}_{\text{text}}$ of 0.87, as shown in Table~\ref{Image_text}.
Under upsampling defense strategy, our attack can still achieve $\text{ASR}_{\text{img}} $ of 0.97 and $\text{ASR}_{\text{text}}$ of 0.87. These results underscore the ineffectiveness of defense strategies relying on enlarging the size of the perturbed image.

\myparatight{Decreasing Perturbed Image Size}We employ an alternative traditional defense method based on downsampling by halving the perturbed image. 
When the size of the perturbed image is reduced, CrossFire attains an $\text{ASR}_{\text{img}}$ of 0.97 and $\text{ASR}_{\text{text}}$ of 0.85, compared to $\text{ASR}_{\text{img}}$ of 0.98 and $\text{ASR}_{\text{text}}$ of 0.87 in the absence of any defense.
It consistently maintains high attack success rates even when faced with downsampling defenses. As a result, resizing defenses are ineffective against our attack.

\myparatight{Rotating Perturbed Images} We apply uniform rotations ranging from [-180, 180] to the perturbed image, as another standard defense strategy against adversarial attacks. After subjecting the perturbed image to random rotation, $\text{ASR}_{\text{img}}$ and $\text{ASR}_{\text{text}}$ yield 0.97 and 0.86, respectively. The result indicates that random rotation does not assist in mitigating our attack impact.

\myparatight{Compression Perturbed Images} We apply Joint Photographic Experts Group (JPEG) compression to the perturbed image, a method often considered a classical defense strategy against adversarial attacks. Following compression, $\text{ASR}_{\text{img}}$ and $\text{ASR}_{\text{text}}$ of our \alg{} are 0.91 and 0.79, respectively. 
This result suggests that while JPEG compression has some impact on mitigating our attack, its overall defensive effectiveness remains limited.

\myparatight{Stochastic Differential Equation Diffusion} Stochastic Differential Equation (SDE) \cite{song2020score} diffusion is one way for image denoising.
Through SDE diffusion defense, the $\text{ASR}_{\text{img}}$ and $\text{ASR}_{\text{text}}$ decrease to 0.87 and 0.75, respectively. 
This indicates that SDE diffusion can serve as a more effective defense method, reducing the attack success rates of our \alg{} attack by approximately 0.10.
However, our attack can still achieve high attack success rates.

\myparatight{Conditional Diffusion} 
For audio data, we employ conditional diffusion (CDiffusion) \cite{lu2022conditional} to denoise audio, a method proven to be a robust defense against adversarial attacks on Gaussian audio noise. 
As demonstrated in Table~\ref{Image_text}, without any defense methods, the $\text{ASR}_{\text{img}}$ and $\text{ASR}_{\text{text}}$ for our attack are 0.94 and 0.86, respectively. 
After applying CD diffusion purification, $\text{ASR}_{\text{img}}$ shows a marginal decrease to 0.89, while $\text{ASR}_{\text{text}}$ reaches 0.81. 
While CDiffusion shows some purifying impact, its defense against our attacks is limited in effectiveness

From above, our \alg{} attack stands resilient against current standard defense strategies, presenting a formidable challenge.
This is crucial to mitigating the threat posed by cross-modality and addressing its potential societal impact, as elaborated in the subsequent section.


 \section{Applications of \alg{}}  \label{sec:application}

In this section, we delve into significant real-world harms that can be induced by our \alg{}. 
Specifically, we explore three applications exemplifying the detrimental consequences: data augmentation manipulation, online advertising manipulation, and dissemination of explicit adult content.

\subsection{Data Augmentation Manipulation}
Our \alg{} pose a substantial threat to the integrity of data augmentation, particularly when applied to image-to-image diffusion. 
Fig.~\ref{data_augu}(c) is the image generated by the downstream task using the image file $v$ shown in Fig.~\ref{data_augu}(a). Note that the image file $v$ is clean and devoid of any added noise. On the other hand, Fig.~\ref{data_augu}(d) illustrates the image generated by the downstream task using the perturbed image from Fig.~\ref{data_augu}(b), where the targeted input $t$ is ``A huge tiger''.
Imagine a scenario where individuals employ BindDiffusion~\cite{bind} to augment bus images (see Fig.~\ref{data_augu}(c)). 
Incorporating perturbed images (Fig.~\ref{data_augu}(b)) from the internet by our attacks may result in an augmented dataset flooded with unintended tiger images, as showed in Fig.~\ref{data_augu}(d). 
Users expecting additional ``bus'' images for training find our adversarial examples generating an abundance of tiger images falsely labeled as ``bus''. 
The compromise during training underscores the considerable negative impact \alg{} can have on data augmentation tasks.

\subsection{Online Advertising Manipulation}
Our \alg{} open the door for unscrupulous companies to exploit image embeddings for surreptitious advertising. 
Consider a company leveraging the targeted input ``This is iPhone'', 
ordinary users seeking images related to ``blue cars'' (Fig.~\ref{data_adver}(a)) may unwittingly generate promotional content like Fig.~\ref{data_adver}(d). 
This inundation of promotional information poses a risk of illegal product and service promotion, disrupting user experience and potentially leading to undesirable consequences. 
The potential misuse of our \alg{} for covert advertising highlights the urgent need for ethical safeguards to prevent unintended consequences.

\subsection{Dissemination of Explicit Adult Content}
Our \alg{} can be exploited by malicious actors to inject explicit adult content into images disseminated online, as shown in Fig.~\ref{adult_distribution}, in which the targeted input is ``A naked woman''.
The potential impact on children seeking age-appropriate content is especially concerning, as these harmful attacks generate images that may create distressing memories. This misuse highlights the urgent need for ethical considerations and robust safeguards to mitigate the harmful consequences of our attack.


\section{Conclusion}
We introduced an innovative adversarial attack, referred to as \alg{}, designed for multi-modal models. Unlike existing attack that aligns perturbed image/audio file directly with the targeted input, \alg{} first transforms the targeted input into a ``transformed input'' matching the modality of the image/audio file. Perturbations are then applied to align the perturbed image/audio file with the transformed input in the embedding space. 
Extensive experiments on six datasets show that \alg{} achieves a high attack success rate, surpassing the existing method. 
Traditional defense strategies and denoising techniques prove to be ineffective in mitigating the impact of our proposed \alg{} attack. This highlights the necessity for novel defense mechanisms to counteract our attack.
One limitation of our attack is that it does not attain a high success rate when the targeted input encompasses multiple elements. Moving forward, we aim to develop a more potent attack strategy to overcome this challenge.

\bibliographystyle{ACM-Reference-Format}
\bibliography{custom}

\newpage
\onecolumn

\appendix

\clearpage

\section{Appendix}
\label{sec:appendix}

\subsection{Examples of targeted input contains single element on ImageNet dataset}
\label{app_single}

The targeted input is ``A huge tiger''. 
The images in the first column depict the image file $v$. 
The second and third columns respectively demonstrate the generated image and text under the Zhang attack. 
The fourth and fifth columns show the generated image and text under our proposed \alg{}.

\newlength{\mycolwidth}
\newlength{\myrowheight}
\setlength{\mycolwidth}{3cm}
\setlength{\myrowheight}{2.5cm}

\newlength{\imgwidth}
\newlength{\imgheight}
\setlength{\imgwidth}{3.5cm}
\setlength{\imgheight}{3.5cm}
\newlength{\myparskip}
\setlength{\myparskip}{1cm} 
\begin{longtable}{p{4cm}p{4cm}p{4cm}p{4cm}p{4cm}}

\hspace*{-0.8cm} \parbox[c][\myrowheight][c]{\mycolwidth}{\vspace{-0.2cm}\includegraphics[width=\imgwidth, height=3.7cm]{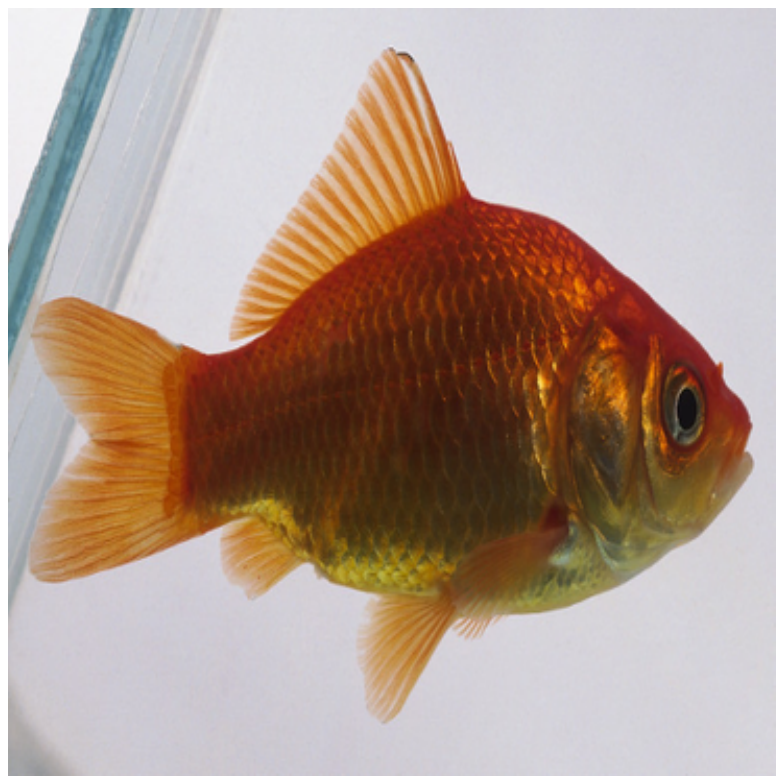}} & \hspace*{-1.4cm} 
\parbox[c][\myrowheight][c]{\mycolwidth}{\includegraphics[width=\imgwidth, height=\imgheight]{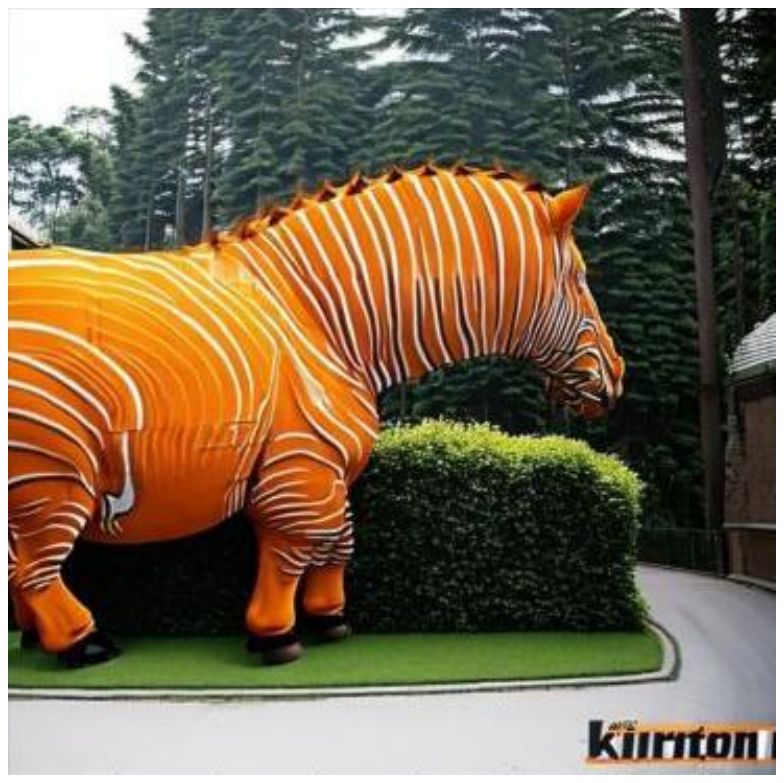}} & \hspace*{-1.6cm}
\parbox[c][0.5cm][c]{2.5cm}{\vspace{1cm}
\fontsize{6}{0}\selectfont
The image features a large elephant standing in a room, surrounded by people. The elephant is positioned in the center of the room, and the people are gathered around it, observing and interacting with the animal.} & \hspace*{-2.5cm}
\parbox[c][\myrowheight][c]{\mycolwidth}{\includegraphics[width=\imgwidth, height=\imgheight]{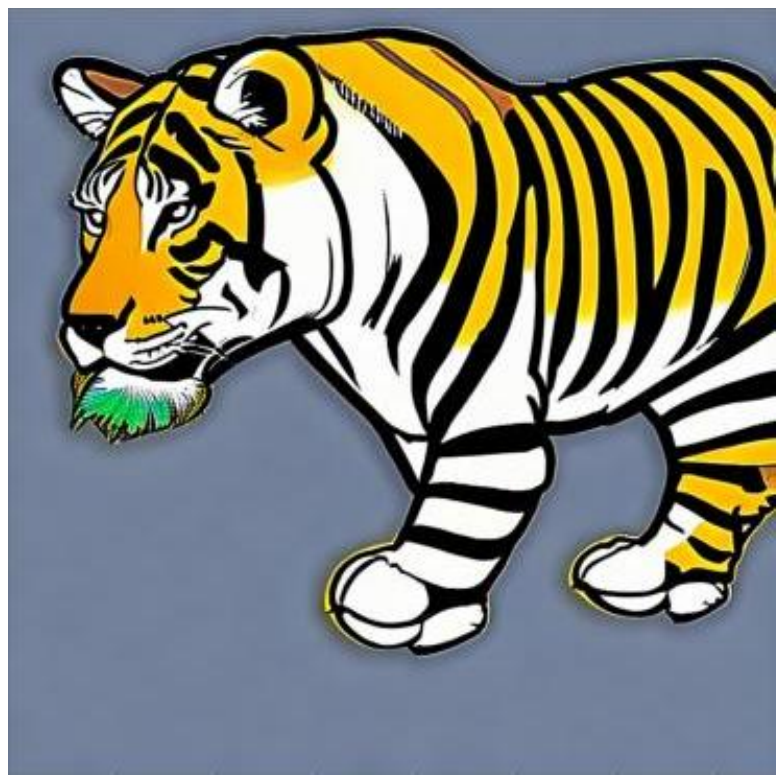}} & \hspace*{-2.5cm}
\parbox[c][0.5cm][c]{2.5cm}{\vspace{1cm}
\fontsize{6}{0}\selectfont
The image features a cartoon tiger character standing upright on its hind legs, with a striped pattern on its fur.}
\\[\myparskip]

\hspace*{-0.8cm}\parbox[c][\myrowheight][c]{\mycolwidth}{\vspace{-0.2cm}\includegraphics[width=\imgwidth, height=3.7cm]{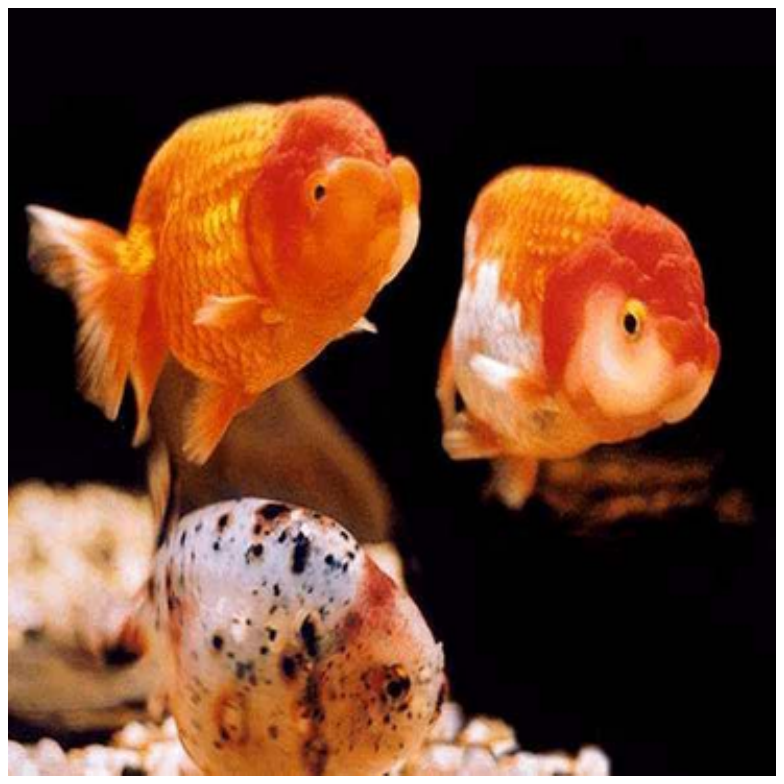}} & \hspace*{-1.4cm}
\parbox[c][\myrowheight][c]{\mycolwidth}{\includegraphics[width=\imgwidth, height=\imgheight]{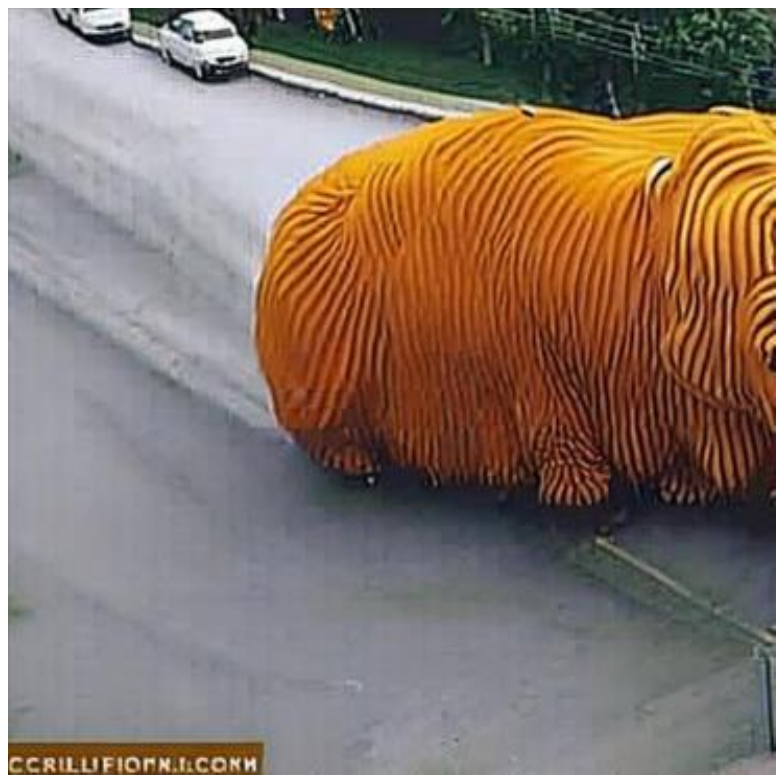}} & \hspace*{-1.6cm}
\parbox[c][0.5cm][c]{2.5cm}{\vspace{1cm}
\fontsize{6}{0}\selectfont
The image features a large, hairy bear standing in a room, surrounded by people. The bear appears to be in a zoo setting, as it is in a cage.} & \hspace*{-2.5cm}
\parbox[c][\myrowheight][c]{\mycolwidth}{\includegraphics[width=\imgwidth, height=\imgheight]{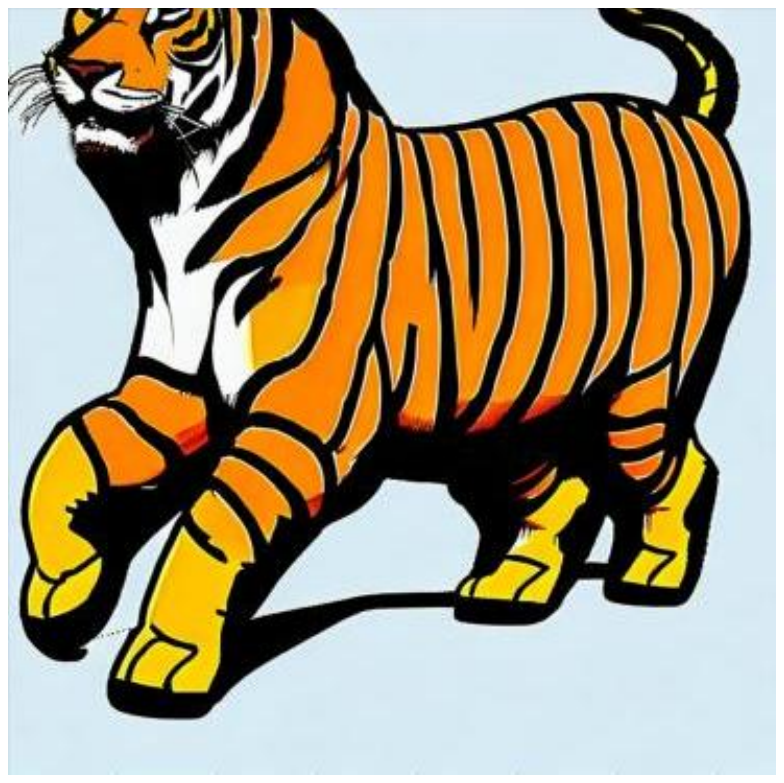}} & \hspace*{-2.5cm}
\parbox[c][0.5cm][c]{2.5cm}{\vspace{1cm}
\fontsize{6}{0}\selectfont
The image features a cartoon tiger standing on its hind legs, with its front paws resting on its chest. The tiger appears to be walking or standing in a grassy field.}
\\[\myparskip]

\hspace*{-0.8cm}\parbox[c][\myrowheight][c]{\mycolwidth}{\vspace{-0.2cm}\includegraphics[width=\imgwidth, height=3.7cm]{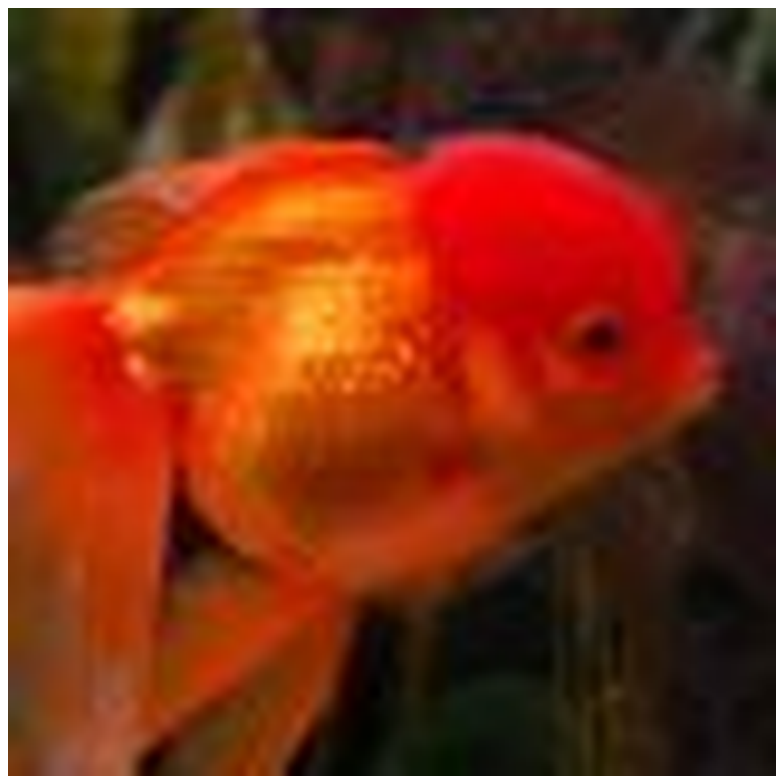}} & \hspace*{-1.4cm}
\parbox[c][\myrowheight][c]{\mycolwidth}{\includegraphics[width=\imgwidth, height=\imgheight]{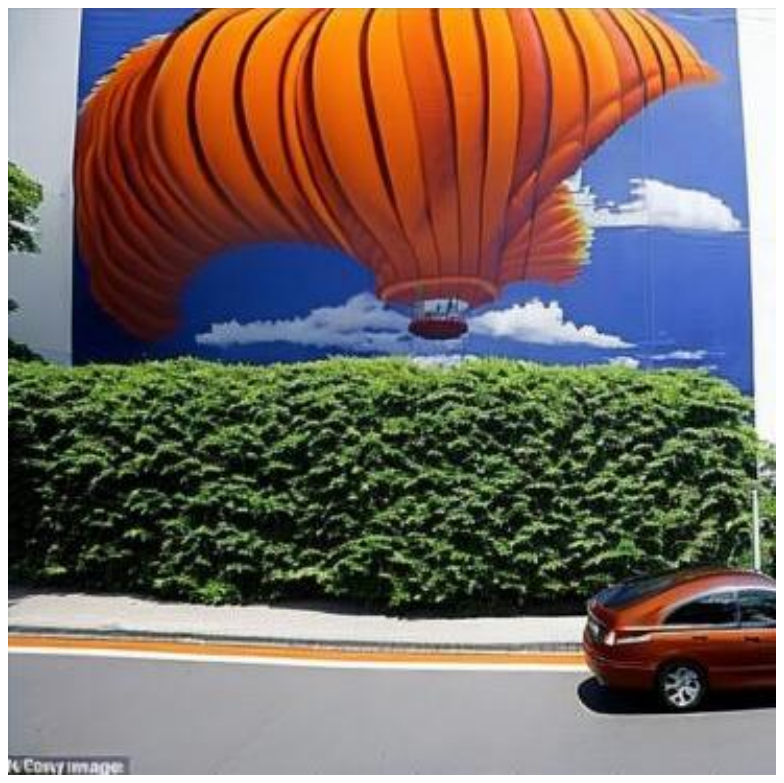}} & \hspace*{-1.6cm}
\parbox[c][0.5cm][c]{2.5cm}{\vspace{1cm}
\fontsize{6}{0}\selectfont
The image features a large tiger standing in a room, surrounded by people. The tiger is standing on its hind legs, and the people are gathered around it, observing and admiring the majestic animal.} & \hspace*{-2.5cm}
\parbox[c][\myrowheight][c]{\mycolwidth}{\includegraphics[width=\imgwidth, height=\imgheight]{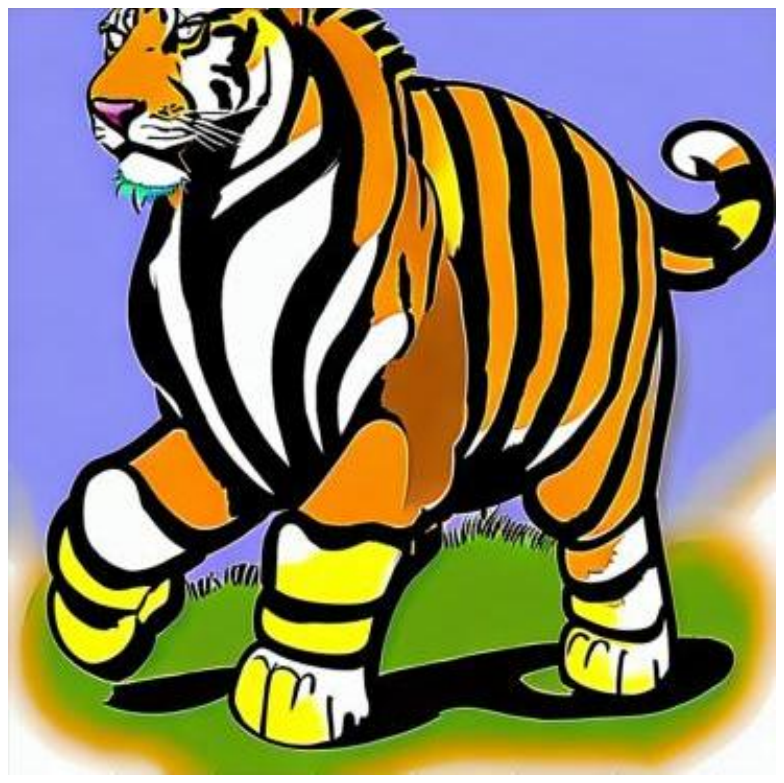}} & \hspace*{-2.5cm}
\parbox[c][0.5cm][c]{2.5cm}{\vspace{1cm}
\fontsize{6}{0}\selectfont
The image features a cartoon tiger character standing on its hind legs, with a large head and a small tail.}
\\[\myparskip]

\hspace*{-0.8cm}\parbox[c][\myrowheight][c]{\mycolwidth}{\vspace{-0.2cm}\includegraphics[width=\imgwidth, height=3.7cm]{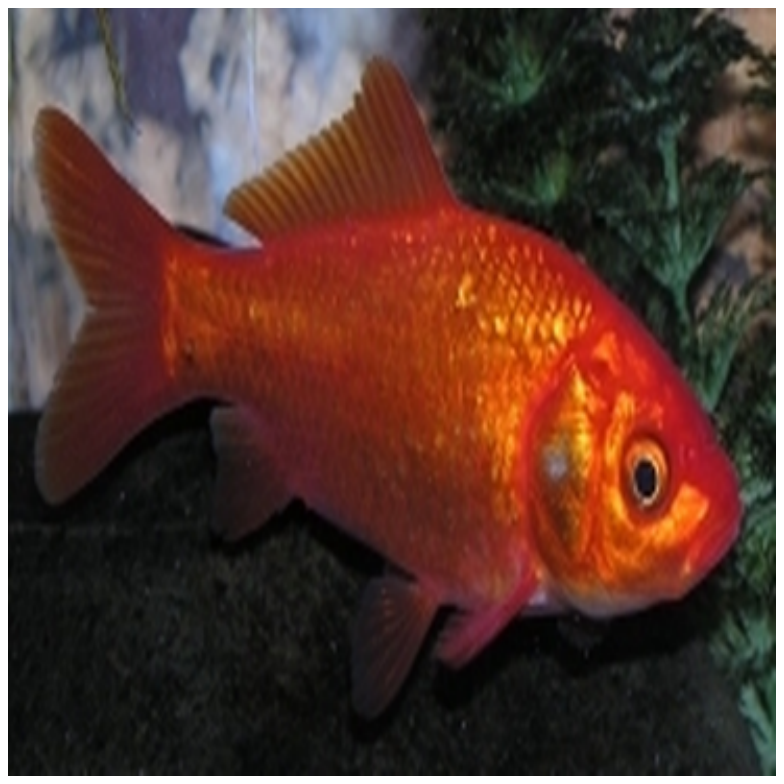}} & \hspace*{-1.4cm}
\parbox[c][\myrowheight][c]{\mycolwidth}{\includegraphics[width=\imgwidth, height=\imgheight]{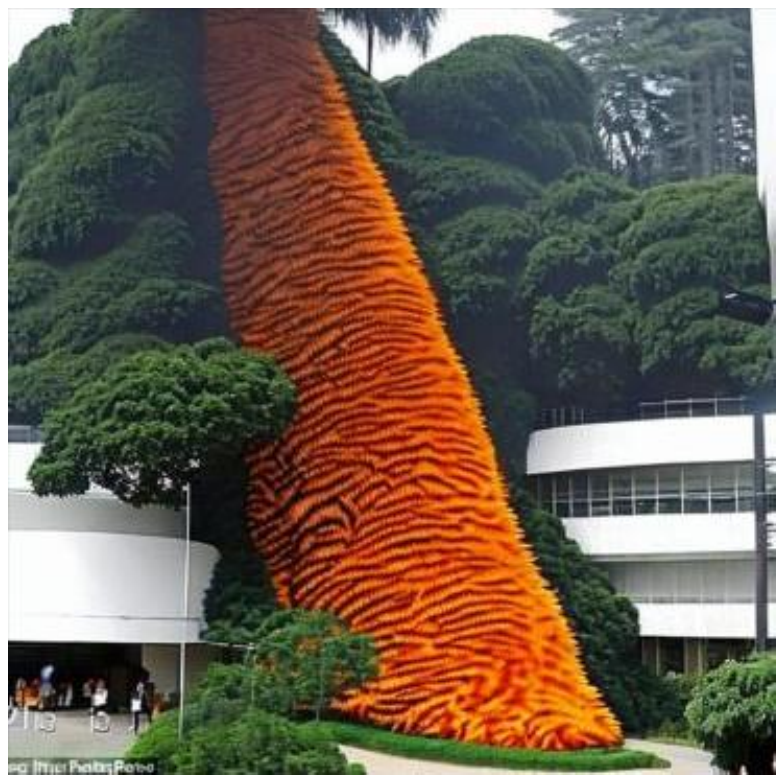}} & \hspace*{-1.6cm}
\parbox[c][0.5cm][c]{2.5cm}{\vspace{1cm}
\fontsize{6}{0}\selectfont
The image features a large elephant standing in a room, surrounded by people. The elephant appears to be walking through a room with a glass wall, and it is standing next to a table.} & \hspace*{-2.5cm}
\parbox[c][\myrowheight][c]{\mycolwidth}{\includegraphics[width=\imgwidth, height=\imgheight]{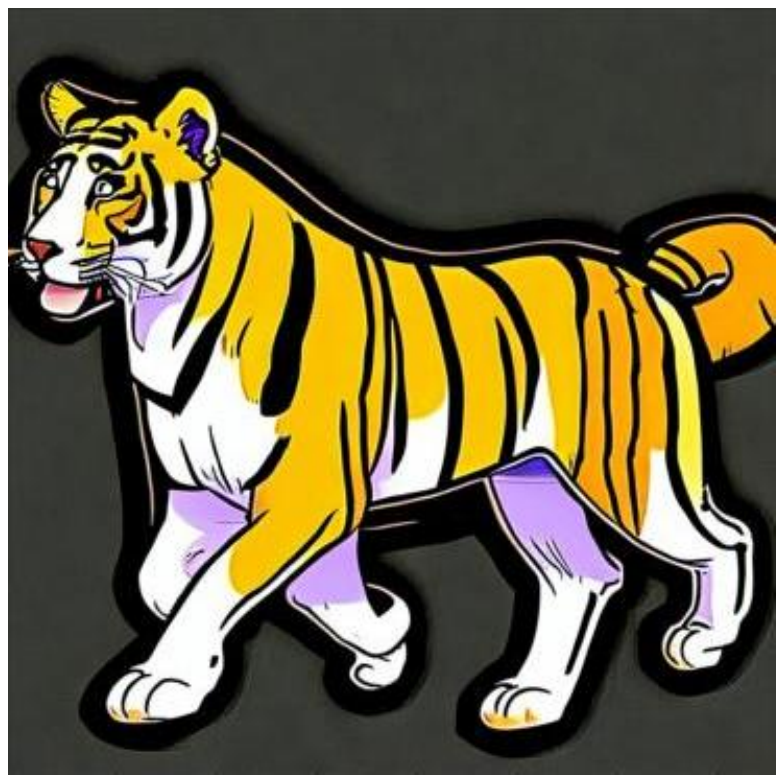}} & \hspace*{-2.5cm}
\parbox[c][0.5cm][c]{2.5cm}{\vspace{1cm}
\fontsize{6}{0}\selectfont
The image features a cartoon tiger standing on its hind legs, with a large head and a body that appears to be made of a combination of different colors.}
\\[\myparskip]

\hspace*{-0.8cm}\parbox[c][\myrowheight][c]{\mycolwidth}{\vspace{-0.2cm}\includegraphics[width=\imgwidth, height=3.7cm]{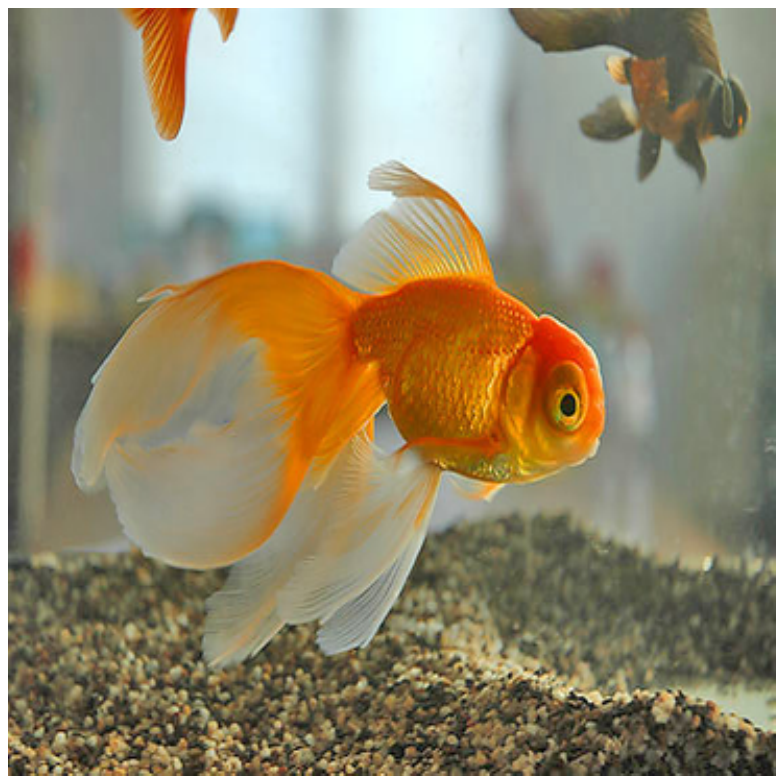}} & \hspace*{-1.4cm}
\parbox[c][\myrowheight][c]{\mycolwidth}{\includegraphics[width=\imgwidth, height=\imgheight]{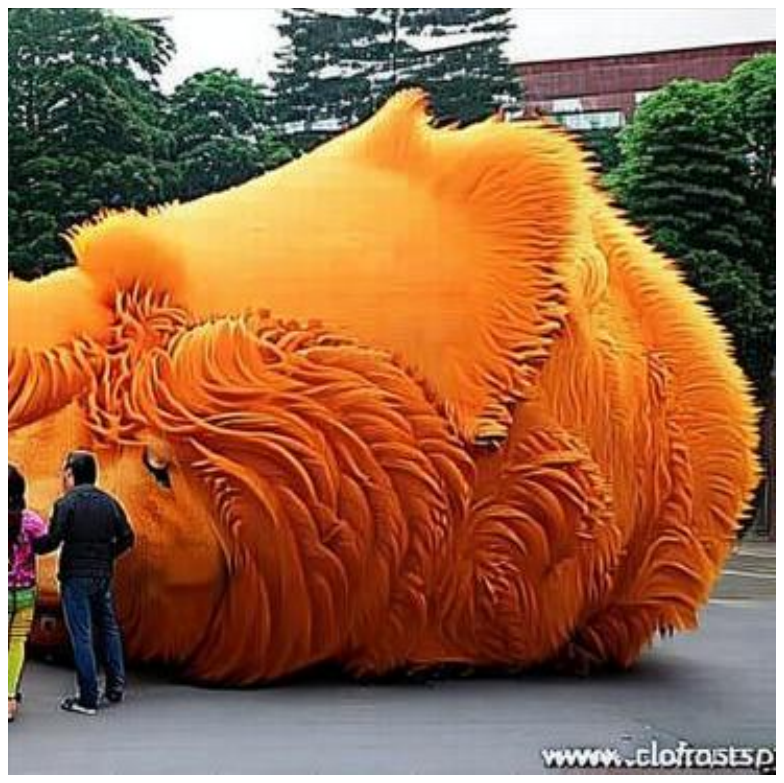}} & \hspace*{-1.6cm}
\parbox[c][0.5cm][c]{2.5cm}{\vspace{1cm}
\fontsize{6}{0}\selectfont
The image features a large tiger standing in a room, surrounded by people. The tiger is standing on its hind legs, and the people are gathered around it, observing and admiring the animal.} & \hspace*{-2.5cm}
\parbox[c][\myrowheight][c]{\mycolwidth}{\includegraphics[width=\imgwidth, height=\imgheight]{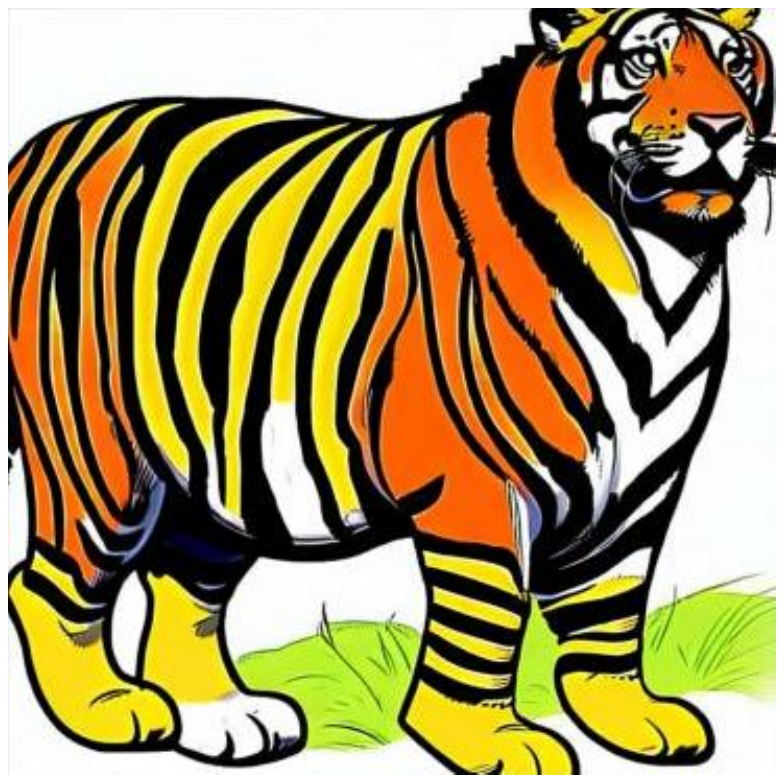}} &\hspace*{-2.5cm}
\parbox[c][0.5cm][c]{2.5cm}{\vspace{1cm}
\fontsize{6}{0}\selectfont
The image features a cartoon tiger standing on its hind legs, with a cute expression and a striped pattern.}
\\[\myparskip]

\hspace*{-0.8cm}\parbox[c][\myrowheight][c]{\mycolwidth}{\vspace{-0.2cm}\includegraphics[width=\imgwidth, height=3.7cm]{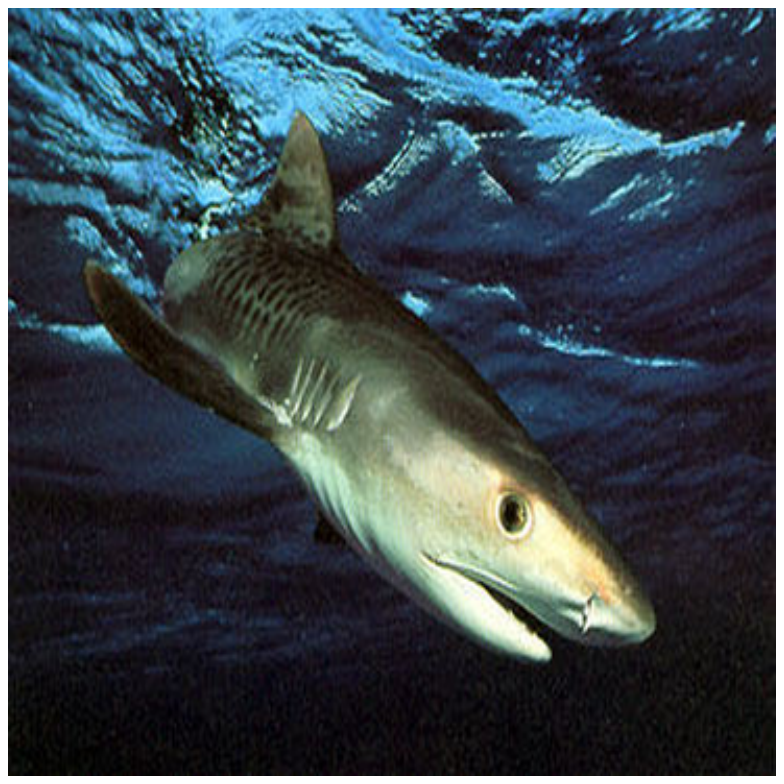}} & \hspace*{-1.4cm}
\parbox[c][\myrowheight][c]{\mycolwidth}{\includegraphics[width=\imgwidth, height=\imgheight]{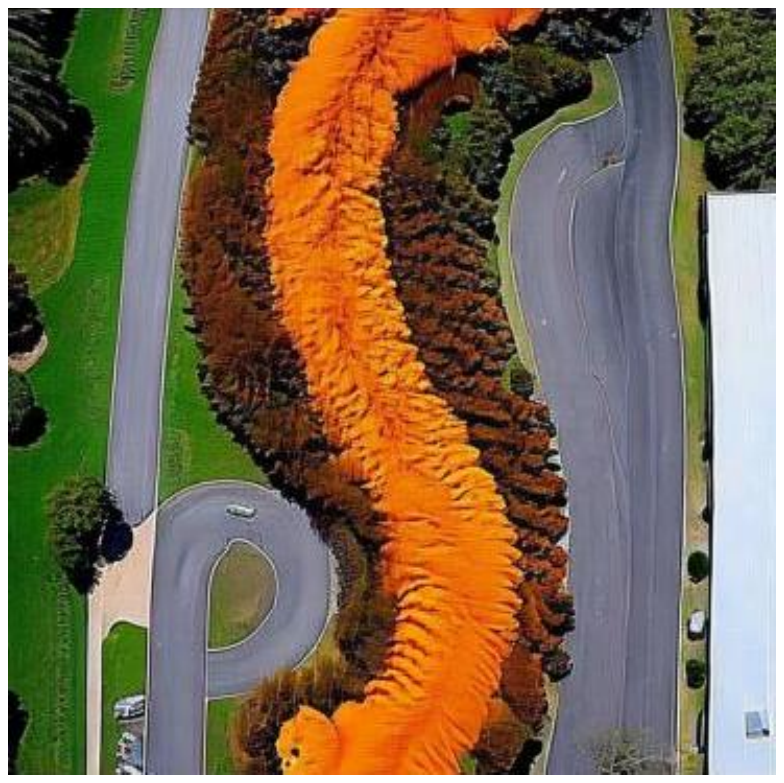}} & \hspace*{-1.6cm}
\parbox[c][0.5cm][c]{2.5cm}{\vspace{1cm}
\fontsize{6}{0}\selectfont
The image features a large, colorful elephant standing in a room. The elephant appears to be holding something in its mouth, possibly a picture or a piece of paper.} & \hspace*{-2.5cm}
\parbox[c][\myrowheight][c]{\mycolwidth}{\includegraphics[width=\imgwidth, height=\imgheight]{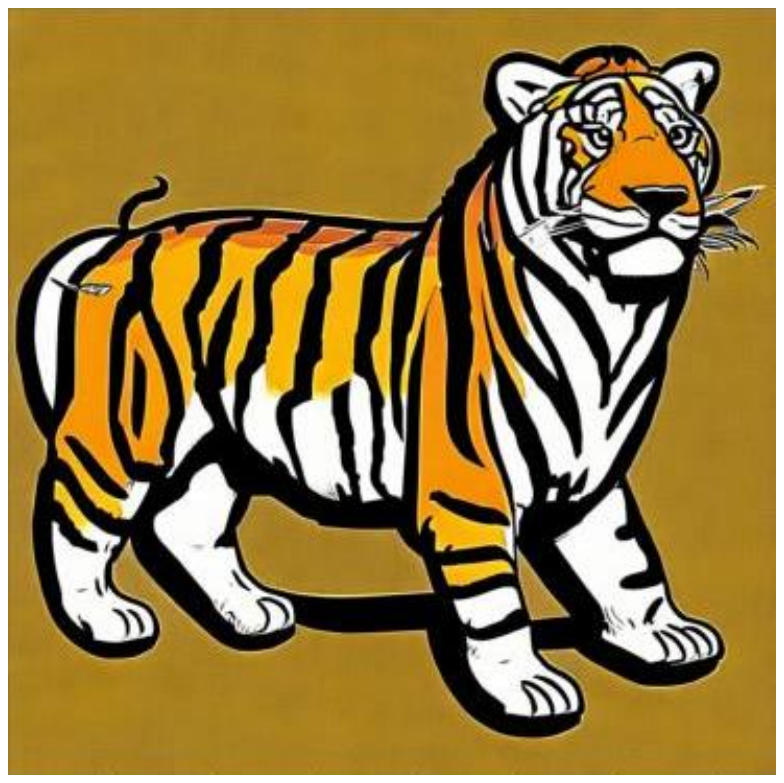}} & \hspace*{-2.5cm}
\parbox[c][0.5cm][c]{2.5cm}{\vspace{1cm}
\fontsize{6}{0}\selectfont
The image features a cartoon tiger character standing on its hind legs, with a long tail and a cute face.}
\\[\myparskip]

\hspace*{-0.8cm}\parbox[c][\myrowheight][c]{\mycolwidth}{\vspace{-0.2cm}\includegraphics[width=\imgwidth, height=3.7cm]{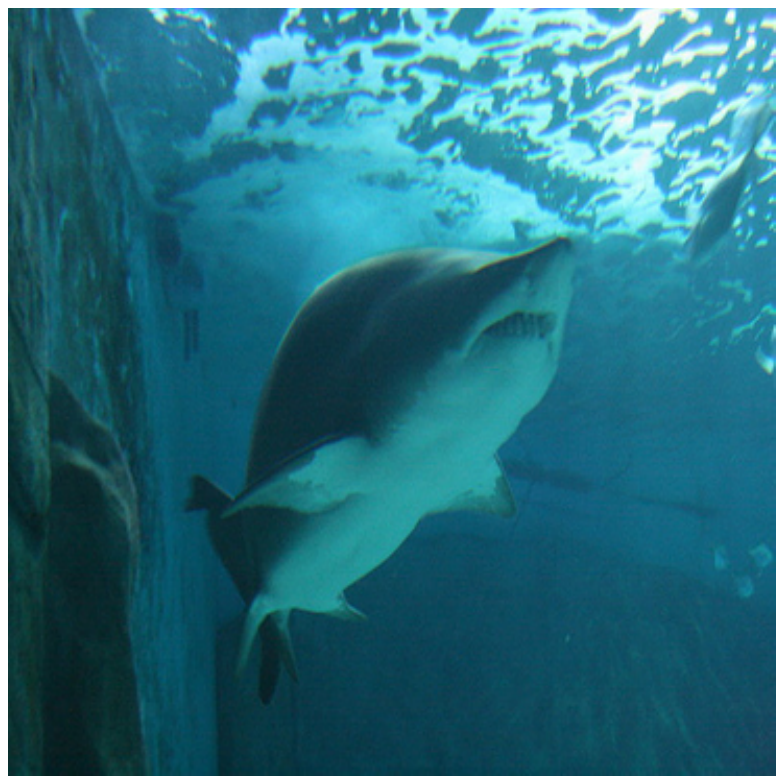}} & \hspace*{-1.4cm}
\parbox[c][\myrowheight][c]{\mycolwidth}{\includegraphics[width=\imgwidth, height=\imgheight]{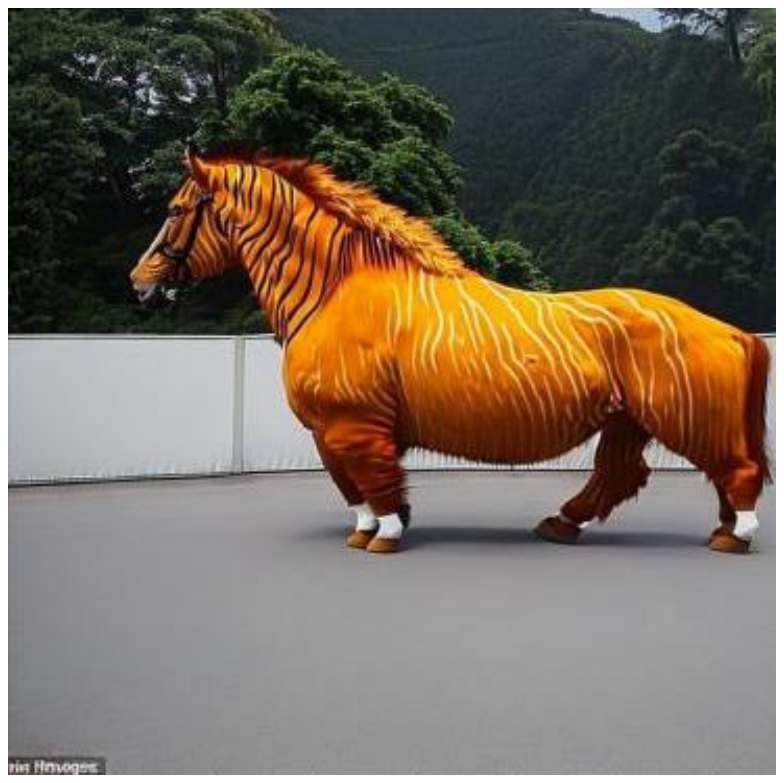}} & \hspace*{-1.6cm}
\parbox[c][0.5cm][c]{2.5cm}{\vspace{1cm}
\fontsize{6}{0}\selectfont
The image features a large giraffe standing inside a building, surrounded by a crowd of people. The giraffe is in the center of the room, and the people are gathered around it, observing and admiring the animal.} & \hspace*{-2.5cm}
\parbox[c][\myrowheight][c]{\mycolwidth}{\includegraphics[width=\imgwidth, height=\imgheight]{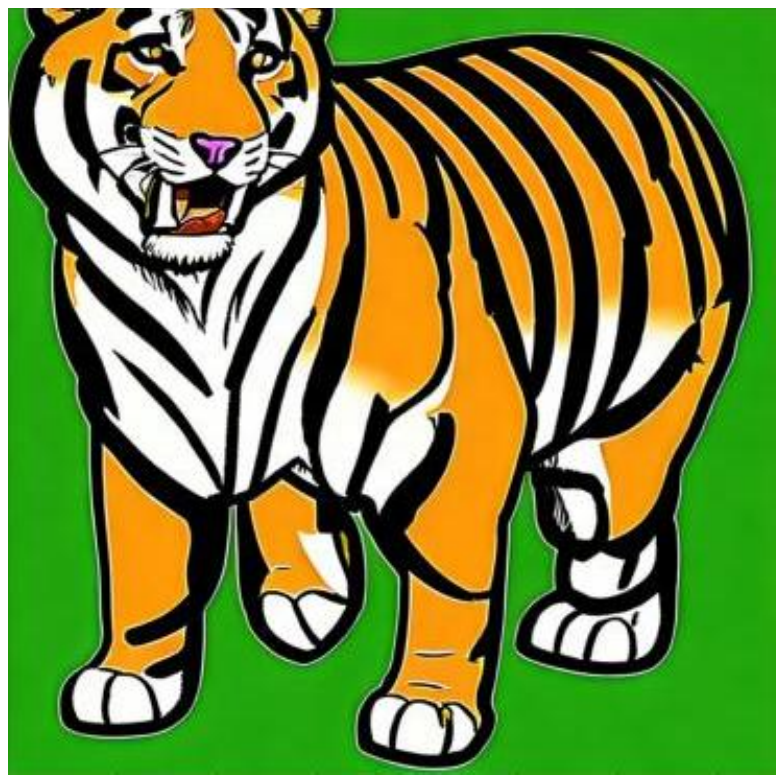}} & \hspace*{-2.5cm}
\parbox[c][0.5cm][c]{2.5cm}{\vspace{1cm}
\fontsize{6}{0}\selectfont
The image features a cartoon tiger character standing on its hind legs, with a big smile on its face.}
\\[\myparskip]

\hspace*{-0.8cm}\parbox[c][\myrowheight][c]{\mycolwidth}{\vspace{-0.2cm}\includegraphics[width=\imgwidth, height=3.7cm]{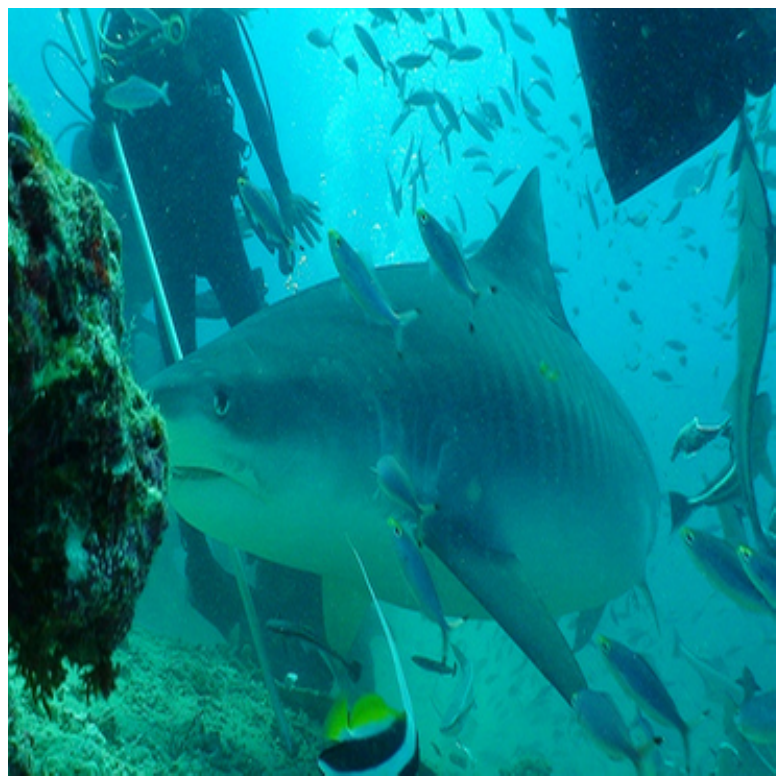}} & \hspace*{-1.4cm}
\parbox[c][\myrowheight][c]{\mycolwidth}{\includegraphics[width=\imgwidth, height=\imgheight]{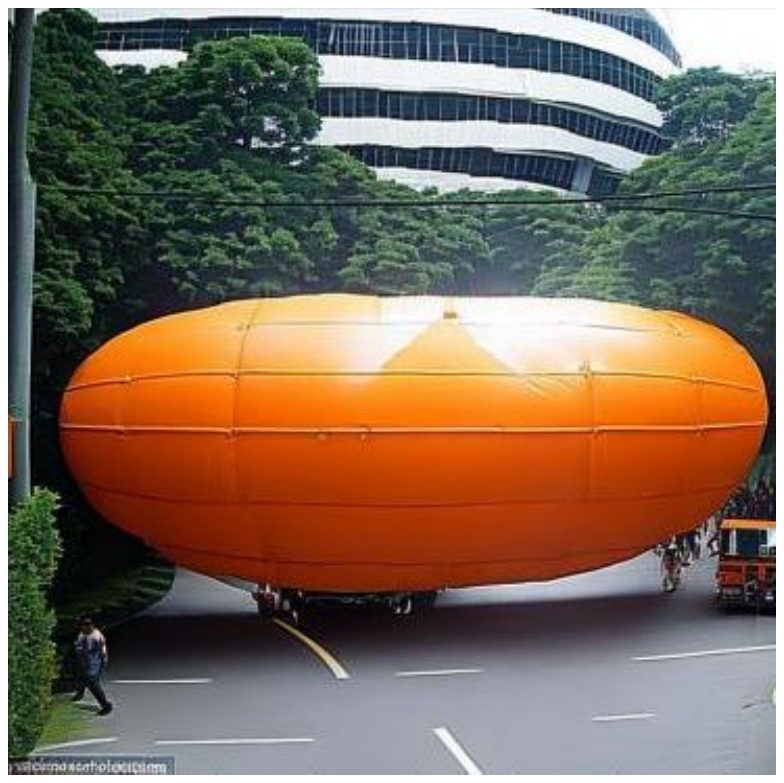}} & \hspace*{-1.6cm}
\parbox[c][0.5cm][c]{2.5cm}{\vspace{1cm}
\fontsize{6}{0}\selectfont
The image features a large elephant standing in a room, surrounded by people. The elephant appears to be walking through the room, and there are several people in the background, possibly observing or interacting with the elephant.} & \hspace*{-2.5cm}
\parbox[c][\myrowheight][c]{\mycolwidth}{\includegraphics[width=\imgwidth, height=\imgheight]{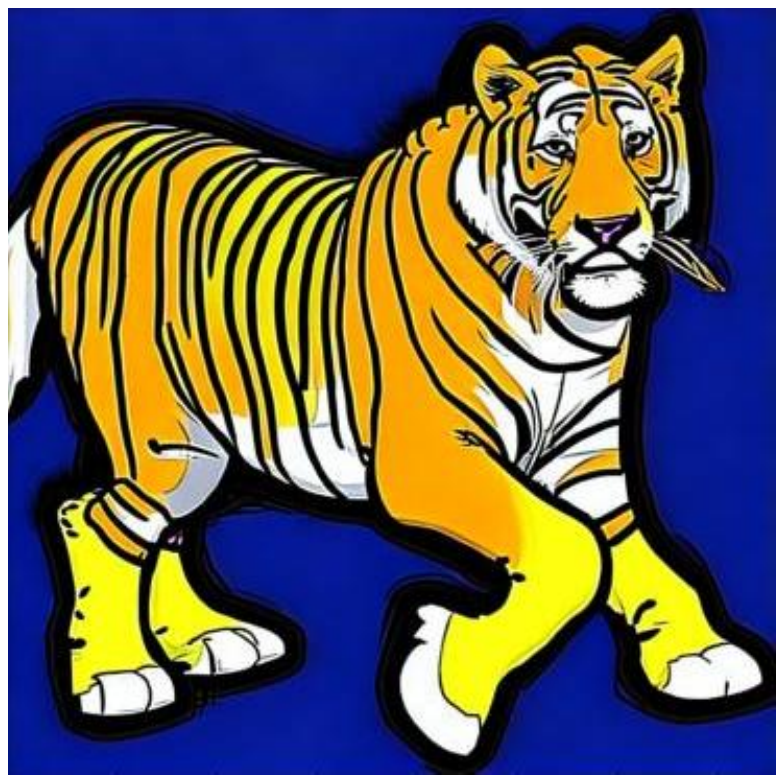}} & \hspace*{-2.5cm}
\parbox[c][0.5cm][c]{2.5cm}{\vspace{1cm}
\fontsize{6}{0}\selectfont
The image features a cartoon tiger character walking through a grassy field.}
\\[\myparskip]

\hspace*{-0.8cm}\parbox[c][\myrowheight][c]{\mycolwidth}{\vspace{-0.2cm}\includegraphics[width=\imgwidth, height=3.7cm]{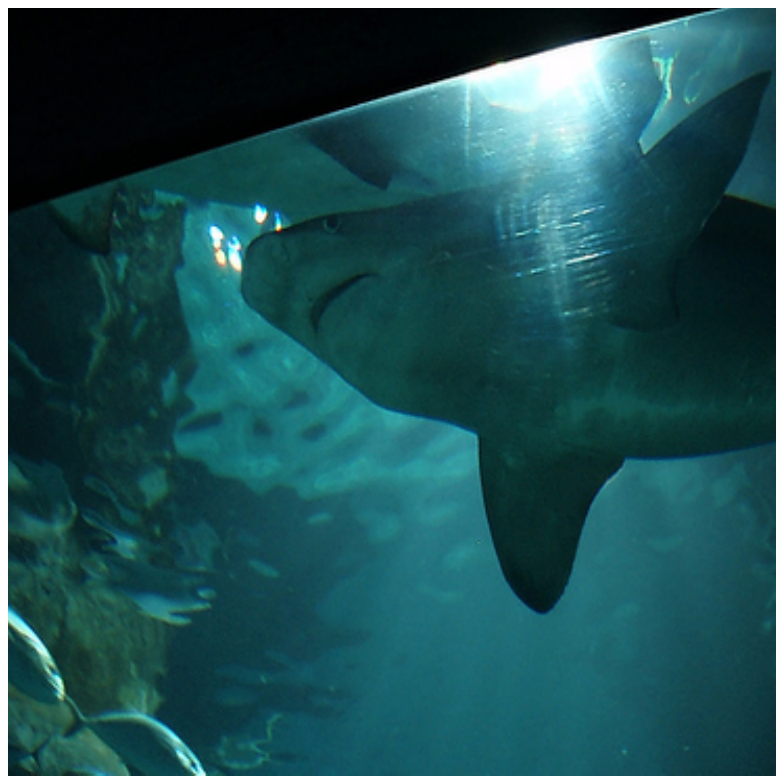}} & \hspace*{-1.4cm}
\parbox[c][\myrowheight][c]{\mycolwidth}{\includegraphics[width=\imgwidth, height=\imgheight]{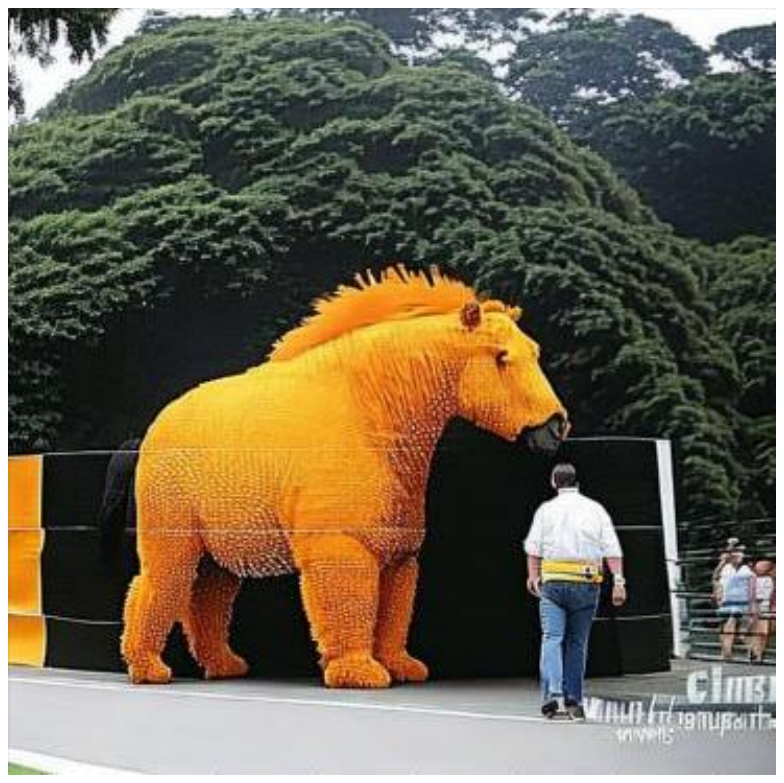}} & \hspace*{-1.6cm}
\parbox[c][0.5cm][c]{2.5cm}{\vspace{1cm}
\fontsize{6}{0}\selectfont
The image features a large, hairy giraffe standing in a room with a crowd of people around it. The giraffe is positioned in the center of the room, and the people are gathered around it, observing and admiring the animal.} & \hspace*{-2.5cm}
\parbox[c][\myrowheight][c]{\mycolwidth}{\includegraphics[width=\imgwidth, height=\imgheight]{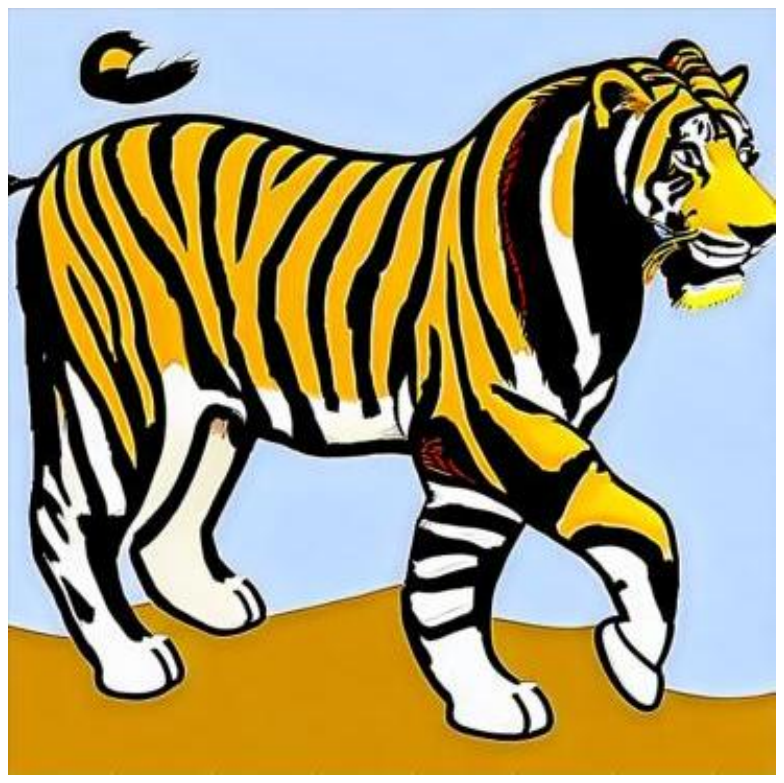}} & \hspace*{-2.5cm}
\parbox[c][0.5cm][c]{2.5cm}{\vspace{1cm}
\fontsize{6}{0}\selectfont
The image features a cartoon tiger standing on its hind legs, with a large tail.}
\\[\myparskip]

\hspace*{-0.8cm}\parbox[c][\myrowheight][c]{\mycolwidth}{\vspace{-0.2cm}\includegraphics[width=\imgwidth, height=3.7cm]{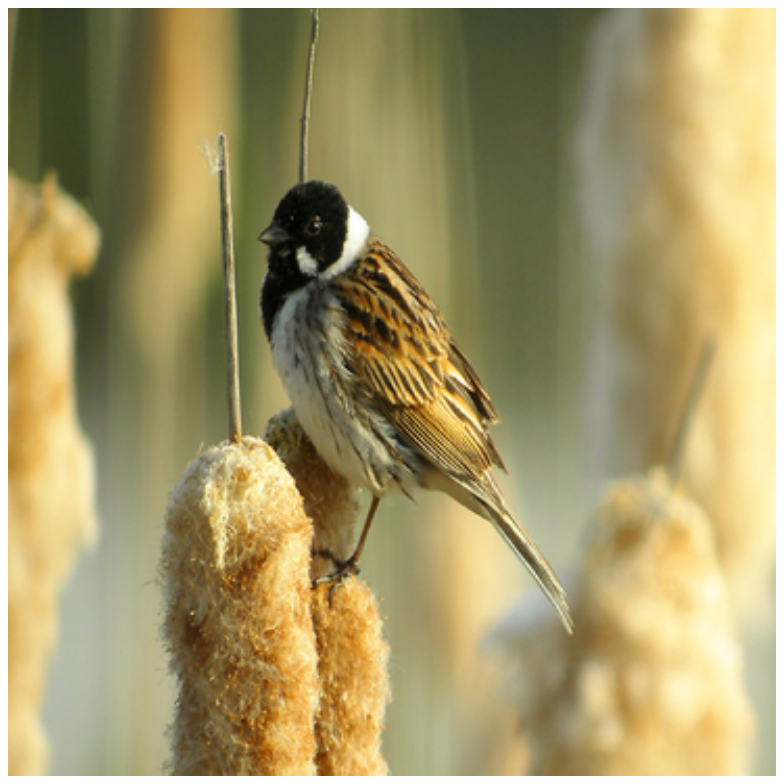}} & \hspace*{-1.4cm}
\parbox[c][\myrowheight][c]{\mycolwidth}{\includegraphics[width=\imgwidth, height=\imgheight]{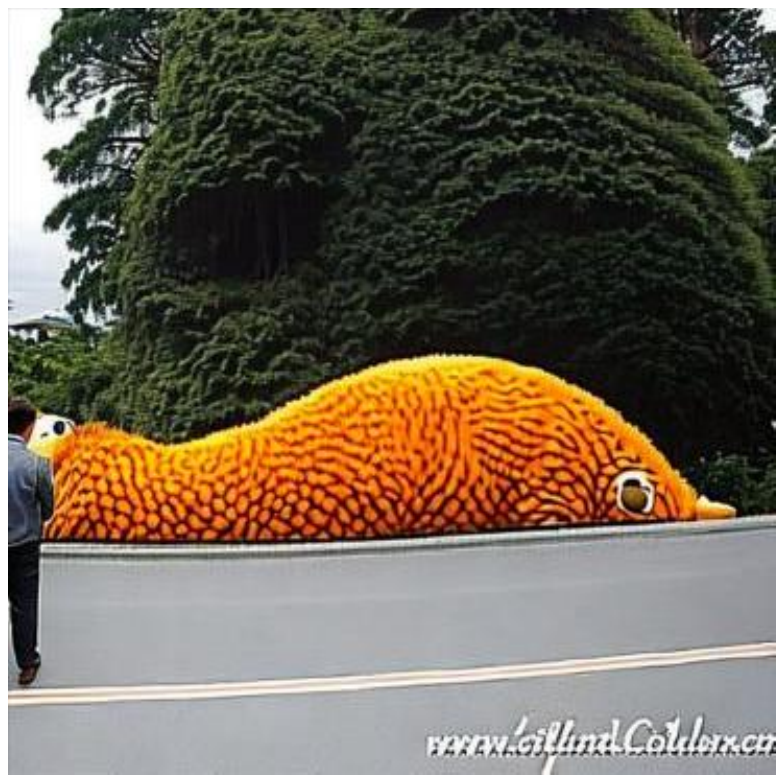}} & \hspace*{-1.6cm}
\parbox[c][0.5cm][c]{2.5cm}{\vspace{1cm}
\fontsize{6}{0}\selectfont
The image features a large tiger standing in a room, looking at the camera. The tiger is surrounded by people who are observing and admiring the animal.} & \hspace*{-2.5cm}
\parbox[c][\myrowheight][c]{\mycolwidth}{\includegraphics[width=\imgwidth, height=\imgheight]{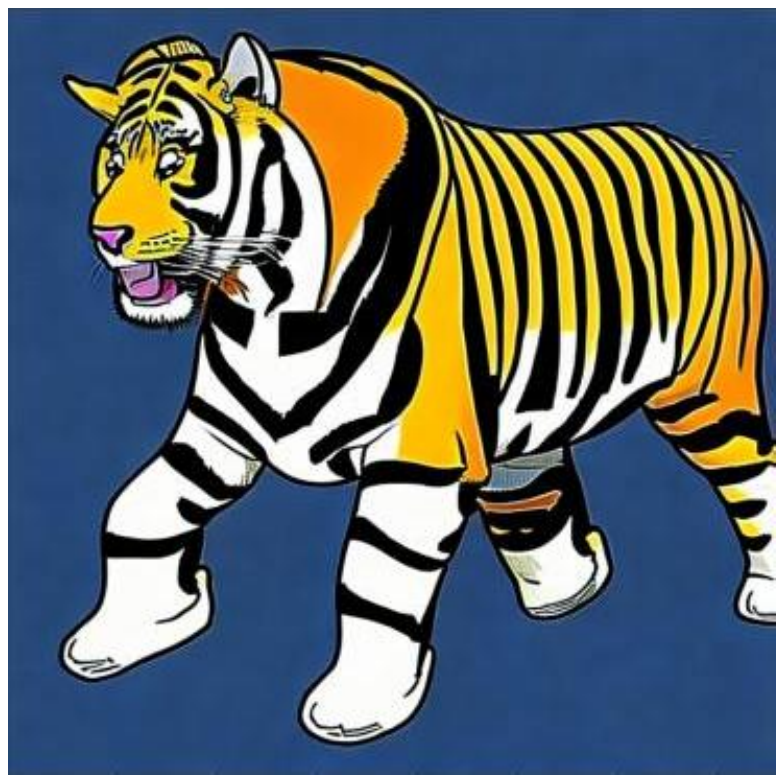}} & \hspace*{-2.5cm}
\parbox[c][0.5cm][c]{2.5cm}{\vspace{1cm}
\fontsize{6}{0}\selectfont
The image features a cartoon tiger character standing on its hind legs, with a striped pattern on its body.}
\\[\myparskip]

\end{longtable}

\subsection{Examples of targeted input contains multiple elements on ImageNet dataset}
\label{app_multiple}

The targeted input is ``A dog is playing ball''. 
The figures in the first column are the image file $v$.
The second and third columns respectively show the generated image and text  under the Zhang attack. 
The fourth and fifth columns respectively display the generated image and text  under our \alg{}.

\begin{longtable}{p{4cm}p{4cm}p{4cm}p{4cm}p{4cm}}
\hspace*{-0.8cm}\parbox[c][\myrowheight][c]{\mycolwidth}{\vspace{-0.2cm}\includegraphics[width=\imgwidth, height=3.7cm]{appendix/ori_image/0.pdf}} & \hspace*{-1.4cm}
\parbox[c][\myrowheight][c]{\mycolwidth}{\includegraphics[width=\imgwidth, height=\imgheight]{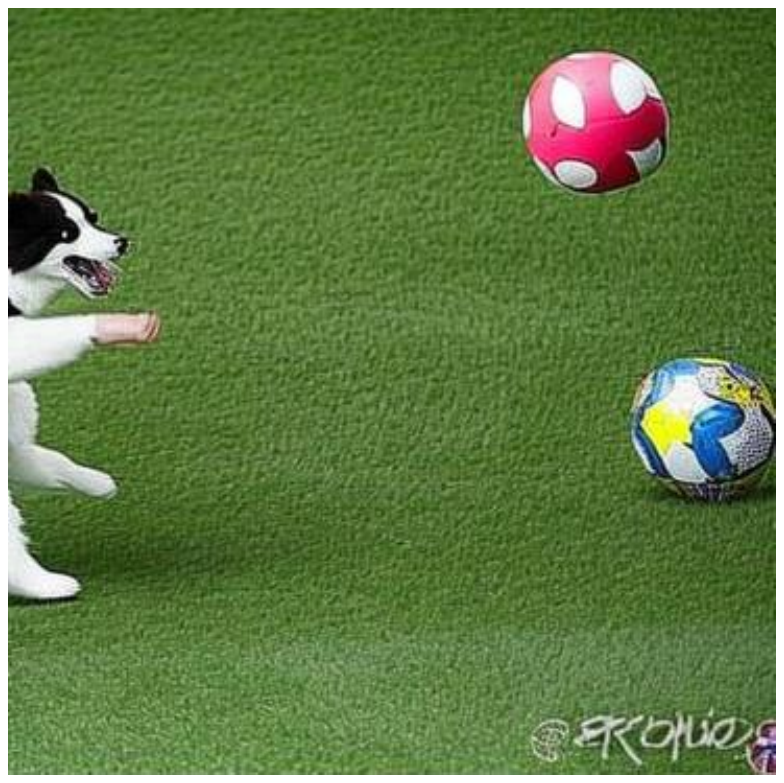}} & \hspace*{-1.6cm}
\parbox[c][0.5cm][c]{2.5cm}{\vspace{1cm}
\fontsize{6}{0}\selectfont
The image shows a dog playing with a frisbee on a grassy field. The dog is jumping in the air to catch the frisbee, which is flying through the air.} & \hspace*{-2.5cm}
\parbox[c][\myrowheight][c]{\mycolwidth}{\includegraphics[width=\imgwidth, height=\imgheight]{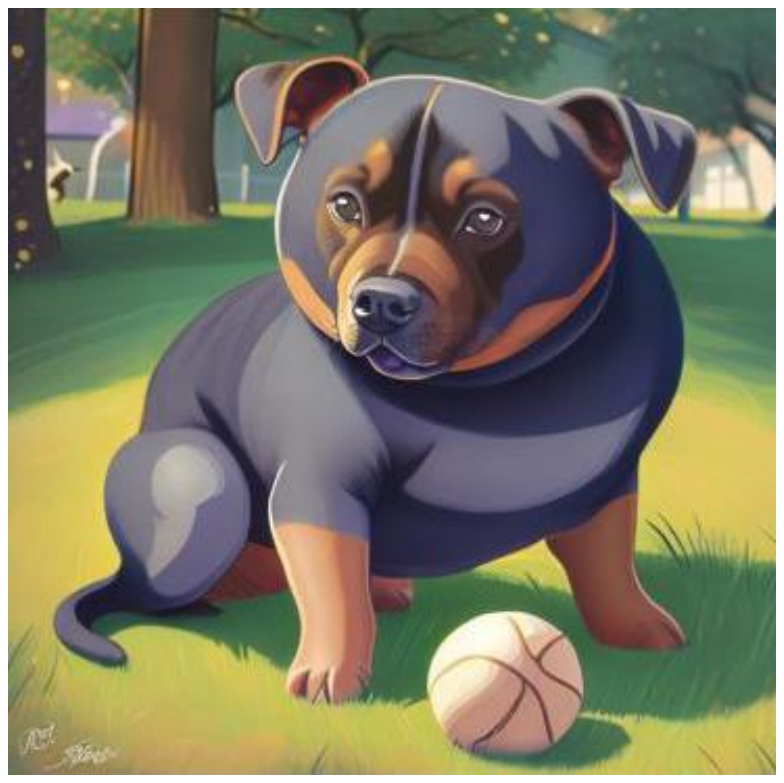}} & \hspace*{-2.5cm}
\parbox[c][0.5cm][c]{2.5cm}{\vspace{1cm}
\fontsize{6}{0}\selectfont
The image features a large brown dog sitting on the grass, holding a tennis ball in its mouth. The dog appears to be enjoying a game of fetch.}
\\[\myparskip]

\hspace*{-0.8cm}\parbox[c][\myrowheight][c]{\mycolwidth}{\vspace{-0.2cm}\includegraphics[width=\imgwidth, height=3.7cm]{appendix/ori_image/1.pdf}} & \hspace*{-1.4cm}
\parbox[c][\myrowheight][c]{\mycolwidth}{\includegraphics[width=\imgwidth, height=\imgheight]{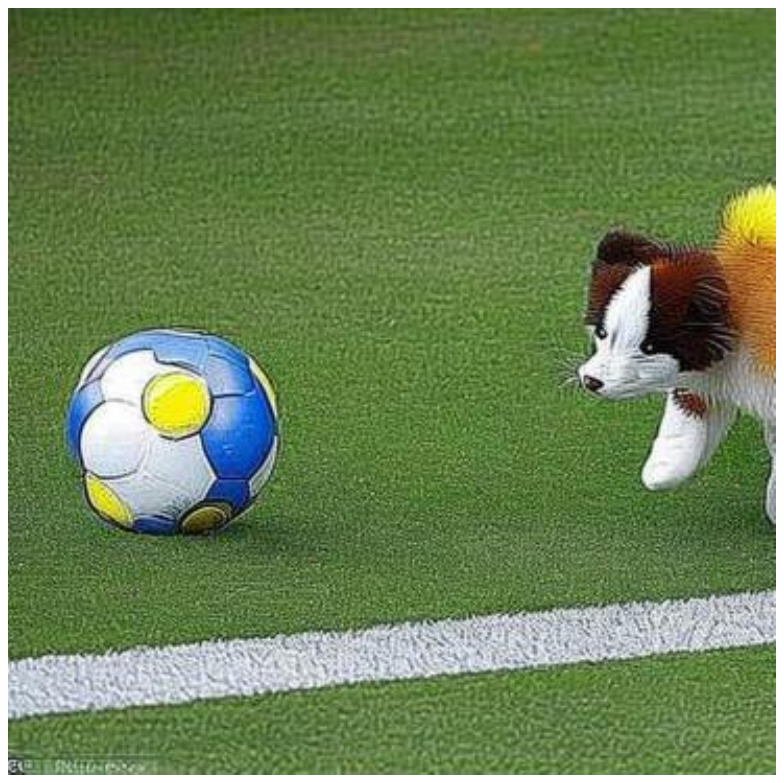}} & \hspace*{-1.6cm}
\parbox[c][0.5cm][c]{2.5cm}{\vspace{1cm}
\fontsize{6}{0}\selectfont
The image shows a dog playing with a frisbee in a yard. The dog is in the process of catching the frisbee, which is located in the upper right part of the scene. The dog appears to be in a playful and active state, enjoying the game.} & \hspace*{-2.5cm}
\parbox[c][\myrowheight][c]{\mycolwidth}{\includegraphics[width=\imgwidth, height=\imgheight]{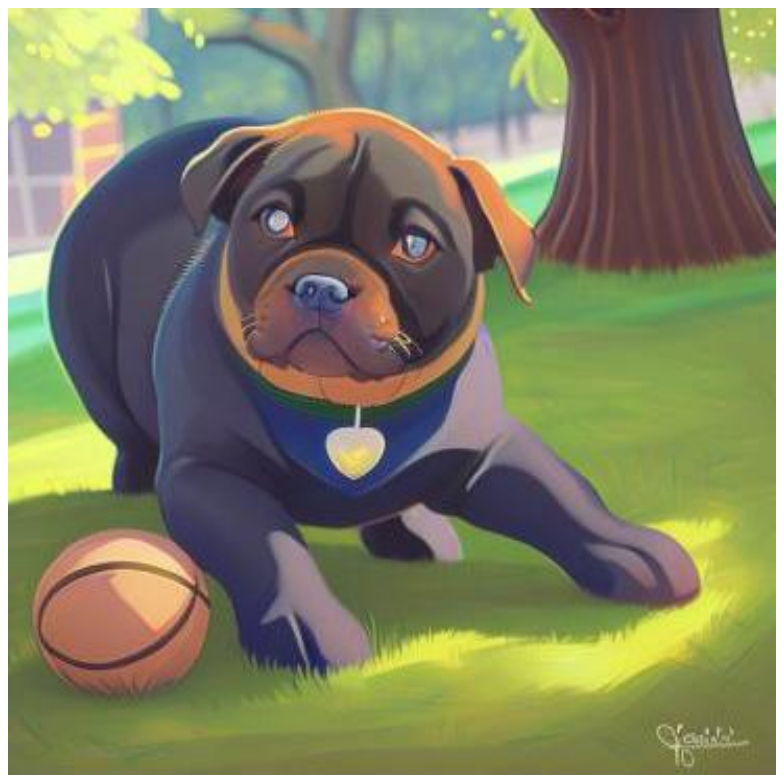}} & \hspace*{-2.5cm}
\parbox[c][0.5cm][c]{2.5cm}{\vspace{1cm}
\fontsize{6}{0}\selectfont
The image features a brown dog sitting on the grass, holding a tennis ball in its mouth. The dog appears to be enjoying its time outdoors, possibly playing with the tennis ball.}
\\[\myparskip]

\hspace*{-0.8cm}\parbox[c][\myrowheight][c]{\mycolwidth}{\vspace{-0.2cm}\includegraphics[width=\imgwidth, height=3.7cm]{appendix/ori_image/2.pdf}} & \hspace*{-1.4cm}
\parbox[c][\myrowheight][c]{\mycolwidth}{\includegraphics[width=\imgwidth, height=\imgheight]{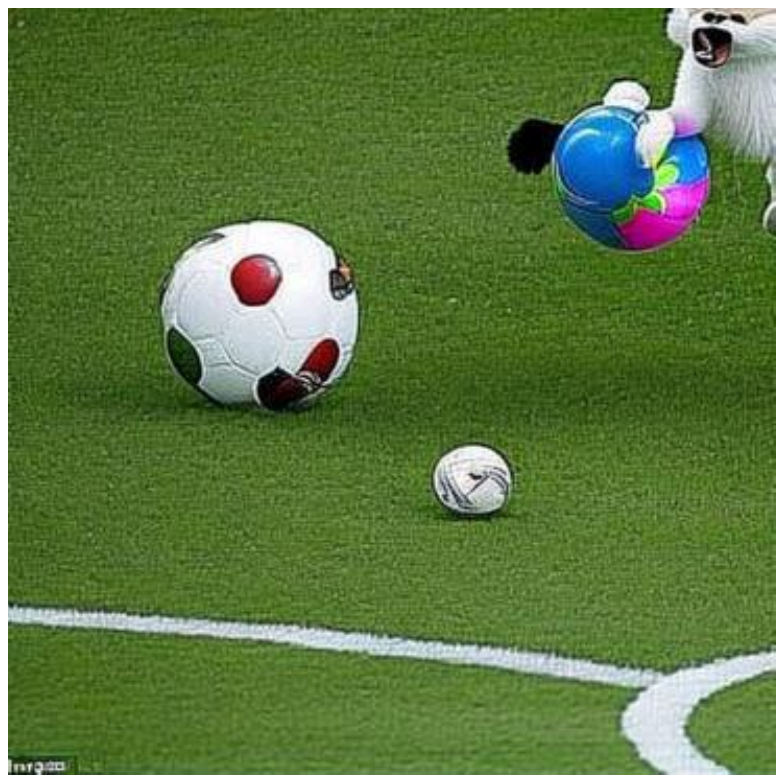}} & \hspace*{-1.6cm}
\parbox[c][0.5cm][c]{2.5cm}{\vspace{1cm}
\fontsize{6}{0}\selectfont
The image shows a dog playing with a frisbee in a yard. The dog is holding the frisbee in its mouth, and there is a person nearby, possibly playing with the dog or watching the game.} & \hspace*{-2.5cm}
\parbox[c][\myrowheight][c]{\mycolwidth}{\includegraphics[width=\imgwidth, height=\imgheight]{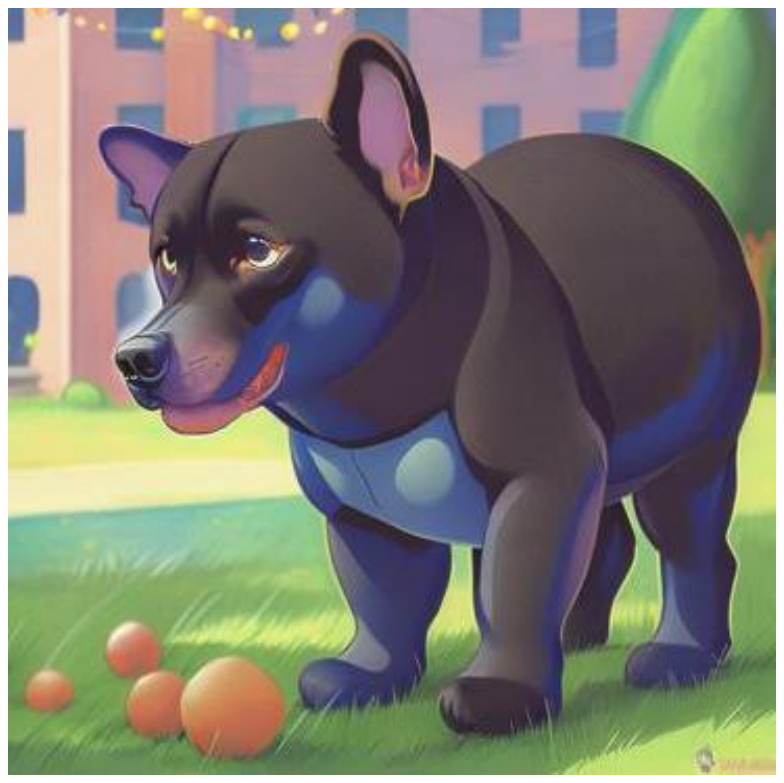}} & \hspace*{-2.5cm}
\parbox[c][0.5cm][c]{2.5cm}{\vspace{1cm}
\fontsize{6}{0}\selectfont
The image features a large brown dog sitting on the grass, holding a tennis ball in its mouth. The dog appears to be enjoying the outdoors and playing with the tennis ball.}
\\[\myparskip]

\hspace*{-0.8cm}\parbox[c][\myrowheight][c]{\mycolwidth}{\vspace{-0.2cm}\includegraphics[width=\imgwidth, height=3.7cm]{appendix/ori_image/3.pdf}} & \hspace*{-1.4cm}
\parbox[c][\myrowheight][c]{\mycolwidth}{\includegraphics[width=\imgwidth, height=\imgheight]{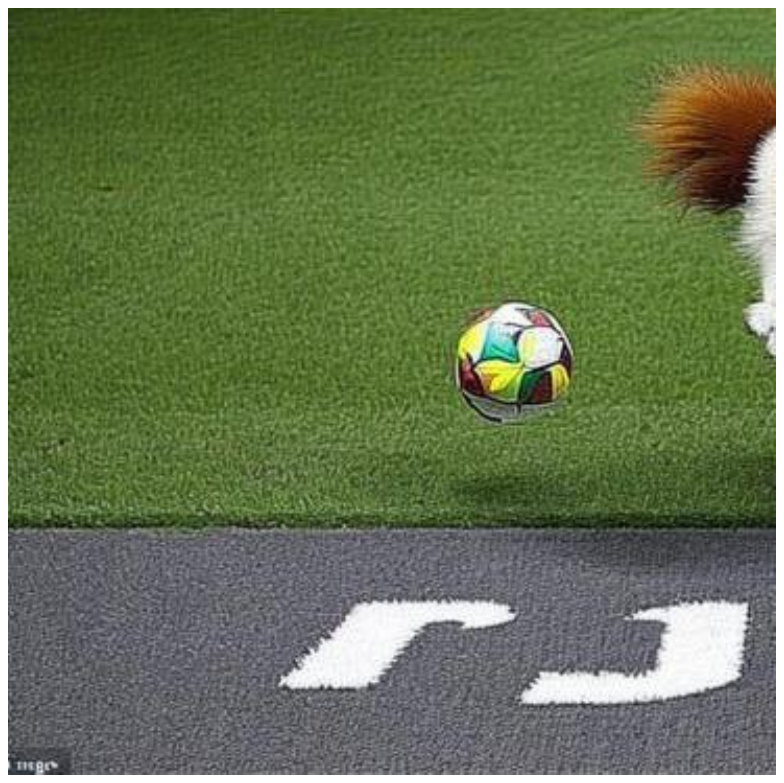}} & \hspace*{-1.6cm}
\parbox[c][0.5cm][c]{2.5cm}{\vspace{1cm}
\fontsize{6}{0}\selectfont
The image shows a dog playing with a frisbee on a grassy field. The dog is jumping in the air to catch the frisbee, which is located near the center of the image.} & \hspace*{-2.5cm}
\parbox[c][\myrowheight][c]{\mycolwidth}{\includegraphics[width=\imgwidth, height=\imgheight]{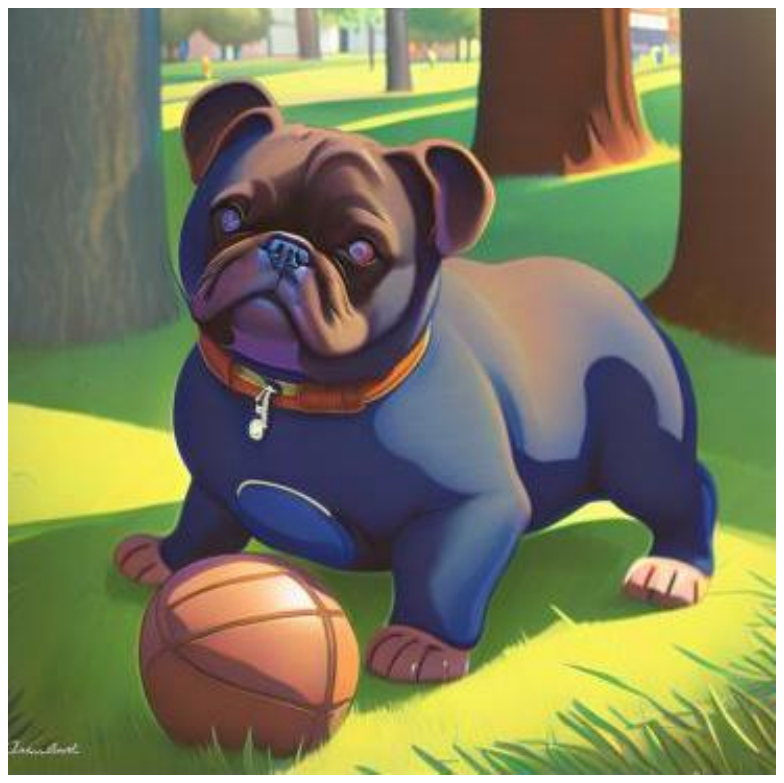}} & \hspace*{-2.5cm}
\parbox[c][0.5cm][c]{2.5cm}{\vspace{1cm}
\fontsize{6}{0}\selectfont
The image features a dog sitting on the grass, holding a tennis ball in its mouth.}
\\[\myparskip]

\hspace*{-0.8cm}\parbox[c][\myrowheight][c]{\mycolwidth}{\vspace{-0.2cm}\includegraphics[width=\imgwidth, height=3.7cm]{appendix/ori_image/4.pdf}} & \hspace*{-1.4cm}
\parbox[c][\myrowheight][c]{\mycolwidth}{\includegraphics[width=\imgwidth, height=\imgheight]{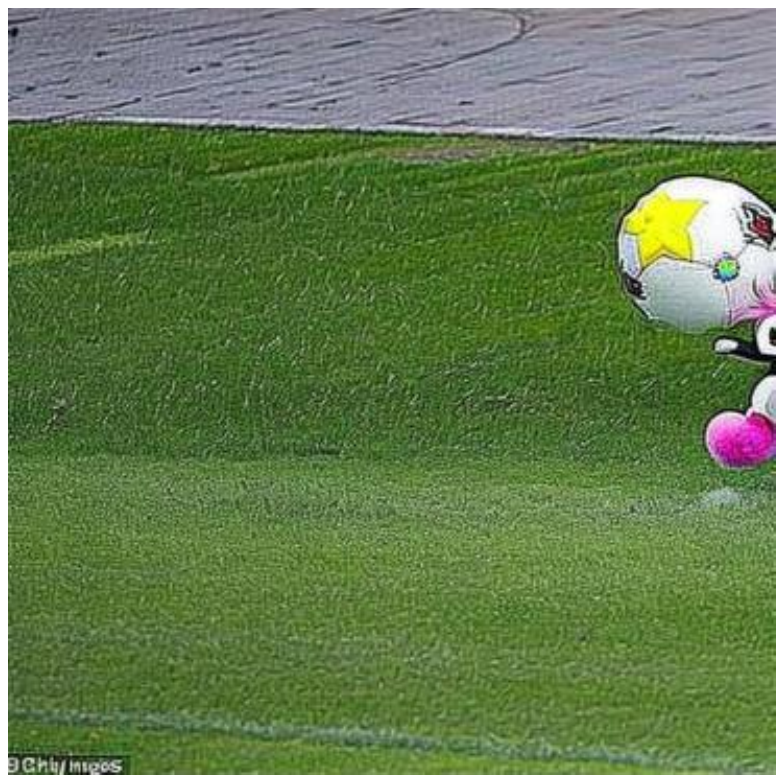}} & \hspace*{-1.6cm}
\parbox[c][0.5cm][c]{2.5cm}{\vspace{1cm}
\fontsize{6}{0}\selectfont
The image features a dog playing with a frisbee in a grassy area. The dog is holding the frisbee in its mouth, and it appears to be enjoying the game.} & \hspace*{-2.5cm}
\parbox[c][\myrowheight][c]{\mycolwidth}{\includegraphics[width=\imgwidth, height=\imgheight]{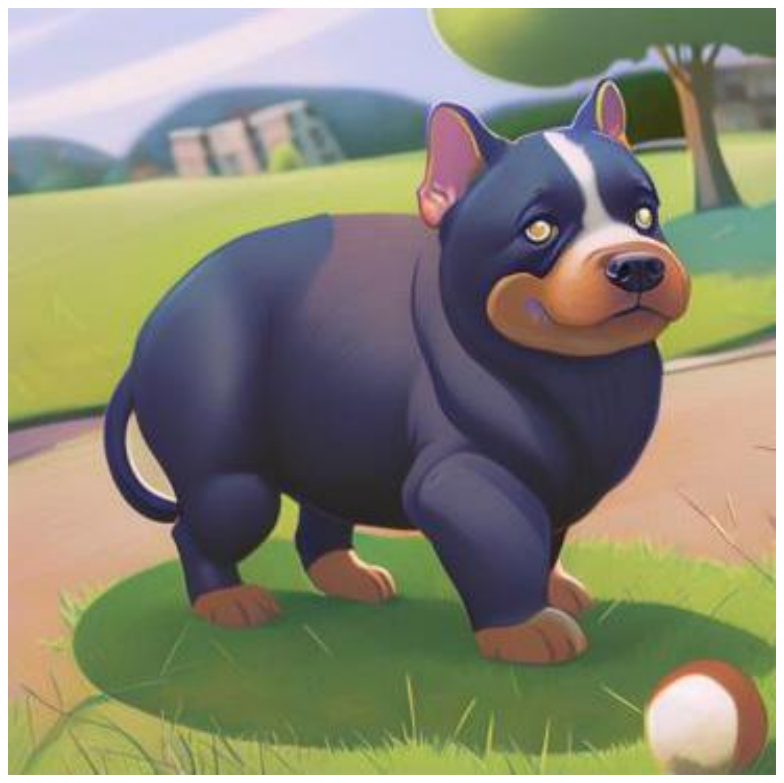}} & \hspace*{-2.5cm}
\parbox[c][0.5cm][c]{2.5cm}{\vspace{1cm}
\fontsize{6}{0}\selectfont
The image features a large brown dog sitting on the grass, holding a tennis ball in its mouth. The dog appears to be enjoying its time outdoors, possibly playing with the tennis ball.}
\\[\myparskip]

\hspace*{-0.8cm}\parbox[c][\myrowheight][c]{\mycolwidth}{\vspace{-0.2cm}\includegraphics[width=\imgwidth, height=3.7cm]{appendix/ori_image/5.pdf}} & \hspace*{-1.4cm}
\parbox[c][\myrowheight][c]{\mycolwidth}{\includegraphics[width=\imgwidth, height=\imgheight]{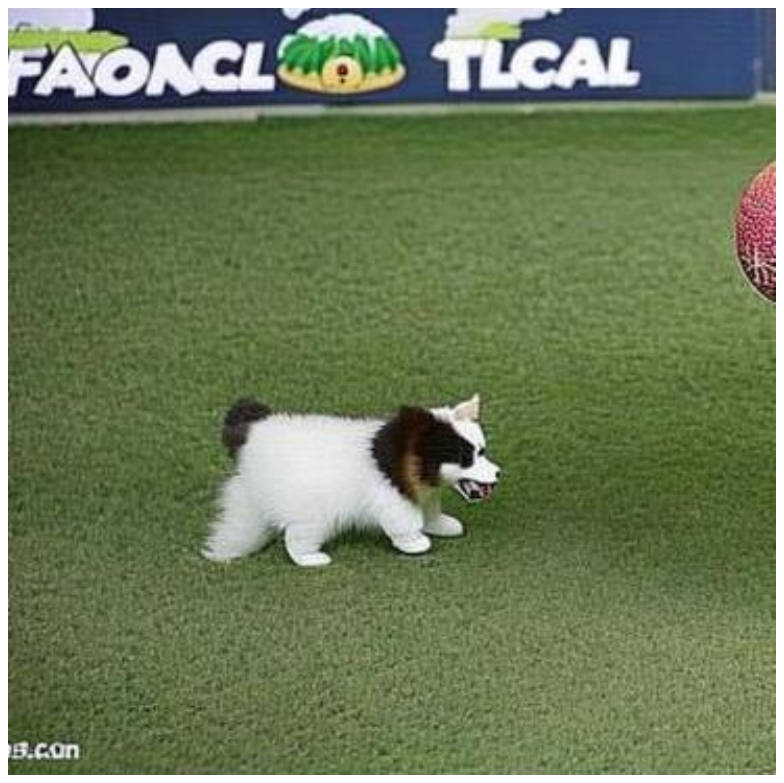}} & \hspace*{-1.6cm}
\parbox[c][0.5cm][c]{2.5cm}{\vspace{1cm}
\fontsize{6}{0}\selectfont
The image shows a dog playing with a frisbee on a grassy field. The dog is jumping up in the air to catch the frisbee, which is flying through the air.} & \hspace*{-2.5cm}
\parbox[c][\myrowheight][c]{\mycolwidth}{\includegraphics[width=\imgwidth, height=\imgheight]{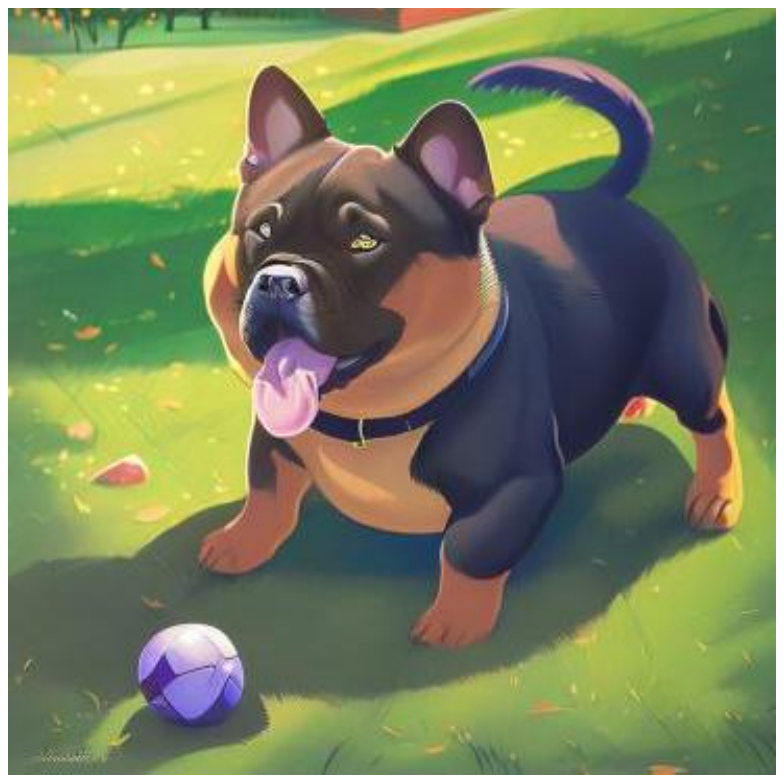}} & \hspace*{-2.5cm}
\parbox[c][0.5cm][c]{2.5cm}{\vspace{1cm}
\fontsize{6}{0}\selectfont
The image features a brown dog sitting on the grass, holding a tennis ball in its mouth. The dog appears to be enjoying a game of fetch.}
\\[\myparskip]

\hspace*{-0.8cm}\parbox[c][\myrowheight][c]{\mycolwidth}{\vspace{-0.2cm}\includegraphics[width=\imgwidth, height=3.7cm]{appendix/ori_image/6.pdf}} & \hspace*{-1.4cm}
\parbox[c][\myrowheight][c]{\mycolwidth}{\includegraphics[width=\imgwidth, height=\imgheight]{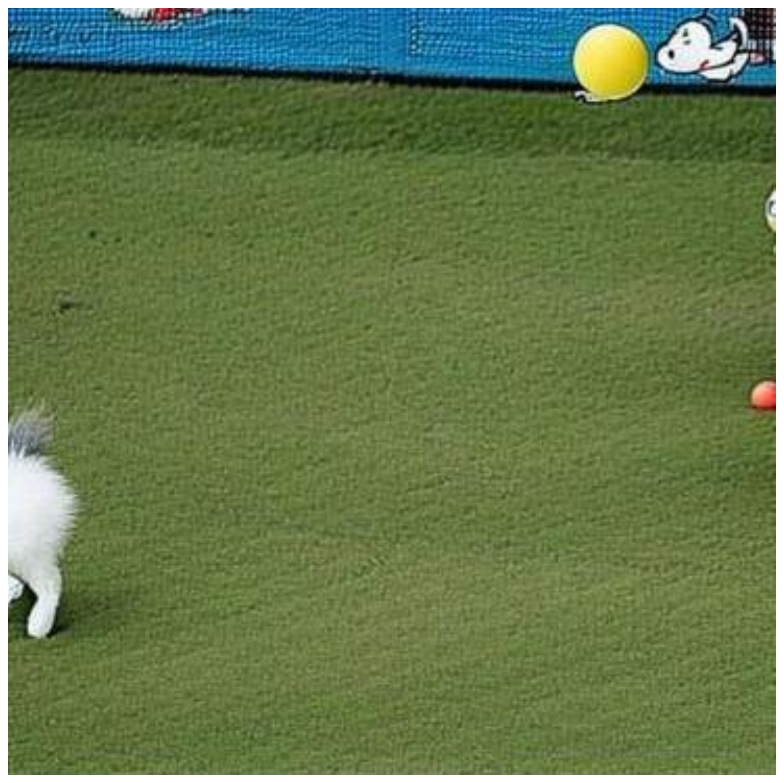}} & \hspace*{-1.6cm}
\parbox[c][0.5cm][c]{2.5cm}{\vspace{1cm}
\fontsize{6}{0}\selectfont
The image shows a dog playing with a frisbee in a grassy field. The dog is in the process of catching the frisbee, which is flying through the air.} & \hspace*{-2.5cm}
\parbox[c][\myrowheight][c]{\mycolwidth}{\includegraphics[width=\imgwidth, height=\imgheight]{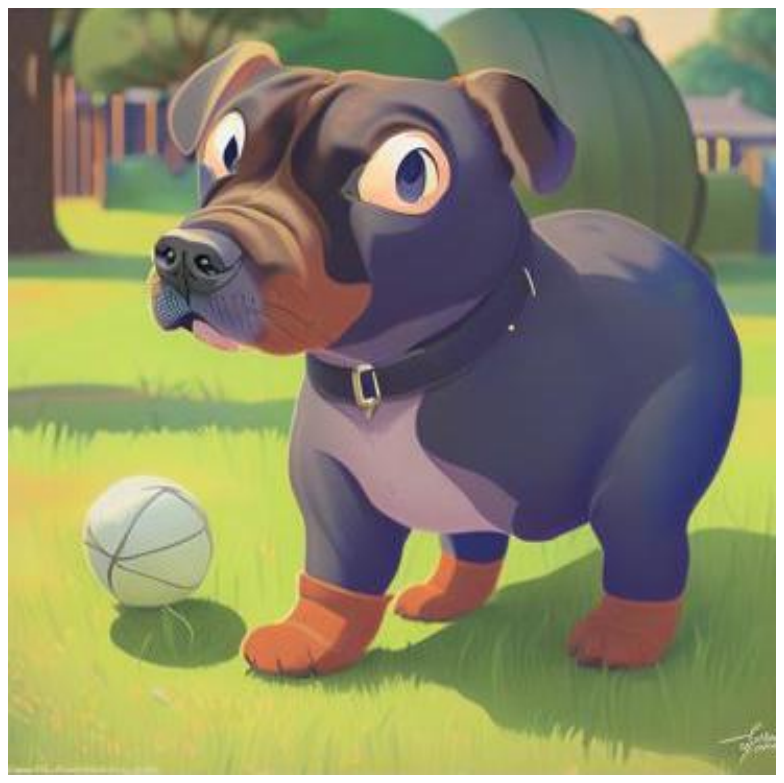}} & \hspace*{-2.5cm}
\parbox[c][0.5cm][c]{2.5cm}{\vspace{1cm}
\fontsize{6}{0}\selectfont
The image features a dog sitting on the grass, holding a tennis ball in its mouth. The dog appears to be enjoying a game of fetch, as it is holding the tennis ball with its mouth.}
\\[\myparskip]

\hspace*{-0.8cm}\parbox[c][\myrowheight][c]{\mycolwidth}{\vspace{-0.2cm}\includegraphics[width=\imgwidth, height=3.7cm]{appendix/ori_image/7.pdf}} & \hspace*{-1.4cm}
\parbox[c][\myrowheight][c]{\mycolwidth}{\includegraphics[width=\imgwidth, height=\imgheight]{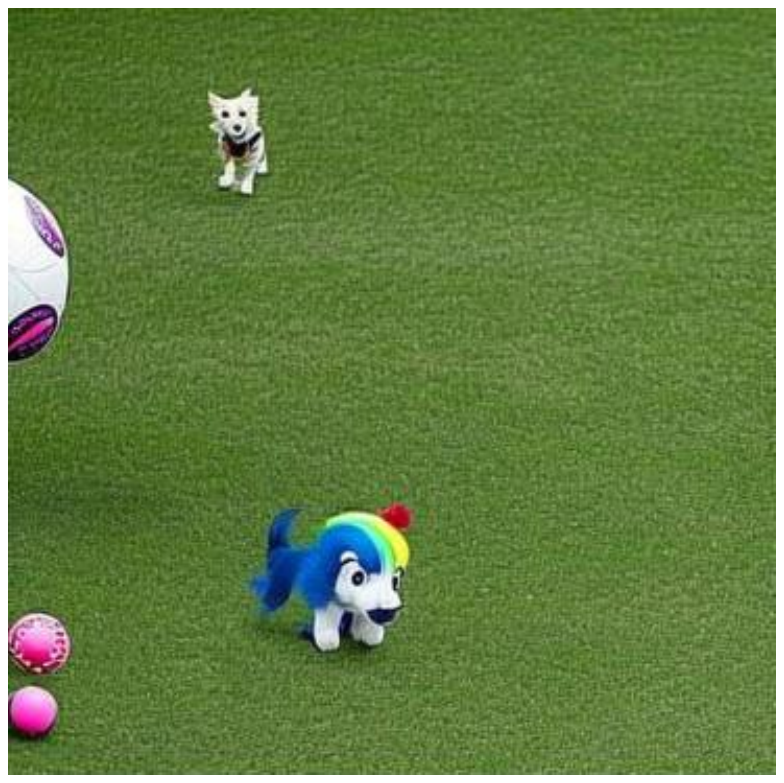}} & \hspace*{-1.6cm}
\parbox[c][0.5cm][c]{2.5cm}{\vspace{1cm}
\fontsize{6}{0}\selectfont
The image features a dog playing with a ball on a grassy field. The dog is running with the ball in its mouth, and there is a person in the background watching the dog's playful activity.} & \hspace*{-2.5cm}
\parbox[c][\myrowheight][c]{\mycolwidth}{\includegraphics[width=\imgwidth, height=\imgheight]{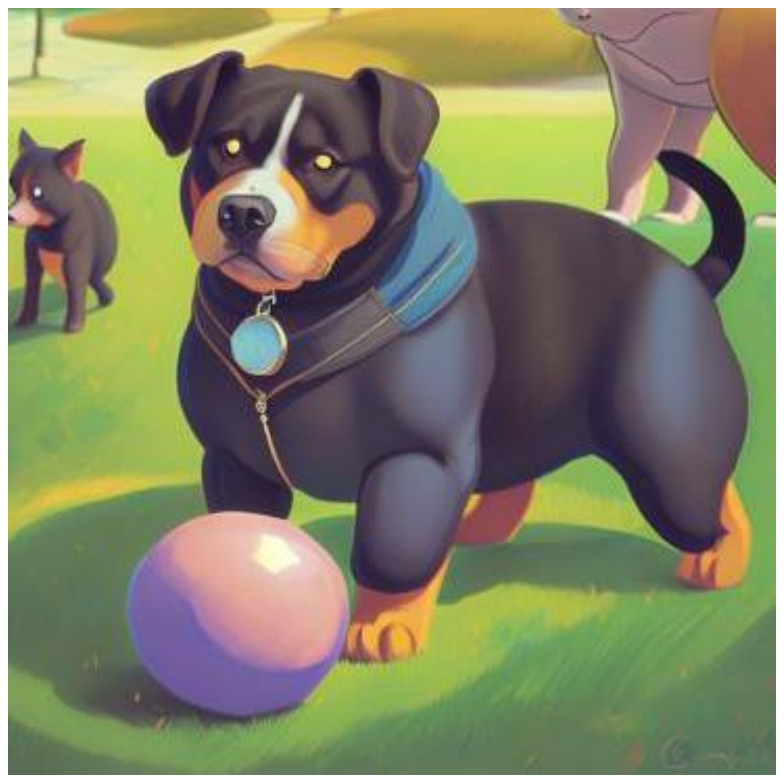}} & \hspace*{-2.5cm}
\parbox[c][0.5cm][c]{2.5cm}{\vspace{1cm}
\fontsize{6}{0}\selectfont
The image features a brown dog sitting on the grass, holding a tennis ball in its mouth.}
\\[\myparskip]

\hspace*{-0.8cm}\parbox[c][\myrowheight][c]{\mycolwidth}{\vspace{-0.2cm}\includegraphics[width=\imgwidth, height=3.7cm]{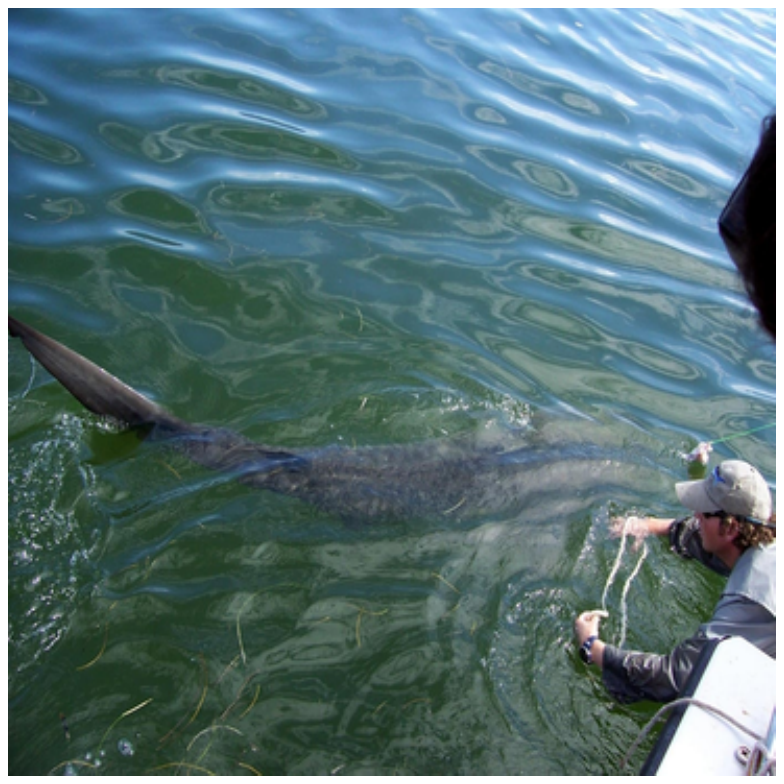}} & \hspace*{-1.4cm}
\parbox[c][\myrowheight][c]{\mycolwidth}{\includegraphics[width=\imgwidth, height=\imgheight]{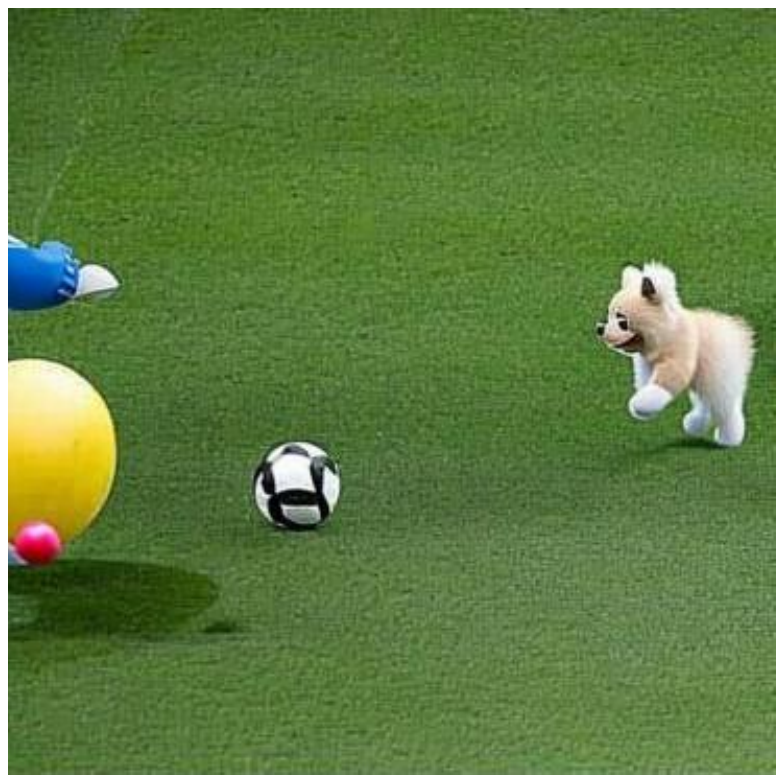}} & \hspace*{-1.6cm}
\parbox[c][0.5cm][c]{2.5cm}{\vspace{1cm}
\fontsize{6}{0}\selectfont
The image features a dog playing with a ball on a grassy field. The dog is jumping up to catch the ball, which is located near the center of the image.} & \hspace*{-2.5cm}
\parbox[c][\myrowheight][c]{\mycolwidth}{\includegraphics[width=\imgwidth, height=\imgheight]{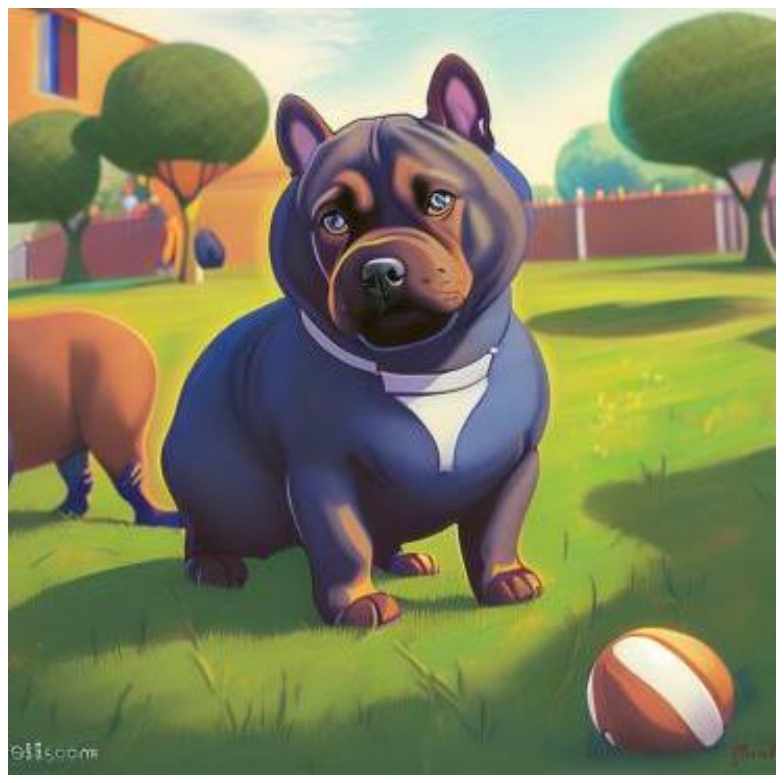}} & \hspace*{-2.5cm}
\parbox[c][0.5cm][c]{2.5cm}{\vspace{1cm}
\fontsize{6}{0}\selectfont
The image features a brown dog sitting on the grass, holding a tennis ball in its mouth. The dog appears to be enjoying a game of fetch or simply posing with the tennis ball.}
\\[\myparskip]

\hspace*{-0.8cm}\parbox[c][\myrowheight][c]{\mycolwidth}{\vspace{-0.2cm}\includegraphics[width=\imgwidth, height=3.7cm]{appendix/ori_image/9.pdf}} & \hspace*{-1.4cm}
\parbox[c][\myrowheight][c]{\mycolwidth}{\includegraphics[width=\imgwidth, height=\imgheight]{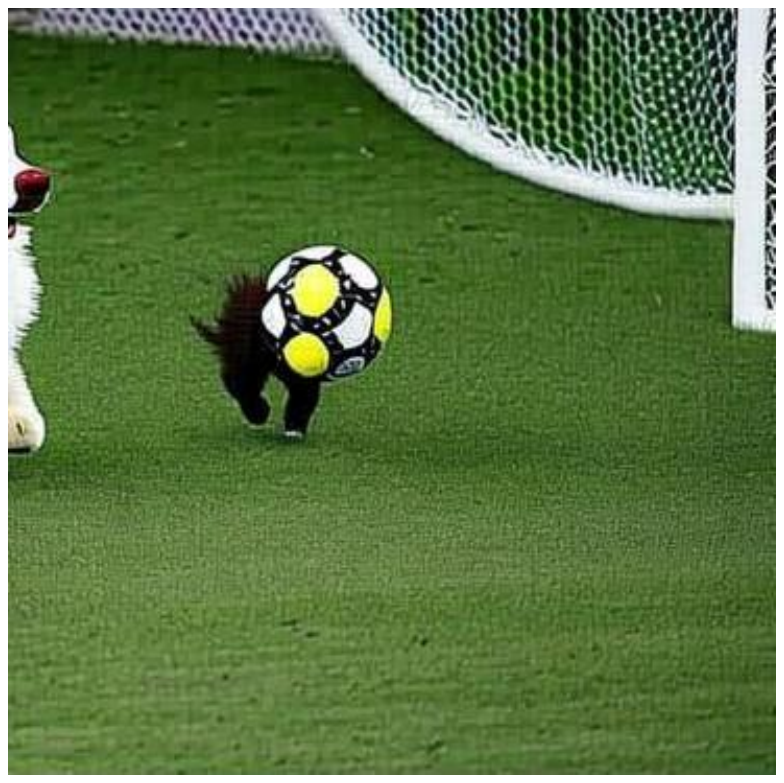}} & \hspace*{-1.6cm}
\parbox[c][0.5cm][c]{2.5cm}{\vspace{1cm}
\fontsize{6}{0}\selectfont
The image shows a dog playing with a frisbee on a grassy field. The dog is jumping in the air to catch the frisbee, which is flying through the air.} & \hspace*{-2.5cm}
\parbox[c][\myrowheight][c]{\mycolwidth}{\includegraphics[width=\imgwidth, height=\imgheight]{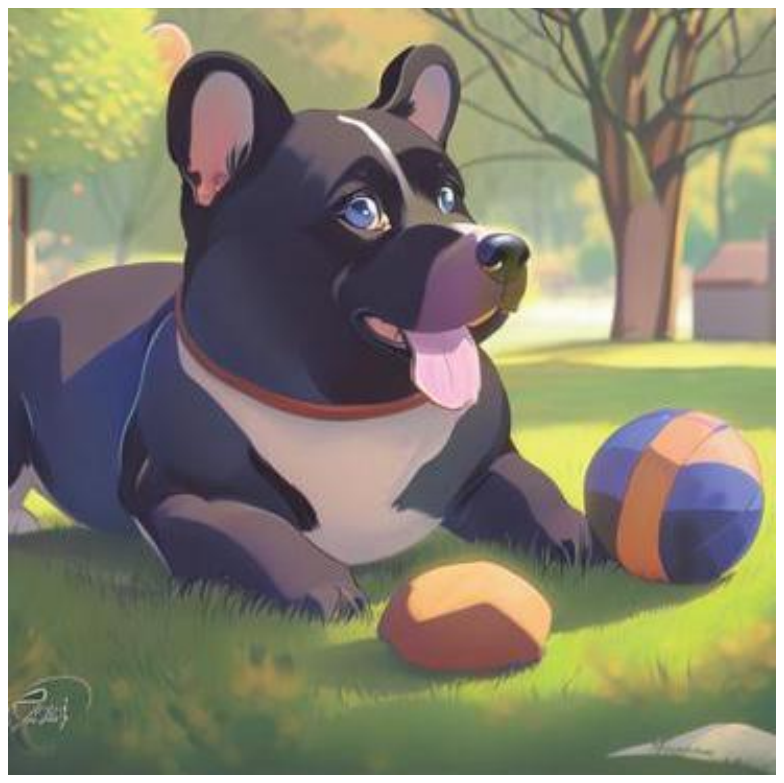}} & \hspace*{-2.5cm}
\parbox[c][0.5cm][c]{2.5cm}{\vspace{1cm}
\fontsize{6}{0}\selectfont
The image features a large brown dog sitting on the grass, holding a ball in its mouth. The dog appears to be enjoying a game of fetch.}
\\[\myparskip]

\end{longtable}
\onecolumn

\end{document}